\documentclass[a4paper,12pt]{article}
\usepackage{amsfonts}
\usepackage{latexsym}
\usepackage{amsmath}
\usepackage{amssymb}
\usepackage{amssymb}
\usepackage{slashed}
\usepackage{upgreek}
\usepackage{mathrsfs}
\usepackage{bbm}

\usepackage{amsthm}

\theoremstyle{remark}

\numberwithin{equation}{section}

\usepackage[utf8]{inputenc}
\usepackage[english]{babel}

\usepackage{amsmath}
\usepackage{amssymb} 
\usepackage{slashed}
\usepackage{mathtools} 
\usepackage{dsfont} 
\usepackage{xcolor}

\usepackage[toc,page]{appendix}

\renewcommand{\*}[1]{{\,^* \!#1}} 
\newcommand{\D}[2]{\left \langle #1, #2 \right \rangle_D} 

\newcommand{\pp}{=\kern-0.40em{\vert}}

\theoremstyle{definition}

\newcommand*{\defeq}{\mathrel{\vcenter{\baselineskip0.5ex \lineskiplimit0pt
                     \hbox{\scriptsize.}\hbox{\scriptsize.}}}%
                     =}

\allowdisplaybreaks

\usepackage{relsize}
\usepackage{graphicx}
\usepackage{comment}

\renewcommand{\thefootnote}{\fnsymbol{footnote}}

\def\appendix#1{\addtocounter{section}{1}\setcounter{equation}{0}
\renewcommand{\thesection}{\Alph{section}}
\section*{Appendix \thesection\protect\indent \parbox[t]{11.15cm}{#1}}
\addcontentsline{toc}{section}{Appendix \thesection\ \ \ #1}}

\font\mybb=msbm10 at 11pt

\def\bb#1{\hbox{\mybb#1}}

\def\bR {\bb{R}}

\def\bC {\bb{C}}

\newcommand{\bea}{\begin{eqnarray}}
\newcommand{\eea}{\end{eqnarray}}
\newcommand{\nn}{\nonumber \\}

\def\rep{\operatorname{Re}}
\def\imp{\operatorname{Im}}

\def\f5{f_{(5)}} 
\def\tf5{\tilde{f}_{(5)}}
\def\g4{f_{(4)}} 
\def\tg4{\tilde{f}_{(4)}}

\usepackage[normalem]{ulem}				
\usepackage{microtype}
\usepackage[colorlinks, citecolor=blue, linkcolor=blue, urlcolor=Maroon, filecolor=Maroon, linktocpage=true]{hyperref} 

\usepackage{chngcntr}

\begin{document}

\begin{center}
\vspace*{-1.0cm}
\begin{flushright}
\end{flushright}


\vspace{2.0cm} {\Large \bf TCFHs, hidden symmetries  and  type II theories} \\[.2cm]

\vskip 2cm
 L. Grimanellis, G. Papadopoulos and J. Phillips
\\
\vskip .6cm

\begin{small}
\textit{Department of Mathematics
\\
King's College London
\\
Strand
\\
 London WC2R 2LS, UK}\\
\texttt{loukas.grimanellis@kcl.ac.uk}
\\
\texttt{george.papadopoulos@kcl.ac.uk}
\\
\texttt{jake.phillips@kcl.ac.uk}
\end{small}
\\*[.6cm]

\end{center}

\vskip 2.5 cm

\begin{abstract}
\noindent
We present the  twisted covariant form hierarchies (TCFH) of type IIA and IIB 10-dimensional supergravities and show that all form bilinears of supersymmetric backgrounds satisfy the conformal Killing-Yano equation with respect to a TCFH connection.  We also compute the Killing-St\"ackel, Killing-Yano and closed conformal Killing-Yano tensors of all spherically symmetric type II brane backgrounds and demonstrate that the geodesic flow on these solutions is completely integrable by giving all  independent charges in involution. We then identify all form bilinears of common sector and D-brane backgrounds which generate hidden symmetries for particle and string probe actions. We also explore the question on whether  charges constructed from form bilinears are sufficient to prove the integrability
of probes on supersymmetric backgrounds.

\end{abstract}

\vskip 1.5 cm


\newpage

\renewcommand{\thefootnote}{\arabic{footnote}}


\section{Introduction}

It has been known for sometime  that some gravitational backgrounds admit Killing-St\"ackel (KS)  and Killing-Yano (KY) tensors, see \cite{carter-b}-\cite{lun}, the reviews \cite{revky} and \cite{frolov} and the references within.  These are used to demonstrate the separability and integrability of  classical equations, such as the geodesic, Hamilton-Jacobi and Dirac  equations, on these backgrounds. A key property of  KS tensors is that they generate  hidden symmetries for   relativistic particles  while KY tensors generate  hidden symmetries \cite{gibbons} for spinning particles \cite{bvh} propagating on gravitational backgrounds.

It has been shown in \cite{gptcfh} that the conditions imposed by the gravitino Killing spinor equation (KSE)   on the (Killing spinor) form bilinears can be arranged as a twisted covariant form hierarchy (TCFH) \cite{jggp}.  This means that there is a connection, ${\cal D}^{\cal F}$, on the space of spacetime forms which depends on the fluxes, ${\cal F}$, of the theory such that the highest weight representation of ${\cal D}^{\cal F}\Omega$ vanishes, where $\Omega$ is a collection of forms of various degrees and ${\cal D}^{\cal F}$ may not be form degree preserving. Equivalently, this condition can be written  as
\bea
{\cal D}_X^{\cal F}\Omega= i_X {\cal P}+ X\wedge {\cal Q}~,
\label{tcfheqn}
\eea
for every spacetime vector field $X$, where  ${\cal P}$ and ${\cal Q}$  are  appropriate multi-forms and $X$ also denotes  the associated 1-form constructed from the vector field $X$ after using  the spacetime metric $g$, $X(Y)=g(X,Y)$. The proof of this result is rather general and includes supergravities on spacetimes of any signature as well as the effective theories of strings which include higher  order curvature corrections. It also puts the conditions imposed by the KSEs on the form bilinears on a firm geometric basis.

One consequence of the TCFH is that the form bilinears satisfy a generalisation of the conformal Killing-Yano (CKY) equation  with respect to the
connection ${\cal D}^{\cal F}$. This can be easily seen after taking the skew-symmetric part and contraction  with respect the metric $g$ of (\ref{tcfheqn}), and so one
identifies ${\cal P}$ with an exterior derivative constructed from ${\cal D}^{\cal F}$ and ${\cal Q}$ with a formal adjoint of ${\cal D}^{\cal F}$ acting on $\Omega$. This raises the
question on whether  the form bilinears  generate hidden symmetries for worldvolume actions which describe the propagation of certain  probes  in supersymmetric backgrounds. This question was first investigated in the context of 5- and 4-dimensional supergravities in \cite{gpeb}.

 The purpose of this paper is twofold. One  is to  present the TCFHs of IIA and IIB supergravities and to discuss some of the properties of the TCFH connections ${\cal D}^{\cal F}$, like for example their holonomy, on generic as well as on some special supersymmetric backgrounds. As a consequence we demonstrate that the form bilinears of these theories satisfy a CKY equation with respect to ${\cal D}^{\cal F}$ in agreement with the general result of \cite{gptcfh}. Another purpose of this paper is to give the KS tensors  of type II branes\footnote{Brane solutions have been instrumental in the understanding of string dualities \cite{hulltown, town}.} \cite{funstring}-\cite{d8} and to use them to prove the complete integrability of the geodesic flow of those solutions that are spherically symmetric, i.e. those that depend on a harmonic function with one centre. In addition, the KY tensors that square to the KS tensors  of these backgrounds will be given and the symmetries of spinning particles propagating on these backgrounds will be explored. Furthermore we shall investigate the conditions required for the TCFH to yield symmetries for particle and string probes propagating in common sector and D-brane backgrounds. Finally we shall compare the results we have obtained from the point of view of KS and KY tensors with those that arise from the TCFHs.

To investigate under which conditions the form bilinears  generate symmetries for certain probe actions propagating in type II supersymmetric backgrounds, we shall match the conditions required for certain probe actions to be invariant under transformations generated by form bilinears with those imposed on them by the TCFHs. For the common sector of type II theories, it is shown that all form bilinears which are covariantly constant with respect to a connection with  torsion given by the NS-NS 3-form field strength generate symmetries for string and spinning particle probes propagating on these backgrounds. Common sector backgrounds also admit form bilinears which are not covariantly constant and instead  satisfy a general TCFH. These form bilinears may not generate symmetries for probes propagating in common sector backgrounds but nevertheless are part of their geometric structure. In particular the form bilinears of the fundamental string  and NS5-brane solutions that are allowed to depend on multi-centre harmonic functions have been computed. It has been found that the type II fundamental string solution admits  $2^7$ covariantly constant independent form bilinears while the type II NS5-brane solution admits $2^5$ covariantly constant independent form bilinears. All these forms generate (hidden) symmetries for probe string and spinning particle actions propagating on these backgrounds.

A similar analysis is presented for all type II D-branes. In particular, the form bilinears of all D-branes are computed. It is found that the requirement for these to generate symmetries for spinning particle probes propagating on these backgrounds is rather restrictive. This is due to the difficulties of constructing probe actions which exhibit appropriate form couplings.  Nevertheless all type II D-branes, which may depend on multi-centre harmonic functions,  admit form bilinears which generate symmetries for spinning particle probe actions. It turns out that all such form bilinears have components only along the worldvolume directions of the D-branes. A comparison of the symmetries we have found  generated by the KS and KY tensors  and those generated by the form bilinears in type II brane backgrounds will be presented in the conclusions.

This paper is organised as follows. In section 2 and 3, we give the TCFHs of IIA and IIB supergravities and discuss some of the properties of the TCFH connections.
In section 4, after a summary of the properties of the KS and KY tensors, we present the KS and KY tensors of all type II branes.  In addition,  we prove the complete integrability of the geodesic flow in all type II branes that depend on a harmonic function with one centre by presenting all the independent conserved charges which are in involution. In section 5, we demonstrate that all covariantly constant  form bilinears with respect to a connection with skew-symmetric torsion generate symmetries for certain probe string and particle actions propagating on common sector backgrounds. In addition,  we explicitly give  all the
covariantly constant  form bilinears for the type II fundamental string and NS5-brane solutions. In sections 6 and 7, we identify the form bilinears which
generate symmetries for spinning particle actions propagating on type II D-brane backgrounds. In section 8, we give our conclusions. In appendices \ref{apa}
and \ref{apb}, we give all the form bilinears of type II common sector branes and type II D-branes, respectively.

\section{The TCFH of (massive) IIA  supergravity }\label{iiatcfhs}

The KSEs of massive IIA supergravity \cite{romans} are given by the vanishing conditions of the supersymmetry variations of the gravitino and dilatino fields evaluated at the locus that all fermions are set to zero. The KSE associated with the gravitino field is a parallel transport equation for the supercovariant connection
$\mathcal{D}$. In the string frame, this is given by
\begin{equation}
	\begin{split}
		\mathcal{D}_M  \defeq \nabla_M &+ \frac{1}{8}H_{MPQ}\Gamma^{PQ}\Gamma_{11} + \frac{1}{8}e^\Phi S \Gamma_M \\ &+ \frac{1}{16}e^\Phi F_{PQ} \Gamma^{PQ}\Gamma_M\Gamma_{11}+\frac{1}{8\cdot 4!}e^\Phi G_{P_1 \dots P_4} \Gamma^{P_1\dots P_4}\Gamma_M~,
\label{iiasc}
	\end{split}	
\end{equation}
see e.g. \cite{eric}, where $H$ is the NS-NS 3-form field strength, $\Phi$ is the dilaton, and $F$ and $G$ are the 2-form and 4-form R-R field strengths, respectively. In addition, $\nabla$ is the Levi-Civita connection induced on the spinor bundle and $S = e^\Phi m$, where $m$ is a constant which is non-zero in  massive IIA and vanishes in the standard IIA supergravity.   Furthermore $\Gamma$ denotes the  gamma matrices which satisfy the Clifford algebra relation $\Gamma_A\Gamma_B+\Gamma_B \Gamma_A=2 \eta_{AB}$ and in our conventions $\Gamma_{11} \defeq -\Gamma_{012\dots 9}$. In what follows, we shall not make a sharp distinction between spacetime and frame indices but we shall always assume that the indices of gamma matrices are frame indices. It turns out that $\mathcal{D}$ is a connection on the spin bundle over the spacetime associated with the Majorana (real) representation of $\mathfrak{spin} (9,1)$. The (reduced) holonomy of ${\cal D}$ for generic backgrounds is $SL(32, \bR)$ \cite{gpdt}, see \cite{hull, duff, gpdtx} for the computation of the holonomy of the supercovariant derivative of 11-dimensional supergravity.

The Killing spinors $\epsilon$ satisfy the gravitino KSE, $\mathcal{D}\epsilon=0$,  as well as the dilatino KSE which is an algebraic equation.  Backgrounds that admit such Killing  spinors are special and both the spacetime metric and fluxes are suitably restricted, see \cite{gpug} where the IIA KSEs have been solved for one Killing spinor. The TCFHs are associated with the gravitino KSE which we shall focus on in what follows.

 Given $N$ Killing spinors $\epsilon^r$, $r=1,\dots, N$, one can construct the form bilinears
 \bea
 \phi^{rs}={1\over k!}\langle \epsilon^r, \Gamma_{A_1\dots A_k}\epsilon^s\rangle_D\, e^{A_1}\wedge\dots\wedge e^{A_k}~,
 \label{fbil}
 \eea
 where $\langle\cdot, \cdot\rangle_D$ denotes that Dirac inner product and $e^A$ is a suitable spacetime frame, $g_{MN}= \eta_{AB} e^A_M e^B_N$.  As
 \bea
 \nabla_M \phi^{rs}_{A_1\dots A_k}=\langle \nabla_M \epsilon^r, \Gamma_{A_1\dots A_k}\epsilon^s\rangle_D+\langle  \epsilon^r, \Gamma_{A_1\dots A_k} \nabla_M\epsilon^s\rangle_D~,
 \eea
 one can use the gravitino KSE, $\mathcal{D}\epsilon=0$, and (\ref{iiasc}) to express the right-hand side of the above equation in terms of the fluxes and form bilinears of the theory.  In \cite{gptcfh} has been shown that these equations can be organised as TCFH.

 Using the reality condition on $\epsilon$, there are form  bilinears which are  either symmetric or skew-symmetric in the exchange
 of spinors $\epsilon^r$ and $\epsilon^s$ in  (\ref{fbil}).  As a consequence the TCFH of the IIA supergravity factorises in two parts.
  A basis in  form bilinears, up to a Hodge duality\footnote{Our convention for the Hodge duality operation is
  ${}^\star{\omega}_{N_1 \dots N_{n-p}} = \frac{1}{p!}\omega_{P_1\dots P_p}\epsilon^{P_1\dots P_p}{}_{N_1\dots N_{n-p}}$ with $\epsilon_{012\dots (n-1)} = -1$, where $n$ is the spacetime dimension.}
   operation,  which are symmetric in the exchange of the two Killing spinors $\epsilon^r$ and $\epsilon^s$ is
\bea
	&&	\tilde{\sigma}^{rs} = \D{\epsilon^r}{\Gamma_{11}\epsilon^s} , \quad k^{rs} = \D{\epsilon^r}{\Gamma_N\epsilon^s} \, e^N , \quad \tilde{k} = \D{\epsilon^r}{\Gamma_N\Gamma_{11}\epsilon^s} \, e^N , \qquad\qquad  \cr
&&
		\omega^{rs} = \frac{1}{2} \D{\epsilon^r}{\Gamma_{NR} \epsilon^s} \,e^N \wedge e^R, \quad
		\tilde{\zeta}^{rs} = \frac{1}{4!} \D{\epsilon^r}{\Gamma_{N_1 \dots N_4}\Gamma_{11} \epsilon^s}\, e^{N_1} \wedge \dots \wedge e^{N_4},
\cr
		&&
\tau^{rs} = \frac{1}{5!} \D{\epsilon^r}{\Gamma_{N_1 \dots N_5}\epsilon^s}\, e^{N_1} \wedge \dots \wedge e^{N_5}~.
\label{symiia}
\eea
A direct computation reveals that the TCFH is
\begin{equation}
{\cal D}^{\cal F}_M\tilde\sigma\defeq	\nabla_M \tilde{\sigma} = -\frac{1}{4}H_{MPQ}\omega^{PQ} + \frac{1}{4}e^\Phi S \tilde{k}_M - \frac{1}{4} e^\Phi F_{MP}k^P - \frac{1}{4 \cdot 5!}{}^\star{G}_{MP_1\dots P_5}\tau^{P_1\dots P_5}~,
\label{iiatcfha}
\end{equation}

\bea
&&{\cal D}_M^{\cal F} k_N \defeq		\nabla_M k_N = -\frac{1}{2}H_{MNP}\tilde{k}^P + \frac{1}{4}e^\Phi S \omega_{MN} + \frac{1}{8} e^\Phi F_{PQ}\tilde{\zeta}^{PQ}{}_{MN} +\frac{1}{4}e^\Phi F_{MN}\tilde{\sigma}  \cr
		&&\qquad+\frac{1}{4\cdot 4!}e^\Phi {}^\star{G}_{MNP_1\dots P_4}\tilde{\zeta}^{P_1 \dots P_4} + \frac{1}{8}e^\Phi G_{MNPQ}\omega^{PQ}~,
\label{iiatcfhb}
\eea

\bea
&&{\cal D}_M^{\cal F}\tilde k_N\defeq		\nabla_M \tilde{k}_N - \frac{1}{2} e^\Phi F_{MP}\omega^P{}_N- \frac{1}{12}e^\Phi G_{MPQR}\tilde{\zeta}^{PQR}{}_N= -\frac{1}{2}H_{MNP}k^P   \cr
			&&\qquad+ \frac{1}{4}e^\Phi g_{MN}S\tilde{\sigma} +\frac{1}{8}e^\Phi g_{MN} F_{PQ}\omega^{PQ} -\frac{1}{2}e^\Phi F_{[M|P|}\omega^P{}_{N]}
\cr
&&\qquad+ \frac{1}{4\cdot 4!}e^\Phi g_{MN} G_{P_1\dots P_4}\tilde{\zeta}^{P_1 \dots P_4}  - \frac{1}{12}e^\Phi G_{[M|PQR|}\tilde{\zeta}^{PQR}{}_{N]}~,
\label{iiatcfhc}
\eea

\bea
&&{\cal D}_M^{\cal F}\omega_{NR}\defeq \nabla_M \omega_{NR} +\frac{1}{4}H_{MPQ}\tilde{\zeta}^{PQ}{}_{NR}+ e^\Phi F_{M[N}\tilde{k}_{R]} - \frac{1}{12} e^\Phi G_{MP_1P_2P_3}\tau^{P_1P_2P_3}{}_{NR} \cr
		&&\qquad =   \frac{1}{2}H_{MNR}\tilde{\sigma}+ \frac{1}{2}e^\Phi S g_{M[N}k_{R]} + \frac{3}{4} e^\Phi F_{[MN}\tilde{k}_{R]} + \frac{1}{2} e^\Phi g_{M[N}F_{R]P}\tilde{k}^P
\cr
		&&\qquad+\frac{1}{4 \cdot 5!} e^\Phi {}^\star{F}_{MNRP_1\dots P_5}\tau^{P_1\dots P_5}   + \frac{1}{2\cdot 4!}e^\Phi g_{M[N} G_{|P_1\dots P_4|}\tau^{P_1 \dots P_4}{}_{R]}
\cr
		&&\qquad- \frac{1}{8}e^\Phi G_{[M|P_1 P_2 P_3|}\tau^{P_1 P_2 P_3}{}_{NR]}  - \frac{1}{4}e^\Phi G_{MNRP} k^P  ~,
\label{iiatcfhd}
	\eea

\bea
	&&{\cal D}_M^{\cal F} \tilde{\zeta}_{N_1 \dots N_4} \defeq    \nabla_M \tilde{\zeta}_{N_1 \dots N_4} +\frac{1}{3}{}^\star{H}_{M[N_1N_2N_3|PQR|}\tilde{\zeta}^{PQR}{}_{N_4]} - 3 H_{M[N_1 N_2}\omega_{N_3 N_4]}  
\cr
 &&\qquad+ \frac{1}{2} e^\Phi F_{MP}\tau^P{}_{N_1 \dots N_4}-\frac{1}{2}e^\Phi {}^\star{G}_{M[N_1N_2|PQR|}\tau^{PQR}{}_{N_3N_4]}+ 2e^\Phi G_{M[N_1N_2N_3}\tilde{k}_{N_4]}   \cr
			&&\qquad 
= -\frac{1}{12}g_{M[N_1}{}^\star{H}_{N_2 N_3 N_4] P_1 \dots P_4}\tilde{\zeta}^{P_1 \dots P_4}+ \frac{5}{12} {}^\star{H}_{[MN_1N_2N_3|PQR|}\tilde{\zeta}^{PQR}{}_{N_4]} 
\cr
			&&\qquad - \frac{1}{4 \cdot 5!}e^\Phi {}^\star{S}_{MN_1\dots N_4P_1\dots P_5}\tau^{P_1\dots P_5} -\frac{1}{2} e^\Phi g_{M[N_1}F_{|PQ|}\tau^{PQ}{}_{N_2N_3N_4]}
 \cr
			&&\qquad + \frac{5}{8} e^\Phi F_{[M|P|}\tau^P{}_{N_1 \dots N_4]}  -\frac{5}{12}e^\Phi {}^\star{G}_{[MN_1N_2|PQR|}\tau^{PQR}{}_{N_3N_4]}+ 3e^\Phi g_{M[N_1}F_{N_2N_3}k_{N_4]}
\cr
			&&\qquad+ \frac{1}{8}e^\Phi g_{M[N_1}{}^\star{G}_{N_2 N_3|P_1 \dots P_4|}\tau^{P_1 \dots P_4}{}_{N_4]} -\frac{1}{4}e^\Phi {}^\star{G}_{MN_1\dots N_4P}k^P 
\cr
&&
\qquad + \frac{5}{4}e^\Phi G_{[MN_1N_2N_3}\tilde{k}_{N_4]}+ e^\Phi g_{M[N_1}G_{N_2N_3N_4]P}\tilde{k}^P ~,
\label{iiatcfhe}
	\eea

\begin{equation}
	\begin{split}
{\cal D}_M^{\cal F}& \tau_{N_1 \dots N_5}\defeq		\nabla_M \tau_{N_1 \dots N_5}+\frac{5}{6} {}^\star{H}_{M[N_1N_2N_3|PQR|}\tau^{PQR}{}_{N_4 N_5]} - \frac{5}{2}e^\Phi F_{M[N_1}\tilde{\zeta}_{N_2 \dots N_5]}
 \\
 &
 +\frac{5}{2}e^\Phi {}^\star{G}_{M[N_1N_2N_3|PQ|}\tilde{\zeta}^{PQ}{}_{N_4N_5]} +5e^\Phi G_{M[N_1N_2N_3}\omega_{N_4N_5]} = \frac{5}{4} {}^\star{H}_{[MN_1N_2N_3|PQR|}\tau^{PQR}{}_{N_4 N_5]}   \\
		& - \frac{5}{12} g_{M[N_1}{}^\star{H}_{N_2N_3N_4|P_1\dots P_4|}\tau^{P_1\dots P_4}{}_{N_5]} +\frac{1}{4\cdot 4!}e^\Phi {}^\star{S}_{MN_1\dots N_5 P_1 \dots P_4}\tilde{\zeta}^{P_1\dots P_4}   \\
		&-\frac{1}{8} e^\Phi {}^\star{F}_{MN_1\dots N_5}{}^{PQ}\omega_{PQ}-5e^\Phi g_{M[N_1}F_{N_2|P|}\tilde{\zeta}^P{}_{N_3N_4N_5]} - \frac{15}{4} e^\Phi F_{[MN_1}\tilde{\zeta}_{N_2\dots N_5]}  \\
		&+ \frac{15}{8}e^\Phi {}^\star{G}_{[MN_1N_2N_3|PQ|}\tilde{\zeta}^{PQ}{}_{N_4N_5]} + \frac{1}{4}e^\Phi {}^\star{G}_{MN_1\dots N_5}\tilde{\sigma} + \frac{5}{6 }e^\Phi g_{M[N_1}{}^\star{G}_{N_2N_3N_4|PQR|}\tilde{\zeta}^{PQR}{}_{N_5]}  \\
		&+ \frac{15}{4} e^\Phi G_{[MN_1N_2N_3}\omega_{N_4N_5]} + 5 e^\Phi g_{M[N_1}G_{N_2N_3N_4|P|}\omega^P{}_{N_5]} ~,
\label{iiatcfhf}
	\end{split}
\end{equation}
where  for simplicity we have suppressed the $r, s$  indices  on the form bilinears which count the different Killing spinors. The connection ${\cal D}^{\cal F}$ is the minimal connection of the TCFH, see \cite{gptcfh} for the definition.  As it has been explained in the introduction, the above TCFH implies that the form bilinears (\ref{symiia}) satisfy a generalisation of the CKY
with respect to the connection ${\cal D}^{\cal F}$. As expected $k$ is Killing, $\nabla_{(M} k_{N)}=0$.

A basis in the form bilinears, up to a Hodge duality operation, which are skew-symmetric in the exchange of the two Killing spinors  is
\bea
&&
\sigma^{rs} = \D{\epsilon^r}{\epsilon^s}~, \quad \tilde\omega^{rs} = \frac{1}{2} \D{\epsilon^r}{\Gamma_{NR}\Gamma_{11}\epsilon^s} \, e^N \wedge e^R~, 
\cr
&& \pi^{rs} = \frac{1}{3!} \D{\epsilon^r}{\Gamma_{NRS} \epsilon^s} \, e^N \wedge e^R \wedge e^S~,\quad \tilde\pi^{rs} = \frac{1}{3!} \D{\epsilon^r}{\Gamma_{NRS}\Gamma_{11}\epsilon^s} \, e^N \wedge e^R \wedge e^S~, 
 \cr
 &&\zeta^{rs} = \frac{1}{4!} \D{\epsilon^r}{\Gamma_{N_1 \dots N_4} \epsilon^s} \, e^{N_1} \wedge \dots \wedge e^{N_4}~.
\label{iiaskew}
\eea
 The associated TCFH with respect to the minimal connection is
\bea
	\mathcal{D}^\mathcal{F}_M\sigma\defeq \nabla_M \sigma = - \frac{1}{4} H_{MPQ} \tilde \omega^{PQ} - \frac{1}{8} e^\Phi F_{PQ} \tilde\pi^{PQ}{}_M + \frac{1}{4!} e^\Phi G_{MPQR} \pi^{PQR} ~,
\eea

\bea
	&&\mathcal{D}^\mathcal{F}_M \tilde\omega_{NR}\defeq 	\nabla_M \tilde\omega_{NR} +\frac{1}{4} H_{MPQ} \zeta^{PQ}{}_{NR} + \frac{1}{2}e^\Phi F_{MP} \pi^P{}_{NR}
 - \frac{1}{2} e^\Phi G_{M[N|PQ|} \tilde\pi^{PQ}{}_{R]}
 \cr
 &&
 \qquad=  \frac{1}{2} H_{MNR} \sigma  + \frac{1}{4} e^\Phi S \tilde\pi_{MNR} -\frac{1}{4}e^\Phi g_{M[N}F_{|PQ|}\pi^{PQ}{}_{R]}  + \frac{3}{4} e^\Phi F_{[M|P|} \pi^P{}_{NR]}  
 \cr
 &&\qquad+ \frac{1}{4!} e^\Phi {}^\star{G}_{MNRP_1P_2P_3}\pi^{P_1P_2P_3} - \frac{1}{12} e^\Phi g_{M[N}G_{R]P_1P_2P_3}\tilde\pi^{P_1P_2P_3} 
 \cr
 &&
 \qquad- \frac{3}{8}e^\Phi G_{[MN|PQ|} \tilde\pi^{PQ}{}_{R]} ~,
\label{iiatcfha1}
\eea

\bea
&&\mathcal{D}^\mathcal{F}_M	\pi_{NRS}\defeq	\nabla_M \pi_{NRS} +\frac{3}{2} H_{M[N|P|} \tilde\pi^P{}_{RS]}
 -\frac{3}{2} e^\Phi F_{M[N} \tilde\omega_{RS]}- \frac{3}{4}e^\Phi G_{M[N|PQ|} \zeta^{PQ}{}_{RS]}
  \cr
		&& \qquad=  \frac{1}{4} e^\Phi S\zeta_{MNRS}+ \frac{1}{4 \cdot 4!} e^\Phi {}^\star{F}_{MNRSP_1 \dots P_4} \zeta^{P_1\dots P_4}
- \frac{3}{2} e^\Phi g_{M[N} F_{R|P|} \tilde \omega^P{}_{S]}
\cr
&&\qquad - \frac{3}{2} e^\Phi F_{[MN} \tilde\omega_{RS]} - \frac{1}{4} e^\Phi G_{MNRS} \sigma -\frac{1}{8} e^\Phi {}^\star{G}_{MNRSPQ} \tilde\omega^{PQ} \cr
		&&\qquad  - \frac{1}{4} e^\Phi g_{M[N} G_{R|P_1P_2P_3|}\zeta^{P_1P_2P_3}{}_{S]} - \frac{3}{4}e^\Phi G_{[MN|PQ|}\zeta^{PQ}{}_{RS]} ~,
\label{iiatcfha2}
\eea

\bea
&&\mathcal{D}^\mathcal{F}_M	\tilde\pi_{NRS}\defeq		\nabla_M \tilde\pi_{NRS}+\frac{3}{2} H_{M[N|P|} \pi^P{}_{RS]} - \frac{1}{2} e^\Phi F_{MP} \zeta^P{}_{NRS}
  +\frac{3}{2} e^\Phi G_{M[NR|P|} \tilde\omega^P{}_{S]}
  \cr
  &&\qquad+ \frac{1}{4}e^\Phi {}^\star{G}_{M[NR|P_1P_2P_3|}\zeta^{P_1P_2P_3}{}_{S]}= + \frac{3}{4} e^\Phi S g_{M[N} \tilde\omega_{RS]} + \frac{1}{2} e^\Phi F_{MP} \zeta^P{}_{NRS} 
  \cr
		&&\qquad
+ \frac{3}{8} e^\Phi g_{M[N} F_{|PQ|} \zeta^{PQ}{}_{RS]} - e^\Phi F_{[M|P|} \zeta^P{}_{NRS]} - \frac{3}{4}e^\Phi g_{M[N}F_{RS]} \sigma
  \cr
		&&\qquad- \frac{3}{8} e^\Phi g_{M[N}G_{RS]PQ} \tilde \omega^{PQ}+ e^\Phi G_{[MNR|P|} \tilde\omega^P{}_{S]} \cr
		&&\qquad - \frac{1}{32} e^\Phi g_{M[N} {}^\star{G}_{RS]P_1 \dots P_4}\zeta^{P_1\dots P_4} + \frac{1}{6} e^\Phi {}^\star{G}_{[MNR|P_1P_2P_3|}\zeta^{P_1P_2P_3}{}_{S]} ~,
\label{iiatcfha3}
\eea

\bea
&&\mathcal{D}^\mathcal{F}_M  \zeta_{N_1 \dots N_4}	\defeq		\nabla_M \zeta_{N_1 \dots N_4} +  \frac{1}{3} {}^\star{H}_{M[N_1N_2N_3|PQR|}\zeta^{PQR}{}_{N_4]}
 - 3 H_{M[N_1N_2} \tilde\omega_{N_3N_4]}
 \cr
 &&
\qquad  +2 e^\Phi F_{M[N_1} \tilde\pi_{N_2 N_3 N_4]}
 +3 e^\Phi G_{M[N_1N_2|P|}\pi^P{}_{N_3N_4]}- e^\Phi {}^\star{G}_{M[N_1N_2N_3|PQ|}\tilde\pi^{PQ}{}_{N_4]}
 \cr
 &&\qquad
 = -\frac{1}{12} g_{M[N_1} {}^\star{H}_{N_2 N_3 N_4] P_1 \dots P_4} \zeta^{P_1\dots P_4} + \frac{5}{12} {}^\star{H}_{[MN_1N_2N_3|PQR|}\zeta^{PQR}{}_{N_4]}  
 \cr
		&&\qquad
+e^\Phi S g_{M[N_1} \pi_{N_2N_3N_4]}
 - \frac{1}{4!} e^\Phi {}^\star{F}_{MN_1 \dots N_4 PQR} \pi^{PQR}  + 3 e^\Phi g_{M[N_1} F_{N_2|P|}\tilde\pi^P{}_{N_3N_4]} 
 \cr
		&&\qquad+ \frac{5}{2} e^\Phi F_{[MN_1}\tilde\pi_{N_2N_3N_4]}
- \frac{1}{6} e^\Phi g_{M[N_1} {}^\star{G}_{N_2N_3N_4]PQR}\tilde\pi^{PQR}  - \frac{5}{8} e^\Phi {}^\star{G}_{[MN_1N_2N_3|PQ|}\tilde\pi^{PQ}{}_{N_4]}
 \cr
		&&\qquad- \frac{3}{2} e^\Phi  g_{M[N_1} G_{N_2N_3|PQ|}\pi^{PQ}{}_{N_4]} + \frac{5}{2} e^\Phi G_{[MN_1N_2|P|} \pi^P{}_{N_3N_4]} ~.
\label{iiatcfha4}
\eea
As in the previous case, a consequence of the TCFH above is that the forms (\ref{iiaskew}) satisfy a generalisation of the CKY equation with respect to the connection $\mathcal{D}^\mathcal{F}$. Later we shall demonstrate that in some cases the forms (\ref{symiia}) and (\ref{iiaskew}) generate symmetries
in string and particle actions probing some IIA backgrounds.

The factorisation of the domain that the minimal TCFH connection $\mathcal{D}^\mathcal{F}$ acts as in (\ref{symiia}) and (\ref{iiaskew}) can be understood as follows.
The  product of two Majorana representations $\Delta_{32}$ in terms of  forms is   $\otimes^2 \Delta_{32}=\Lambda^*(\bR^{9,1})$. Therefore the form bilinears of all spinor span all spacetime forms. Therefore generically the TCFH connection acts on the space of all spacetime forms.  However we have seen that the TCFH connection preserves the forms which are symmetric (skew-symmetric) in the exchange of the two Killing spinors, i.e. it preserves that symmetrised
$S^2\,(\Delta_{32})$ and skew-symmetrised $\Lambda^2\,(\Delta_{32})$ subspaces of the product. As $\mathrm{dim}\, S^2 ( \Delta_{32})=528$ and $\mathrm{dim}\, \Lambda^2 ( \Delta_{32})=496$, the holonomy  of $\mathcal{D}^\mathcal{F}$ is included\footnote{In fact the (reduced) holonomy of $\mathcal{D}^\mathcal{F}$ is included in $GL(527)\times GL(495)$ as it acts with partial derivatives on the scalars $\tilde \sigma$ and $\sigma$ but the holonomy of other TCFH connections, like that of the maximal connection, will be included in $GL(528)\times GL(496)$.} in $GL(528)\times GL(496)$. Of course the holonomy  of $\mathcal{D}^\mathcal{F}$ reduces for special backgrounds.

\section{The TCFH of IIB supergravity}

The KSEs of IIB supergravity \cite{js} are again associated with the vanishing conditions of the gravitino and dilatino supersymmetry variations. The gravitino KSE is a parallel transport equation for the supercovariant derivative ${\cal D}$ of the theory.  In the string frame, this can be expressed \cite{Bergshoeff:2005ac} as

\begin{align}
{\cal D}_M&\defeq \nabla_M - \frac{1}{8}\,\Gamma^{N_1N_2}\,H_{MN_1N_2}\,\sigma_3 - \frac{1}{4}\,e^{\Phi}\,\Gamma^{N}{}_{M}\,G^{(1)}_N\,(i\sigma_2) - \frac{1}{4}\,e^{\Phi}\,G^{(1)}_M\,(i\sigma_2) \nn
&-\frac{1}{24}\,e^{\Phi}\,\Gamma^{N_1N_2N_3}{}_{M}\,G^{(3)}_{N_1N_2N_3}\,\sigma_1 - \frac{1}{8}\,e^{\Phi}\,\Gamma^{N_1N_2}\Gamma^{(3)}_{MN_1N_2}\,\sigma_1 \nn
&- \frac{1}{96}\,e^{\Phi}\,\Gamma^{N_1\dots N_4}\,G^{(5)}_{MN_1\dots N_4}\,(i\sigma_2)~,
\end{align}
where $H$ and $G^{(n)}$  are the 3-form and  $n$-form, for $n=1,3, 5$, NS-NS and R-R field strengths  of the theory, respectively, $\Phi$ is that dilaton and $\sigma^i$, $i=1,2,3$ are the Pauli matrices. The field strength $G^{(5)}$ is anti-self-dual\footnote{Our Hodge duality conventions are as in the IIA theory.}. ${\cal D}$ is a connection of the spin bundle over the spacetime associated to two copies, $\oplus^2 \Delta^+_{16}$, of the positive chirality Majorana-Weyl representation, $\Delta^+_{16}$, of $\mathfrak{spin}(9,1)$. The (reduced) holonomy of ${\cal D}$ for generic IIB backgrounds
is included in $SL(32,\bR)$ \cite{gpdt}. The KSEs of IIB supergravity have been solved for one Killing spinor in \cite{ugjggp}.

As expected from the general result in \cite{gptcfh}, the conditions imposed on the form bilinears by the gravitino KSE, ${\cal D}\epsilon=0$, can be organised as a TCFH.
Given any two spinors $\epsilon^r$ and $\epsilon^s$, the form bilinears are given by

\bea
&&k^{rs} = \delta_{ab}\,\left\langle \epsilon^{ra}, \Gamma_P\,\epsilon^{sb} \right\rangle_D\, e^P~, \qquad
k^{(i)rs} = \delta_{ab}\,\left\langle \epsilon^{ra}, \Gamma_P\,(\sigma^{i}\,\epsilon^s)^b \right\rangle_D\, e^P ~,
\cr
&&
\pi^{rs} = \frac{1}{3!}\delta_{ab}\,\left\langle \epsilon^{ra}, \Gamma_{P_1P_2P_3}\,\epsilon^{sb} \right\rangle_D\, e^{P_1}\wedge e^{P_2} \wedge e^{P_3}~,
\cr
&&
\pi^{(i)rs} = \frac{1}{3!}\,\delta_{ab}\,\left\langle \epsilon^{ra}, \Gamma_{P_1P_2P_3}\,(\sigma^i\,\epsilon^s)^b \right\rangle_D\, e^{P_1}\wedge e^{P_2} \wedge e^{P_3}~,
\cr
&&
\tau^{rs} = \frac{1}{5!}\,\delta_{ab}\,\left\langle \epsilon^{ra}, \Gamma_{P_1P_2P_3P_4P_5}\,\epsilon^{sb} \right\rangle_D\, e^{P_1}\wedge \dots \wedge e^{P_5}~,
\cr
&&
\tau^{(i)rs} = \frac{1}{5!}\,\delta_{ab}\,\left\langle \epsilon^{ra}, \Gamma_{P_1P_2P_3P_4P_5}\,(\sigma^{i}\epsilon^s)^b \right\rangle_D\, e^{P_1}\wedge \dots \wedge e^{P_5}~,
\eea
where $\left\langle \sigma^i\,\alpha, \beta \right\rangle_D = \left\langle \alpha, \sigma^i\,\beta \right\rangle_D$ as the Pauli matrices are hermitian and $a,b=1,2$. Note that the forms $k$, $k^{(1)}$, $k^{(3)}$, $\pi^{(2)}$, $\tau$, $\tau^{(1)}$ and $\tau^{(3)}$
are symmetric in the exchange of $\epsilon^r$ and $\epsilon^s$ while the rest are skew symmetric.

The forms $k^{(2)}$, $\pi^{(2)}$ and $\tau^{(2)}$ are purely imaginary while the rest are real. One could multiply them with the imaginary unit $i$ so they become real but in such case   the expression for the TCFH below would have been more involved. So we shall not do this here but later when we consider applications, we shall replace $k^{(2)}$, $\pi^{(2)}$ and $\tau^{(2)}$  with $ik^{(2)}$, $i\pi^{(2)}$ and $i\tau^{(2)}$.

Using the gravitino KSE, ${\cal D}\epsilon=0$, one can show that the TCFH of IIB supergravity expressed in terms of the minimal connection ${\cal D}^{\cal F}$  is
\bea
{\cal D}^{\cal F}_M k^{}_{P} &\defeq& \nabla_M\, k^{}_P = \frac{1}{2}\,H_{MP}{}^N\,k^{(3)}_N + \frac{i}{2}\, e^{\Phi}\,G^{(1)N}\,\pi^{(2)}_{NMP} + \frac{1}{12}\,e^{\Phi}\,G^{(3)N_1N_2N_3}\,\tau^{(1)}_{N_1N_2N_3MP} \cr
&&+ \frac{1}{2}\,e^{\Phi}\,G^{(3)}_{MP}{}^N\, k^{(1)}_{N} + \frac{i}{12}\,e^{\Phi}\,G^{(5)}_{MP}{}^{N_1N_2N_3}\,\pi^{(2)}_{N_1N_2N_3}~,
\label{iibtcfh1} 
\eea
\bea
{\cal D}^{\cal F}_M k^{(i)}_{P} &\defeq& \nabla_M\, k^{(i)}_P + \frac{i}{4}\,\varepsilon_{3ij}\,H_{M}{}^{N_1N_2}\pi^{(j)}_{PN_1N_2} - \varepsilon_{2ij}\,e^{\Phi}\,G_{M}^{(1)}\,k^{(j)}_{P} \cr
&& +\frac{i}{2}\,\varepsilon_{1ij}\,e^{\Phi}\,G^{(3)N_1N_2}_M\,\pi^{(j)}_{PN_1N_2} - \frac{1}{48}\,\varepsilon_{2ij}\,e^{\Phi}\,G_{M}^{(5)N_1\dots N_4}\,\tau^{(j)}_{PN_1\dots N_4} \cr
&&= \frac{1}{2}\,\delta_{i3}\,H_{MP}{}^N\,k^{}_N + \frac{i}{2}\,\delta_{i2}\,e^{\Phi}\,G^{(1)N}\,\pi^{}_{MPN} - \varepsilon_{2ij}\,e^{\Phi}\,G^{(1)}_{[M}\,k^{(j)}_{P]} \cr
&&-\frac{1}{2}\,\varepsilon_{2ij}\,e^{\Phi}\,g_{MP}\,G^{(1)N}\,k^{(j)}_N + \frac{1}{12}\,\delta_{i1}\,e^{\Phi}\,G^{(3)N_1N_2N_3}\,\tau^{}_{MPN_1N_2N_3} \cr
&&+\frac{i}{12}\,\varepsilon_{1ij}\,e^{\Phi}\,g_{MP}\,G^{(3)N_1N_2N_3}\,\pi^{(j)}_{N_1N_2N_3} + \frac{i}{2}\,\varepsilon_{1ij}\,e^{\Phi}\,G^{(3)N_1N_2}{}_{[M}\,\pi^{(j)}_{P]N_1N_2} 
\cr
&&+\frac{1}{2}\,\delta_{i1}\,e^{\Phi}\,G^{(3)}_{MP}{}^N\,k^{}_N + \frac{i}{12}\,\delta_{i2}\,e^{\Phi}\,G^{(5)}_{MP}{}^{N_1N_2N_3}\,\pi^{}_{N_1N_2N_3}~, 
\label{iibtcfh2}
\eea
\bea
{\cal D}^{\cal F}_M \pi^{}_{P_1P_2P_3} &\defeq& \nabla_M\,\pi^{}_{P_1P_2P_3} - \frac{3}{2}\,H_{M[P_1}{}^N\,\pi^{(3)}_{P_2P_3]N} - 3\,e^{\Phi}\,G^{(3)}_{M[P_1}{}^N\,\pi^{(1)}_{P_2P_3]N} \cr
&&-\frac{i}{4}\,e^{\Phi}\,G^{(5)}_{M[P_1}{}^{N_1N_2N_3}\,\tau^{(2)}_{P_2P_3]N_1N_2N_3} \cr
&&= \frac{i}{2}\,e^{\Phi}\,G^{(1)N}\,\tau^{(2)}_{MP_1P_2P_3N} + 3i\,e^{\Phi}\,g_{M[P_1}\,G^{(1)}_{P_2}\,k^{(2)}_{P_3]} \cr
&&- \frac{1}{12}\,e^{\Phi}\,{}^{\star}G^{(7)}_{MP_1P_2P_3}{}^{N_1N_2N_3}\,\pi^{(1)}_{N_1N_2N_3}+\frac{3}{2}\,e^{\Phi}\,g_{M[P_1}\,G^{(3)}_{P_2}{}^{N_1N_2}\,\pi^{(1)}_{P_3]N_1N_2} \cr
&&+ 3\,e^{\Phi}\,G^{(3)}_{[P_1P_2}{}^N\,\pi^{(1)}_{P_3M]N} -\frac{i}{2}\,e^{\Phi}\,G^{(5)}_{MP_1P_2P_3}{}^N\,k^{(2)}_N~,
\label{iibtcfh3}
\eea
\bea
{\cal D}^{\cal F}_M \pi^{(i)}_{P_1P_2P_3} &\defeq& \nabla_M\,\pi^{(i)}_{P_1P_2P_3} - \frac{3}{2}\,\delta_{i3}\,H_{M[P_1}{}^N\,\pi^{}_{P_2P_3]N} + \frac{i}{4}\,\varepsilon_{3ij}\,H_{MN_1N_2}\,\tau^{(j)}_{P_1P_2P_3}{}^{N_1N_2} \cr
&&-\frac{3i}{2}\,\varepsilon_{3ij}\,H_{M[P_1P_2}\,k^{(j)}_{P_3]} - \varepsilon_{2ij}\,e^{\Phi}\,G^{(1)}_M\,\pi^{(j)}_{P_1P_2P_3} - 3\,\delta_{i1}\,e^{\Phi}\,G^{(3)}_{M[P_1}{}^N\,\pi^{}_{P_2P_3]N} 
\cr
&&+\frac{i}{2}\,\varepsilon_{1ij}\,e^{\Phi}\,G^{(3)}_{M}{}^{N_1N_2}\,\tau^{(j)}_{P_1P_2P_3N_1N_2} - 3i\,\varepsilon_{1ij}\,e^{\Phi}\,G^{(3)}_{M[P_1P_2}\,k^{(j)}_{P_3]} \cr
&&+\frac{3}{2}\,\varepsilon_{2ij}\,e^{\Phi}\,G^{(5)}_{M[P_1P_2}{}^{N_1N_2}\,\pi^{(j)}_{P_3]N_1N_2} -\frac{i}{4}\,\delta_{i2}\,e^{\Phi}\,G^{(5)}_{M[P_1}{}^{N_1N_2N_3}\,\tau^{}_{P_2P_3]N_1N_2N_3} \cr
&&= \frac{i}{2}\,\delta_{i2}\,e^{\Phi}\,G^{(1)N}\,\tau^{}_{MP_1P_2P_3N} + 3i\,\delta_{i2}\,e^{\Phi}\,g_{M[P_1}\,G^{(1)}_{P_2}\,k^{}_{P_3]} + 2\,\varepsilon_{2ij}\,e^{\Phi}\,G^{(1)}_{[P_1}\,\pi^{(j)}_{P_2P_3M]} \cr
&&-\frac{3}{2}\,\varepsilon_{2ij}\,e^{\Phi}\,G^{(1)N}\,g_{M[P_1}\,\pi^{(j)}_{P_2P_3]N} - \frac{1}{12}\,\delta_{i1}\,e^{\Phi}\,{}^{\star}G^{(7)}_{MP_1P_2P_3}{}^{N_1N_2N_3}\,\pi^{}_{N_1N_2N_3} \cr
&&+\frac{3}{2}\,\delta_{i1}\,e^{\Phi}\,g_{M[P_1}\,G^{(3)}_{P_2}{}^{N_1N_2}\,\pi^{}_{P_3]N_1N_2} + 3\,\delta_{i1}\,e^{\Phi}\,G^{(3)}_{[P_1P_2}{}^N\,\pi^{}_{P_3M]N} 
\cr
&&+\frac{i}{4}\,\varepsilon_{1ij}\,e^{\Phi}\,G^{(3)N_1N_2N_3}\,g_{M[P_1}\,\tau^{(j)}_{P_2P_3]N_1N_2N_3} - i\,\varepsilon_{1ij}e^{\Phi}\,G^{(3)}_{[P_1}{}^{N_1N_2}\,\tau^{(j)}_{P_2P_3M]N_1N_2} \cr
&&-\frac{3i}{2}\,\varepsilon_{1ij}\,e^{\Phi}\,g_{M[P_1}\,G^{(3)}_{P_2P_3]}{}^N\,k^{(j)}_N + 2i\,\varepsilon_{1ij}\,e^{\Phi}\,G^{(3)}_{[P_1P_2P_3}\,k^{(j)}_{M]} \cr
&&-\frac{i}{2}\,\delta_{i2}\,e^{\Phi}\,G^{(5)}_{MP_1P_2P_3}{}^N\,k^{}_N + \frac{1}{4}\,\varepsilon_{2ij}\,e^{\Phi}\,g_{M[P_1}\,G^{(5)}_{P_2P_3]}{}^{N_1N_2N_3}\,\pi^{(j)}_{N_1N_2N_3} \cr
&&-\varepsilon_{2ij}\,e^{\Phi}\,G^{(5)}_{[P_1P_2P_3}{}^{N_1N_2}\,\pi^{(j)}_{M]N_1N_2}~,
\label{iibtcfh4} 
\eea
\bea
{\cal D}^{\cal F}_M \tau^{}_{P_1 \dots P_5} &\defeq& \nabla_M\,\tau^{}_{P_1 \dots P_5} + \frac{5}{2}\,H_M{}^N{}_{[P_1}\,\tau^{(3)}_{P_2\dots P_5]N} - 5\,e^{\Phi}\,G^{(3)}_{M[P_1}{}^N\,\tau^{(1)}_{P_2\dots P_5]N} \cr
&&+10i\,e^{\Phi}\,G^{(5)}_{M[P_1P_2P_3}{}^N\,\pi^{(2)}_{P_4P_5]N} \cr
&&=-\frac{i}{12}\,e^{\Phi}\,{}^{\star}G^{(9)}_{MP_1\dots P_5}{}^{N_1N_2N_3}\,\pi^{(2)}_{N_1N_2N_3}\, + 10i\,e^{\Phi}\,g_{M[P_1}\,G^{(1)}_{P_2}\,\pi^{(2)}_{P_3P_4P_5]} \cr
&&+ \frac{1}{2}\,e^{\Phi}\,{}^{\star}G^{(7)}_{MP_1\dots P_5}{}^N\,k^{(1)}_N +\frac{15}{2}\,e^{\Phi}\,G^{(3)}_{[P_1P_2}{}^N\,\tau^{(1)}_{P_3P_4P_5M]N} \cr
&&+ 5\,e^{\Phi}\,g_{M[P_1}\,G^{(3)}_{P_2}{}^{N_1N_2}\,\tau^{(1)}_{P_3P_4P_5]N_1N_2} -10\,e^{\Phi}\,g_{M[P_1}\,G^{(3)}_{P_2P_3P_4}\,k^{(1)}_{P_5]} \cr
&&-5i\,e^{\Phi}\,g_{M[P_1}\,G^{(5)}_{P_2P_3P_4}{}^{N_1N_2}\,\pi^{(2)}_{P_5]N_1N_2} -\frac{15i}{2}\,e^{\Phi}\,G^{(5)}_{[P_1\dots P_4}{}^N\,\pi^{(2)}_{P_5M]} ~, 
\label{iibtcfh5}
\eea
\bea
{\cal D}^{\cal F}_M \tau^{(i)}_{P_1\dots P_5} &\defeq& \nabla_M\,\tau^{(i)}_{P_1\dots P_5} +\frac{5}{2}\,\delta_{i3}\,H_M{}^N{}_{[P_1}\,\tau^{}_{P_2\dots P_5]N} +\frac{5i}{4}\,\varepsilon_{3ij}\,{}^{\star}H_{M[P_1\dots P_4}{}^{N_1N_2}\,\pi^{(j)}_{P_5]N_1N_2} \cr
&&-5i\,\varepsilon_{3ij}\,H_{M[P_1P_2}\,\pi^{(j)}_{P_3P_4P_5]} -\varepsilon_{2ij}\,e^{\Phi}\,G^{(1)}_M\,\tau^{(j)}_{P_1\dots P_5} - 5\,\delta_{i1}\,e^{\Phi}\,G^{(3)}_{M[P_1}{}^N\,\tau^{}_{P_2 \dots P_5]N} \cr
&&+\frac{5i}{2}\,\varepsilon_{1ij}\,e^{\Phi}\,{}^{\star}G^{(7)}_{M[P_1 \dots P_4}{}^{N_1N_2}\,\pi^{(j)}_{P_5]N_1N_2} -10i\,\varepsilon_{1ij}\,e^{\Phi}\,G^{(3)}_{M[P_1P_2}\,\pi^{(j)}_{P_3P_4P_5]} \cr
&&- 5\,\varepsilon_{2ij}\,e^{\Phi}\,G^{(5)}_{M[P_1\dots P_4}\,k^{(j)}_{P_5]} +\frac{5}{2}\,\varepsilon_{2ij}\,e^{\Phi}\,G^{(5)}_{M[P_1P_2}{}^{N_1N_2}\,\tau^{(j)}_{P_3P_4P_5]N_1N_2} \cr
&&+10i\,\delta_{i2}\,e^{\Phi}\,G^{(5)}_{M[P_1P_2P_3}{}^N\,\pi^{}_{P_4P_5]N} \cr
&&=  \frac{5i}{12}\varepsilon_{3ij}\,g_{M[P_1}\,{}^{\star}H_{P_2\dots P_5]}{}^{N_1N_2N_3}\,\pi^{(j)}_{N_1N_2N_3} -\frac{3i}{2}\,\varepsilon_{3ij}\,{}^{\star}H_{[P_1\dots P_5}{}^{N_1N_2}\,\pi^{(j)}_{M]N_1N_2} \cr
&&- \frac{i}{12}\,\delta_{i2}\,e^{\Phi}\,{}^{\star}G^{(9)}_{MP_1\dots P_5}{}^{N_1N_2N_3}\,\pi^{}_{N_1N_2N_3} +10i\,\delta_{i2}\,e^{\Phi}\,g_{M[P_1}\,G^{(1)}_{P_2}\,\pi^{}_{P_3P_4P_5]} \cr
&&- \frac{5}{2}\,\varepsilon_{2ij}\,e^{\Phi}\,G^{(1)N}\,g_{M[P_1}\,\tau^{(j)}_{P_2\dots P_5]N} +3\,\varepsilon_{2ij}\,e^{\Phi}\,G^{(1)}_{[P_1}\,\tau^{(j)}_{P_2 \dots P_5M]} \cr
&&+\frac{1}{2} \delta_{i1}\,e^{\Phi}\,{}^{\star}G^{(7)}_{MP_1\dots P_5}{}^N\,k^{}_N +5\,\delta_{i1}\,e^{\Phi}\,g_{M[P_1}\,G^{(3)}_{P_2}{}^{N_1N_2}\,\tau^{}_{P_3P_4P_5]N_1N_2} \cr
&&+ \frac{15}{2}\delta_{i1}\,e^{\Phi}\,G^{(3)}_{[P_1P_2}{}^N\,\tau^{}_{P_3P_4P_5M]N} -10\delta_{i1}\,e^{\Phi}\,g_{M[P_1}\,G^{(3)}_{P_2P_3P_4}\,k^{}_{P_5]} \cr
&&+ 10i\,\varepsilon_{1ij}\,e^{\Phi}\,G^{(3)}_{[P_1P_2P_3}\,\pi^{(j)}_{P_4P_5M]} -15i\,\varepsilon_{1ij}\,e^{\Phi}\,g_{M[P_1}\,G^{(3)}_{P_2P_3}{}^N\,\pi^{(j)}_{P_4P_5]N} \cr
&&+ \frac{5i}{12}\,\varepsilon_{1ij}e^{\Phi} \,g_{M[P_1}\,{}^{\star}G^{(7)}_{P_2\dots P_5]}{}^{N_1N_2N_3}\,\pi^{(j)}_{N_1N_2N_3} -\frac{3i}{2}\,\varepsilon_{1ij}\,e^{\Phi}\,{}^{\star}G^{(7)}_{[P_1\dots P_5}{}^{N_1N_2}\,\pi^{(j)}_{M]N_1N_2} \cr
&&- \frac{5}{2}\,\varepsilon_{2ij}\,e^{\Phi}\,g_{M[P_1}\,G^{(5)}_{P_2\dots P_5]}{}^N\,k^{(j)}_N +3\,\varepsilon_{2ij}\,e^{\Phi}\,G^{(5)}_{[P_1\dots P_5}\,k^{(j)}_{M]} \cr
&&- 5i\,\delta_{i2}\,e^{\Phi}\,g_{M[P_1}G^{(5)}_{P_2P_3P_4}{}^{N_1N_2}\,\pi^{}_{P_5]N_1N_2} -\frac{15i}{2}\delta_{i2}\,e^{\Phi}\,G^{(5)}_{[P_1\dots P_4}{}^N\,\pi^{}_{P_5M]N}~,
\label{iibtcfh6}
\eea
where  for simplicity we have suppressed the $r,s$ indices on the form bilinears that label the Killing spinors.
Although it is not manifest from the expression of TCFH above, the TCFH preserves the form bilinears that are either symmetric or skew-symmetric  in the exchange of the two Killing spinors. Moreover all terms of the IIB TCFH can be arranged to be real. The imaginary unit that appears in some terms can be eliminated  after replacing the purely imaginary forms $k^{(2)}$, $\pi^{(2)}$ and $\tau^{(2)}$  with the real forms $ik^{(2)}$, $i\pi^{(2)}$ and $i\tau^{(2)}$.
A consequence of the TCFH is that all form bilinears satisfy a generalisation of the CKY equation with respect to the minimal connection
${\cal D}^{\cal F}$. In particular $k$ is Killing, $\nabla_{(M}\,k^{rs}_{P)}=0$, as expected.

To understand the factorisation of the domain that ${\cal D}^{\cal F}$ acts note  that the product of two Majorana-Weyl representations $\Delta_{16}^+$ of $\mathfrak{spin}(9,1)$ decomposes as
\bea
\otimes^2  \Delta_{16}^+=
\Lambda^1(\bR^{9,1})\oplus \Lambda^3(\bR^{9,1})\oplus \Lambda^{5-}(\bR^{9,1})~,
\eea
 where $\Lambda^{5-}(\bR^{9,1})$ is the space of anti-self-dual 5-forms on $\bR^{9,1}$.
The Killing spinors lie in two copies of $\Delta_{16}^+$, i.e. $\Delta^+_{32}=\oplus^2 \Delta_{16}^+$. Therefore the space of all IIB form  bilinears  is identified with the product
 $\otimes^2 \Delta^+_{32}$.  This can be decomposed   in terms of spacetime  forms as indicated above.  Indeed notice that $\mathrm{dim} \,\big(\otimes^2 \Delta^+_{32}\big)=32\cdot 32=4 [\mathrm{dim}\,\big(\Lambda^1(\bR^{9,1})\big)+ \mathrm{dim}\,\big(\Lambda^3(\bR^{9,1})\big)+\mathrm{dim}\,\big(\Lambda^{5-}(\bR^{9,1})\big)]$. The minimal connection  $\mathcal{D}^\mathcal{F}$ of the TCFH preserves
 the symmetric $S^2\big(\Delta^+_{32}\big)$ and skew-symmetric  $\Lambda^2\big(\Delta^+_{32}\big)$  subspaces of $\otimes^2 \Delta^+_{32}$. As a consequence
  the (reduced) holonomy of  $\mathcal{D}^\mathcal{F}$  for a generic background is included in $GL(528)\times GL(496)$ as in IIA case investigated in the previous section.

\section{Particles and integrability of type II branes }

Before we proceed to investigate the symmetries of particle and string probes generated by the TCFHs of type II theories, we shall summarise some of the properties of KS, KY and CKY tensors and their applications to generating symmetries for particle actions, for more detailed studies,  see reviews  \cite{revky} and \cite{frolov} and the references within. We shall also present some of the particle actions that are invariant under the symmetries generated by such tensors. Then we shall construct the KS and KY tensors of type II brane solutions which to our knowledge have not presented before. We shall use these to argue that the geodesic flow of some of these solutions is completely integrable and we shall give the associated independent conserved charges in involution.

\subsection{Killing-St\"ackel and Killing-Yano tensors}

\subsubsection{Definitions and outline of properties}

A  rank $k$ conformal Killing-St\"ackel ($k$-CKS) tensor   is a symmetric (0,$k$) tensor $d$ on a $n$-dimensional  spacetime $M$ with metric $g$ which satisfies the equation
\bea
 \nabla_{(M} d_{N_1N_2\cdots N_k)}= g_{(MN_1} q_{N_2\cdots N_k)}~,
 \label{cks}
\eea
where $q$ is a symmetric $(0, k-1)$ tensor, $g$ is the spacetime metric and $\nabla$ is the Levi-Civita connection of  $g$. For $k=1$, the equation reduces to that of a conformal Killing vector field. If $q$ vanishes, $q=0$, then $d$ will be a Killing-St\"ackel (KS) tensor.

Furthermore observe that if $d$ and $e$ are  $k-$ and $\ell-$ CKS (KS) tensors on $M$,  then
\bea
(d\otimes_s e)_{N_1\cdots N_{k+\ell}}\defeq d_{(N_1\cdots N_k} e_{N_{k+1}\cdots N_{k+\ell})}~,
\eea
is a  $(k+\ell)-$CKS (KS) tensor on $M$.

KS tensors are associated with conserved charges of test particle systems. Indeed consider the action
\bea
A={1\over2}\int\, d\tau\, g_{MN}\, \dot x^M\, \dot x^N~,
\label{gact}
\eea
which describes the geodesic flow\footnote{When viewing the geodesic flow as a dynamical system, $M$ is identified with its configuration space.} on a spacetime (manifold) $M$ with metric $g$, where $\dot x$ denotes the derivative of the coordinate $x$ with respect to the affine parameter $\tau$. It is straightforward to show that if the spacetime $M$ admits a KS tensor $d$, then
\bea
Q(d)= d_{N_1N_2\cdots N_k}\, \dot x^{N_1}\, \dot x^{N_2}\cdots \dot x^{N_k}~,
\label{ccd}
\eea
  is conserved along the geodesic flow, i.e. $\dot Q(d)=0$ subject to the geodesic equations with affine parameter $\tau$.  This charge generates the infinitesimal  transformation
\bea
\delta x^M= \epsilon\, d^M{}_{N_1\cdots N_{k-1}} \dot x^{N_1} \cdots \dot x^{N_{k-1}}~,
\eea
which is a symmetry of the action (\ref{gact}) with infinitesimal parameter $\epsilon$.

A rank $k$ conformal Killing-Yano ($k$-CKY) tensor is a $k$-form, $\alpha$, which satisfies the condition
\bea
\nabla_M \alpha_{N_1N_2\cdots N_k}={1\over k+1} d\alpha_{MN_1\dots N_k}-{k\over n-k+1} g_{M[N_1} \delta\alpha_{N_2\cdots N_k]}~.
\label{cky}
\eea
If  $\alpha$ is co-closed, $\delta\alpha=0$, then $\alpha$ is a Killing-Yano (KY) form while if $\alpha$ is closed, $d\alpha=0$, $\alpha$ is a closed conformal Killing-Yano (CCKY) form.  It turns out that if $\alpha$ is KY, then the Hodge dual, ${}^\star\alpha$,  of $\alpha$ is CCKY form.

Furthermore, if $\alpha$ and $\beta$ are $k$-CKY ($k$-KY) forms, then
\bea
\alpha_{(M }{}^{L_1\cdots L_{k-1}} \beta_{N) L_1\cdots L_{k-1}}~,
\eea
is a 2-CKS (2-KS) tensor. In addition, if $\alpha$ and $\beta$ are CCKY forms of rank $k$ and $\ell$,  respectively, then $\alpha\wedge \beta$ is a ($k+\ell$)-CCKY form.

KY forms generate symmetries \cite{gibbons} for spinning particle actions \cite{bvh}.  These are supersymmetric extensions of (\ref{gact}). Such an action is
\bea
A=-{i\over2} \int\, d\tau\, d\theta\, g_{MN}\, D x^M\, \dot x^N~,
\label{sgact}
\eea
where $x$ are superfields $x=x(\tau, \theta)$, $\tau$ is the even and $\theta$ is the odd coordinate of the worldline superspace, and the superspace derivative $D$ satisfies $D^2=i\partial_\tau$.
In particular, the KY form $\alpha$ generates the infinitesimal symmetry
\bea
\delta x^M=\epsilon\, \alpha^M{}_{N_1\cdots N_{k-1}} Dx^{N_1}\cdots Dx^{N_{k-1}}~,
\label{svar}
\eea
for the action (\ref{sgact}), where $\epsilon$ is an infinitesimal parameter.  The associated conserved charge is
\bea
&&Q(\alpha)=(k+1) \alpha_{N_1N_2\cdots N_k} \partial_\tau x^{N_1} Dx^{N_2}\cdots Dx^{N_k}
\cr
&&
\qquad\qquad\qquad
-{i\over k+1} (d\alpha)_{N_1N_2\cdots N_{k+1}} Dx^{N_1} Dx^{N_2}\cdots Dx^{N_{k+1}}~.
\label{ccalpha}
\eea
Observe that $Q(\alpha)$ is conserved, $DQ(\alpha)=0$, subject to the equations of motion of (\ref{sgact}).

Note that if the KY form $\alpha$ is closed, $d\alpha=0$, and so $\alpha$ is covariantly constant (or equivalently parallel) with respect to the Levi-Civita connection, then
\bea
\tilde Q(\alpha)= \alpha_{N_1N_2\cdots N_k} D x^{N_1} Dx^{N_2}\cdots Dx^{N_k}~,
\eea
is also conserved subject to the field equations of (\ref{sgact}), $\partial_\tau \tilde Q(\alpha)=0$. This gives the conservation of two charges
$\tilde Q(\alpha)$ and $D \tilde Q(\alpha)$. The latter is proportional to that in (\ref{ccalpha}) with $d\alpha=0$.

There are several generalisations of  CKY tensors \cite{gggpks, kubiznak, houri1, houri2, kygp, howe2}.  One of the most common ones is to replace the Levi-Civita connection that appears in the
definition  (\ref{cky}) with another connection, for example a connection with skew-symmetric torsion. Some of the properties mentioned above extend to the generalised  KY tensors.  For an application of the KY forms to G-structures see \cite{Ggp, sat}.

\subsubsection{Integrability and separability}

A dynamical system with a $2n$-dimensional phase space $P$ is completely integrable according to Liouville  provided it admits $n$ independent constants of motion, $Q^r$, $r=1,\dots,n$, including the Hamiltonian $H$, in involution. Independence  means that  the  map $Q: P\rightarrow \bR^n$  is of rank $n$, where  $Q=(Q^1, \dots, Q^n)$, and  in involution means that the Poisson bracket algebra of the constants of motion $Q^r$ vanishes
\bea
\{Q^r, Q^s\}_{\mathrm{PB}}=0~.
\eea

Returning to the particle system described by the action (\ref{gact}),
the conserved charges  (\ref{ccd}) can be written as functions  on phase space, $T^*M$,  as
\bea
Q(d)= d^{N_1\cdots N_k} p_{N_1} \dots p_{N_k}~,
\label{kscharge}
\eea
where $p_M$ is the conjugate momentum of $x^M$. It turns out that if $Q(d)$ and $Q(e)$ are conserved charges associated with KS tensors $d$ and $e$, then
$\{Q(d), Q(e)\}_{\mathrm{PB}}$ is associated with the KS tensor given in terms of the Nijenhuis-Schouten bracket
\bea
([d, e]_{\mathrm{NS}})^{N_1\cdots N_{k+\ell-1}}=k d^{M(N_1\cdots N_{k-1}} \partial_{M} e^{N_k\cdots N_{k+\ell-1})}-\ell e^{M(N_1\cdots N_{\ell-1}} \partial_{M} d^{N_k\cdots N_{k+\ell-1})}~,
\eea
of $d$ and $e$.  Therefore, one has
\bea
\{Q(d), Q(e)\}_{\mathrm{PB}}= Q([d, e]_{\mathrm{NS}})~.
\label{nsb}
\eea
Observe that if $d$ is a vector, then $[d, e]_{\mathrm{NS}}={\cal L}_d e$, i.e. the Nijenhuis-Schouten bracket is the Lie derivative of $e$ with respect to the vector field $d$.  So two charges are in involution provided that the Nijenhuis-Schouten bracket of the associated KS tensors vanishes.

Completely integrable systems are  special. There are difficulties in both finding conserved charges in involution and in proving that they are independent. For example if $Q(d)$ and $Q(e)$ are conserved charges, $Q(d) Q(e)$ is not an independent conserved charge,  as its inclusion in the map $Q: P\rightarrow \bR^n$ does not alter its rank. However for the geodesic flow described by the action (\ref{gact}) that we shall investigate below, there is a simplifying feature. The spacetimes we shall be considering  admit a non-abelian  group of isometries.
For every isometry generated by a Killing vector field $K_r$, there is an associated conserved charge
\bea
Q_r= K_r^M p_M~.
\eea
Of course these charges may not be in involution. However note that the charges $Q_r$ written in phase space do not depend on the spacetime metric.  They only depend on the way that the isometry group acts on the spacetime. Typically there are many metrics for which $Q_r$ are constants of motion for the action (\ref{gact}).  Of course any polynomial of $Q_r$ is also conserved and  is independent from the metric of the particle system. We shall refer to these charges as {\it orbital} to emphasise their independence from the spacetime metric.  In many occasions, it is possible to find polynomials of $Q_r$ which are independent and are in involution. Suppose that one can find  $n-1$ such independent (polynomial) orbital charges in involution and the Hamiltonian,
 \bea
H={1\over 2} g^{MN} p_M p_N~,
\eea
is independent from the orbital charges. Then  the geodesic flow  is completely integrable  because the orbital charges will Poisson commute with the Hamiltonian. Of course the Hamiltonian depends on the spacetime metric.  To distinguish the conserved charges which depend on the spacetime metric from the orbital ones we shall refer to former as {\it Hamiltonian}. We shall demonstrate that this strategy of proving complete integrability of a geodesic flow based on non-abelian isometries is particularly effective whenever the non-abelian group of isometries has a principal orbit in a spacetime of codimension of  at most one. The complete integrability of geodesics flows on homogeneous manifolds has  been extensively  investigated in the mathematics literature, see e.g. \cite{thimm}.

\subsubsection{An example}

Before we proceed to investigate the KS and KY  tensors and the integrability of the geodesic flow on some type II backgrounds, let us present an example. The standard example is that of the Kerr black hole. However more suitable for the examples that follow is to consider $\bR^{2n}$ with a conformally flat metric
\bea
g= h(|y|)\delta_{ij} dy^i dy^j~,
\label{cflmetr}
\eea
where $|y|$ is the length of the coordinate $y$ with respect to the Euclidean norm and $h>0$.

A direct computation reveals that the following tensors
\bea
d_{i_1\dots i_k}=  h^{k}(|y|)~   y^{j_1}\dots y^{j_q} a_{j_1\dots j_q, i_1\dots i_k}~,
\eea
are KS tensors provided that the coefficients $a$ are constant and satisfy
\bea
a_{(j_1\dots j_q, i_1)\dots i_k}=a_{j_1\dots (j_q, i_1\dots i_k)}=0~.
\eea
For each of these KS tensors, there is an associated conserved charge $Q(d)$ given in (\ref{kscharge}) of the geodesic flow on $\bR^{2n}$ with metric (\ref{cflmetr}).  These generate an infinite dimensional symmetry algebra for the action (\ref{gact}) with metric (\ref{cflmetr}) which is isomorphic to the Poisson algebra of $Q(d)$'s  up to terms proportional to the equations of motion, i.e. the algebra of symmetry transformations is isomorphic on-shell to the Poisson bracket algebra of the charges.  The conserved charges $Q(d)$ may neither be independent nor  in involution.

Next let us turn to find the KY and CCKY tensors on $\bR^{2n}$ with metric (\ref{cflmetr}).  After some computation, one finds that
 \bea
\alpha= h^{{k\over2}} i_Y \varphi~,~~~~\beta= h^{k+2\over2} Y\wedge \varphi~,
\eea
are KY and CCKY forms, respectively, for any constant $k$-form $\varphi$ on $\bR^{2n}$, where $Y$ is either the vector field $Y=y^i\partial_i$ or the one-form $Y=y_i dy^i$; it is clear from the context what $Y$ denotes in each case.

For each KY tensor above, one can construct the infinitesimal variation   (\ref{svar}) which is a symmetry of the action (\ref{sgact}). However the commutator of two such infinitesimal transformations does not close to an infinitesimal transformation of the same type. Typically, the right-hand side of the commutator will involve a term polynomial in $Dx$ as well as a term which is linear in the velocity $\dot x$. A systematic exploration of such commutators in a related context can be found
in \cite{phgp}.

Next let us turn to investigate the integrability of the geodesic flow of the metric (\ref{cflmetr}). The geodesic equations  can be easily integrated in angular coordinates. However it is instructive to provide a symmetry argument for the complete integrability of the geodesic equations.
The isometry group of the above backgrounds is $SO(2n)$.  The Killing vector fields are
\bea
k_{ij}=y_i\partial_j-y_j\partial_i~,~~~i<j~,
\label{rotvf}
\eea
where $y_i=y^i$. The associated conserved charges are
\bea
Q_{ij}=Q(k_{ij})= y_i  p_j-y_j p_i~.
\label{rotcc}
\eea
Notice that all these conserved charges are orbital as they do not depend on the metric (\ref{cflmetr}).  As ${\cal L}_{k_{ij}} g=0$, one can show that $Q_{ij}$ commute with the Hamiltonian $H= {1\over2} h^{-1} \delta^{ij} p_i p_j$, i.e. $\{H, Q_{ij}\}_{\mathrm{PB}}=0$.

The conserved charges $Q_{ij}$ are not in involution as $\{Q(k_{ij}), Q({k_{pq}})\}_{\mathrm{PB}}=Q([k_{ij}, k_{pq}])$.  However using these, one can verify that the $2n-1$ orbital conserved charges
\bea
D_m={1\over4} \sum_{i,j\geq 2n+1-m} (Q_{ij})^2~,~~~m\geq 2, \dots, 2n~,
\label{casichso2n}
\eea
are in involution.  These together with the Hamiltonian $H= {1\over2} h^{-1} \delta^{ij} p_i p_j$ give $2n$ charges in involution. Therefore the geodesic flow of the metric
 (\ref{cflmetr}) is completely (Liouville) integrable.

An alternative way to think about the complete integrability of the geodesic flow on $\bR^{2n}$ with metric (\ref{cflmetr}) is to consider it as a motion along the round sphere $S^{2n-1}$ in $\bR^{2n}$ and as a motion along the radial direction
$r=|y|$.  For this write the metric (\ref{cflmetr}) as
\bea
g= h(r) (dr^2+ r^2 g(S^{2n-1}))~,
\eea
where $g(S^{2n-1})$ is the metric on the round $S^{2n-1}$ sphere. It is well known that the vector fields (\ref{rotvf}) are tangential to $S^{2n-1}$ and leave the round metric on $S^{2n-1}$ invariant. The associated conserved charges are as in (\ref{rotcc}) and they are functions of  $T^*S^{2n-1}$, i.e. they do not depend on the radial direction $p_r$ of the momentum $p$. One can proceed to define (\ref{casichso2n}) and in turn show that the geodesic flow on $S^{2n-1}$ is completely integrable. Notice that $D_{2n}$ is the Hamiltonian of the geodesic flow on $S^{2n-1}$. All these charges including the Hamiltonian on $S^{2n-1}$ are orbital as they do not depend on the metric (\ref{cflmetr}).
As there are $2n-1$ independent charges in involution associated with the geodesic flow on $S^{2n-1}$, the addition of the Hamiltonian $H= {1\over2} h^{-1} \delta^{ij} p_i p_j$ of the geodesic flow on $\bR^{2n}$ gives
$2n$ independent conserved charges in involution proving the complete integrability of the geodesic flow of the metric (\ref{cflmetr}).

This construction can be reversed engineered and generalised.  In particular  consider a metric  on a n-dimensional manifold $M^n$
\bea
g(M^n)=dz^2+ g(N^{n-1})( z)
\eea
where $z$ is a coordinate and $g(N^{n-1})(z)$ is a metric on the submanifold $N^{n-1}$ of $M^n$ which may depend on $z$.  Suppose now there is a group of isometries on  $M^n$  which has as a principal orbit $N^{n-1}$. Clearly the associated conserved charges $Q=K^M p_M$, for each Killing vector field $K$, will be functions on $T^*N$. If one is able to find  orbital conserved charges $D_m$, $m=1,\dots, n-1$  in involution, then the geodesic flow on $M^n$ will be completely integrable after the inclusion of the Hamiltonian $H$
of the geodesic flow on $M^n$ as an additional conserved charge.  This is because $H$ is a function on $T^*M^n$ and so it is independent from $D_m$ which are functions on $T^*N^{n-1}$. Moreover $\{D_m, H\}_{\mathrm{PB}}=0$ as
$D_m$ are constructed as polynomials of the conserved charges associated with the isometries on $M^n$. This argument will be repeatedly used  to prove complete integrability of geodesic flows of brane backgrounds  and clearly can be adapted to all manifolds which have a principal orbit of codimension at most one with respect to a group action.

\subsection{D-branes}

\subsubsection{The KS and CCKY tensors of D-branes}

The metric of type II Dp-branes in the string frame \cite{d5, d5hs, d3, d7, d8a, d8} is
\bea
g=h^{-{1\over2}} \sum_{a,b=0}^p \eta_{ab}d\sigma^a d\sigma^b+ h^{{1\over2}} \sum_{i,j=1}^{9-p}\delta_{ij} dy^i dy^j~,
\label{dbrane}
\eea
 where $p=0,\dots, 8$ with $p$  even (odd) for IIA (IIB) D-branes,  $\sigma^a$ are the worldvolume coordinates, $y^i$ are the transverse coordinates and $h=h(y)$ is a harmonic function $\delta^{ij} \partial_i\partial_j h=0$.  Apart from the metric, the solutions depend on a non-vanishing dilaton field and an appropriate form field strength which we suppress. For  planar branes located at different points $y_s$ in $\bR^{9-p}$, one takes for $p\leq 6$
  \bea
  h=1+ \sum_{s} {q_s\over |y-y_s|^{7-p}}~,
  \label{mhf}
  \eea
  where $|\cdot|$ is the Euclidean norm in $\bR^{9-p}$ and $q_s$ is a constant proportional to the charge density of the branes.
  The solution is invariant under the action of the Poincar\'e group, $SO(p,1)\ltimes \bR^{p,1}$, acting on the worldvolume coordinates $\sigma^a$. If the harmonic function is chosen such that $h=h(|y|)$\footnote{The harmonic function is $h=1+{q\over |y|^{7-p}}$ for $p=0, \dots, 6$, $h=1+ q \log |y|$ for $=7$ and $h=1+ q |y|$ for $p=8$.}, then the solution will be invariant under the action of $SO(9-p)$ group acting  on the transverse coordinates $y$.

  Considering the Dp-branes (\ref{dbrane}) with $h=h(|y|)$, the KS tensors which are invariant under the worldvolume symmetry of the solution are
\bea
d_{a_1\dots a_{2m}i_1\dots i_k}=  h^{{1\over4} (k-m)}(|y|)\,\,   y^{j_1}\dots y^{j_q} a_{j_1\dots j_q, i_1\dots i_k} \eta_{(a_1a_2}\dots \eta_{a_{2m-1} a_{2m})}~,
\eea
provided that the constant coefficients $a$ satisfy
\bea
a_{(j_1\dots j_q, i_1)\dots i_k}=a_{j_1\dots (j_q, i_1\dots i_k)}=0~.
\eea
Each of these KS tensors will generate a symmetry of the relativistic particle action (\ref{gact}). As a result each such action on a D-brane background admits
an infinite number of symmetries. The  algebra of the associated transformations is on-shell isomorphic to that of the Poisson bracket algebra of the associated charges.

To investigate the symmetries of the spinning particles (\ref{sgact}) propagating on D-branes, it suffices to find the KY  tensors of these backgrounds. For this, one begins with an ansatz which respects the worldvolume isometries of the solutions.  As the KY tensors are dual to CCKY ones, let us focus on the latter. It turns out that
\bea
\beta(\varphi)=h^{k+1-p\over 4}(|y|)\,\, Y\wedge  \varphi\wedge d{\mathrm{vol}}(\bR^{p,1})~,
\eea
is a CCKY tensor for any constant $k$-form $\varphi$ on $\bR^{8-p}$, where $d{\mathrm{vol}}(\bR^{p,1})$ is the volume form of $\bR^{p,1}$ with respect to the flat metric and $Y=\delta_{ij} y^i dy^j$. Therefore Dp-branes admit $2^{8-p}$ linearly independent KY forms each generating a symmetry of the action (\ref{sgact}) of spinning particle probes in these backgrounds.
The associated conserved charges are given in (\ref{ccalpha}).

\subsubsection{Complete integrability of geodesic flow}

The geodesic flow on all Dp-brane backgrounds with $h=h(|y|)$ is completely integrable. Of course one can separate the geodesic equation in angular variables. Here we shall give all the charges which are in involution. As we have already mentioned, the isometry group of  such a Dp-brane solution is $SO(p,1)\ltimes \bR^{p,1}\times SO(9-p)$. Such a group has a codimension one principal orbit $\bR^{p,1}\times S^{8-p}$ in the Dp-brane background.  In particular, the Killing vectors generated by the translations along the worldvolume coordinates are $k_a=\partial_a$ and those generated by $SO(9-p)$ rotations on the transverse coordinates are
\bea
k_{ij}=y_i \partial_j-y_j\partial_i~,~~~i<j~,
\label{rotvfx}
\eea
where $y_i=y^i$. The associated conserved charges written in terms of the momenta are
 \bea
Q_a=p_a~,~~~ Q_{ij}=Q(k_{ij})=  y_i p_j-y_j p_i~.
 \eea
 These charges are  not in  involution.  However, one can verify that the 9 conserved charges
 \bea
 Q_a~,~~~D_m={1\over4} \sum_{i,j\geq 10-p-m} (Q_{ij})^2~,~~~m\geq 2, \dots, 9-p~,
 \eea
 are all orbital, independent and  in involution.  These together with the Hamiltonian of (\ref{gact}) yield 10 charges in involution and the geodesic flow on all such Dp-brane solutions  is completely integrable.

\subsection{Common sector branes}

\subsubsection{KS and KY tensors of common sector branes}

The metric of the fundamental string solution \cite{funstring} is
\bea
g=h^{-1} \eta_{ab} d\sigma^a d\sigma^b+ \delta_{ij} dy^i dy^i~,
\label{fstring}
\eea
where $a,b=0,1$ and $i,j=1,\dots 8$ and $h$ is a harmonic function on $\bR^8$, $\delta^{ij} \partial_i\partial_j h=0$. We have suppressed the  other two fields of the solution the dilaton and 3-form field strength.

As for D-branes consider the fundamental string solution with $h=h(|y|)=1+{q\over |y|^6}$. Such a solution admits the same isometry group as that of D1-brane. Then one can demonstrate that
the KS tensors that preserve the worldvolume symmetry of the fundamental string are
\bea
d_{a_1\dots a_{2m}i_1\dots i_k}=  h^{-m}(|y|)   y^{j_1}\dots y^{j_q} a_{j_1\dots j_q, i_1\dots i_k} \eta_{(a_1a_2}\dots \eta_{a_{2m-1} a_{2m})}~,
\eea
provided that the constant coefficients satisfy
$
a_{(j_1\dots j_q, i_1)\dots i_k}=a_{j_1\dots (j_q, i_1\dots i_k)}=0
$.
As a result a relativistic particle whose dynamics is described by the action (\ref{gact}) on such a background admits  an  infinite number of symmetries generated by these KS tensors.

After some computation, one can verify that  CCKY forms of the fundamental string solution are
\bea
\beta(\varphi)=h^{-1}(|y|)\, Y\wedge  \varphi\wedge d\sigma^0\wedge d\sigma^1~,
\eea
for any constant $k$-form $\varphi$ on $\bR^8$, where $Y=\delta_{ij} y^i dy^j$. These give rise to $2^7$ linearly independent dual  KY forms which generate  symmetries for  a spinning particle with action  (\ref{sgact}) propagating on this background.

The metric of the NS5-brane solution \cite{ns5, callan} is
\bea
g= \eta_{ab} d\sigma^a d\sigma^b+ h \delta_{ij} dy^i dy^j~,
\label{ns5}
\eea
where  $a,b=0,\dots, 5$,  $i,j=1,2,3,4$ and $h$ is a harmonic function on $\bR^4$. We have again suppressed the dilaton and 3-form fields of the solution.  For $h=h(|y|)=1+{q\over |y|^2}$, the solution has the same isometry group as that of the D5-brane.

As for the fundamental string solution above,  the KS tensors that preserve the worldvolume symmetry of the NS5-brane are
\bea
d_{a_1\dots a_{2m}i_1\dots i_k}=  h^{k}(|y|)\,\,   y^{j_1}\dots y^{j_q} a_{j_1\dots j_q, i_1\dots i_k} \eta_{(a_1a_2}\dots \eta_{a_{2m-1} a_{2m})}~,
\eea
provided that the constant tensors $a$ satisfy
$
a_{(j_1\dots j_q, i_1)\dots i_k}=a_{j_1\dots (j_q, i_1\dots i_k)}=0$.
Therefore the action (\ref{gact}) of a relativistic particle action propagating in this background admits an infinite number of symmetries generated by these KS tensors.

 The  CCKY forms of the NS5-brane are
\bea
\beta(\varphi)=h^{{k+2\over2}}(|y|)\,\, Y\wedge  \varphi\wedge d{\mathrm{vol}}(\bR^{5,1})~,
\eea
for any constant $k$-form $\varphi$ on $\bR^4$, where $Y=\delta_{ij} y^i dy^j$ and $d{\mathrm{vol}}(\bR^{5,1})$ is the volume form of the worldvolume of the NS5-branes with respect to the flat metric. These give rise to $2^3$ linearly independent  dual KY forms that generated the symmetries of a spinning particle with action
(\ref{sgact}) propagating on the background.

\subsubsection{Complete integrability of geodesic flow}

Consider  a relativistic particle propagating on the fundamental string solution with $h=h(|y|)$. The worldsheet translations and transverse coordinate $SO(8)$ rotations  give rise to   the conserved charges
\bea
Q_a=p_a~,~~~a=0,1~;\qquad Q_{ij}= y_i p_j-y_j p_i~,~~~i,j=1,\dots,8~,
\eea
respectively. From these one can construct  the following nine independent, orbital, conserved charges
\bea
Q_a~,~~~D_m={1\over4} \sum_{i,j\geq 9-m} (Q_{ij})^2~,~~~m=2, \dots, 8~,
\eea
which are independent and  in involution. These together with the Hamiltonian of the relativistic particle (\ref{gact}) lead to the integrability of the geodesic flow on the fundamental string background.

Similarly, the conserved charges of a relativistic particle propagating on a NS5-brane background associated  with the worldvolume translations and transverse $SO(4)$ rotations are
\bea
Q_a= p_a~,~~~a=0,\dots, 5~;~~~Q_{ij}=  (y_i p_j-y_j p_i)~,~~~i,j=1,2,3,4~.
\eea
These give rise to nine  independent, orbital, conserved charges
\bea
Q_a~,~~~D_m={1\over4} \sum_{i,j\geq  5-m}  (Q_{ij})^2~,~~~m=2,\dots, 4~,
\eea
which are independent and  in involution. These together with the Hamiltonian of the relativistic particle imply the complete integrability of the geodesic flow of NS5-brane.

\section{Common sector and TCFHs}

The simplest sector to explore the TCFH of type II supergravities is the common sector. For this sector, all fields vanish apart from the metric, dilaton and
the NS-NS 3-form field strength $H$, $dH=0$. A direct inspection of the TCFH of type II supergravities reveals that some of the spinor bilinears are covariantly constant with respect to a connection with skew-symmetric torsion while some others satisfy a more general TCFH.  The former are well known, especially in the context of string compactifications,  and have been extensively investigated  in the sigma model approach to string theory. They  generate additional supersymmetries of the worldvolume actions as well as
W-type of symmetries \cite{phgp}. Here we shall demonstrate that string probes on all common sector supersymmetric solutions admit W-type of symmetries generated by the form bilinears.

\subsection{Probes}

Before we proceed with the details of describing how the TCFHs generate symmetries for probes in supersymmetric backgrounds, we shall first describe the probe actions that we shall be considering. The main focus will be on string and particle probes. The dynamics of string probes propagating on a spacetime with metric $g$ and a 2-form gauge potential $b$ \cite{ zumino, lag, ghull, howesierra} is described by the action
\bea
A=\int d^2\rho\, d^2\theta\, (g+b)_{MN}\, D_+x^M\, D_-x^N~,
\label{sact}
\eea
 where $x=x(\rho, \theta)$ are real superfields that depend on the worldsheet superspace with commuting  $(\rho^0, \rho^1)$ and
anti-commuting $(\theta^+, \theta^-)$ real coordinates. The action above has been given as in  \cite{howegp2x} where one  defines the lightcone coordinates, $\rho^{\pp}= \rho^0+\rho^1$, $\rho^{=}=-\rho^0+\rho^1$,  and the algebra of superspace derivatives is   $D_-^2=i\partial_{=}$, $D_+^2=i\partial_{{\pp}}$ and $D_+ D_-+D_- D_+=0$\,. Note that the sign labelling of the worldsheet superspace  coordinates denotes $\mathfrak{spin}(1,1)$  chirality.

The infinitesimal symmetries of (\ref{sact}) that we shall be considering  are  given by
\bea
\delta x^M= \epsilon^{(+)} \beta^M{}_{P_1\dots P_k}\, D_+x^{P_1}\dots D_+x^{P_k}~,
\label{ssym1}
\eea
where  $\beta$ is  a spacetime ($k+1$)-form and $\epsilon^{(+)}$ is an infinitesimal parameter; the superscript $(+)$ indicates that the weight of the infinitesimal parameter $\epsilon$ is such that the right-hand side of (\ref{ssym1}) is a $\mathfrak{spin}(1,1)$ scalar.   The action  (\ref{sact}) is invariant under such transformations provided that
\bea
\nabla^{(+)}_M \beta_{P_1\dots P_{k+1}}=0~,
\label{conp}
\eea
where
\bea
\nabla^{(\pm)}=\nabla\pm {1\over 2}C~,
\eea
with $C=db$, i.e. $\nabla^{(\pm)}_M Y^N=\nabla_M Y^N\pm{1\over2} C^N{}_{MR} Y^R$. Therefore $\beta$ generates a symmetry provided it is a $\nabla^{(+)}$-covariantly constant form.

One can also consider symmetries of (\ref{sact}) generated by the infinitesimal transformation
\bea
\delta x^M= \epsilon^{(-)} \beta^M{}_{P_1\dots P_k} D_-x^{P_1}\dots D_-x^{P_k}~,
\label{ssym2}
\eea
where $\epsilon^{(-)}$ is an infinitesimal parameter.
The condition for invariance of the action in such a case is
\bea
\nabla^{(-)}_M \beta_{P_1\dots P_{k+1}}=0~,
\label{conm}
\eea
i.e. $\beta$ is a $\nabla^{(-)}$-covariantly constant form. In many examples that follow the spacetime will admit several $\nabla^{(\pm)}$-covariantly
constant forms which generate symmetries of the string probe action (\ref{sact}). All $\nabla^{(+)}$-covariantly constant forms of the common sector backgrounds coincide with those of heterotic supersymmetric backgrounds.  In turn these can be computed using  the classification results of \cite{hetgpug} for all heterotic background Killing spinors. The
$\nabla^{(-)}$-covariantly constant forms of common sector backgrounds can also be read from the classification results of \cite{hetgpug}.
One can easily investigate the commutators of these symmetries (\ref{ssym1}) and (\ref{ssym2}). In general these symmetries are of W-type and have been previously explored in  \cite{phgp} both in the context  of string compactifications and special geometric structures.

Actions of spinning particle probes are also invariant under the symmetries generated by either $\nabla^{(+)}-$ or $\nabla^{(-)}-$ covariantly constant forms   $\beta$. One such  worldline probe action is
\bea
A=\int\, d\tau\, d^2\theta\, (g+b)_{MN} D_+x^M D_-x^N~,
\label{pact}
\eea
which in addition to the metric exhibits a 2-form coupling $b$, where the superfields $x^M=x^M(\tau, \theta)$ depend on the worldline superspace with commuting $\tau$ and anti-commuting $(\theta^+, \theta^-)$ real coordinates; see \cite{colesgp} for a systematic description of spinning particle actions with form and other couplings. The algebra of the worldline superspace derivatives is $D_+^2=D_-^2=i\partial_\tau$ and $D_+D_-+D_-D_+=0$. The signs  on $\theta^\pm$ are just labels - there is no chirality in one dimension. The infinitesimal variation of the superfields is as in either (\ref{ssym1}) or (\ref{ssym2}),  but now the fields are worldline superfields and the superspace derivatives are those of the worldline superspace. The conditions for invariance of the action above are given in  either (\ref{conp}) or (\ref{conm}), respectively.

Another class of spinning particle probes we shall be considering are described by the action \cite{colesgp}
\bea
A=-{1\over2} \int\, d\tau\, d\theta\, \big(i g_{MN} Dx^M \partial_\tau x^N+{1\over6} C_{MNR} Dx^M Dx^N Dx^R\big)~,
\label{1part}
\eea
where $g$ is the spacetime metric and $C$ is a 3-form on the spacetime - $C$ is not a necessarily closed 3-form. Moreover
$x^M$ is a superfield that depends on the worldline superspace coordinates $(\tau, \theta)$ and $D^2=i\partial_\tau$. Given a ($k$+1)-form $\beta$
one can construct the infinitesimal transformation
\bea
\delta x^M= \alpha\,\, \beta^M{}_{P_1\dots P_k} Dx^{P_1}\dots Dx^{P_k}~,
\label{ssym3}
\eea
where $\alpha$ is an infinitesimal parameter. The conditions required for this action to be invariant under the transformation (\ref{ssym3}) can be arranged in two different ways. One way is to require, as in previous cases, that $\beta$ is $\nabla^{(+)}$-covariantly constant.  An alternative way to arrange  the conditions for invariance  of (\ref{1part})   is
\bea
&&\nabla^{(+)}_M \beta_{P_1\dots P_{k+1}}=\nabla^{(+)}_{[M} \beta_{P_1\dots P_{k+1}]}~,
\cr
&&di_\beta C+(-1)^k {k+2\over2} i_\beta dC=0~.
\label{addcon}
\eea
These conditions and an explanation of the notation can be found in \cite{kygp}.
Therefore this set of conditions implies that $\beta$ is a $\nabla^{(+)}$-KY form. For $C=0$, one obtains that $\beta$ is a KY form as for the  spinning particles
described by the action (\ref{sgact}).

\subsection{IIA common sector}

\subsubsection{The TCFH}

The TCFH of the common sector can be written as
\bea
\nabla_M \phi_{N_1 \dots N_p} -\frac{p}{2}  H^P{}_{M[N_1} \tilde\phi_{|P|\dots N_p]}=0~,~~\nabla_M \tilde\phi_{N_1 \dots N_p} -\frac{p}{2}  H^P{}_{M[N_1} \phi_{|P|\dots N_p]}=0~,
\label{iiacovf}
\eea
for $\phi= k, \pi, \tau$ and
\begin{equation}
\nabla_M \tilde{\sigma} = -\frac{1}{4}H_{MPQ}\omega^{PQ} ~,~~~		\nabla_M \omega_{NR} +\frac{1}{4}H_{MPQ}\tilde{\zeta}^{PQ}{}_{NR} = \frac{1}{2}H_{MNR}\tilde{\sigma} ~  ,
\label{iaom}
\end{equation}

\bea
&&\nabla_M \tilde{\zeta}_{N_1 \dots N_4} +\frac{1}{3} {}^\star{H}_{M[N_1N_2N_3|PQR|}\tilde{\zeta}^{PQR}{}_{N_4]} - 3 H_{M[N_1 N_2}\,\omega_{N_3 N_4]} =
\cr
&&\quad-\frac{1}{12}g_{M[N_1}{}^\star{H}_{N_2 N_3 N_4] P_1 \dots P_4}\tilde{\zeta}^{P_1 \dots P_4} + \frac{5}{12} {}^\star{H}_{[MN_1N_2N_3|PQR|}\tilde{\zeta}^{PQR}{}_{N_4]}~,
\eea

\begin{equation}
	\nabla_M \sigma = - \frac{1}{4} H_{MPQ} \tilde \omega^{PQ} ~ ,~~~
		\nabla_M \tilde\omega_{NR}+\frac{1}{4} H_{MPQ} \zeta^{PQ}{}_{NR}  = \frac{1}{2} H_{MNR} \sigma ~ ,
\label{iiatom}
\end{equation}

\bea
&&
\nabla_M \zeta_{N_1 \dots N_4} +  \frac{1}{3} {}^\star{H}_{M[N_1N_2N_3|PQR|}\zeta^{PQR}{}_{N_4]}- 3 H_{M[N_1N_2} \tilde\omega_{N_3N_4]}= \cr
&&
\qquad -\frac{1}{12} g_{M[N_1} {}^\star{H}_{N_2 N_3 N_4] P_1 \dots P_4} \zeta^{P_1\dots P_4}  + \frac{5}{12} {}^\star{H}_{[MN_1N_2N_3|PQR|}\zeta^{PQR}{}_{N_4]}  ~.
\label{iiiatom}
\eea
These can be easily derived from the general IIA TCFH in section \ref{iiatcfhs} upon setting all other fields apart from the metric, dilaton and NS-NS 3-form to zero.

It is clear from the TCFH that $k^{\pm rs}=k^{rs}\pm \tilde k^{rs}$, $\pi^{\pm rs}=\pi^{rs}\pm \tilde \pi^{rs}$  and $\tau^{\pm rs}= \tau^{rs}\pm \tilde \tau^{rs}$ are covariantly constant
\bea
\nabla^{(\pm)}k^{\pm rs}=\nabla^{(\pm)} \pi^{\pm rs}=\nabla^{(\pm)} \tau^{\pm rs}=0~,
\label{ccf}
\eea
 with respect to the connections
\bea
\nabla^{(\pm)}=\nabla\pm {1\over2} H~.
\label{nablapm}
\eea
These are the forms that have mostly been explored in the literature. Although the rest do not satisfy such a straightforward condition they are nevertheless part of the
geometric structure of the common sector backgrounds. A consequence of the TCFH above is that
the reduced holonomy of the minimal connection ${\cal D}^{\cal F}$ of a generic common sector background is included in $\times^6 SO(9,1)\times^2 GL(255)$.

\subsubsection{Probe hidden symmetries generated by the TCFH}

After identifying the 3-form coupling $C=db$ of the probe actions (\ref{pact}) and  (\ref{sact}) with the  3-form field strength $H$ of common sector backgrounds,  $C=H$, the conditions
on the form bilinears $k^{\pm rs}$, $\pi^{\pm rs}$  and $\tau^{\pm rs}$ imposed by the TCFH (\ref{ccf})  coincide with those in (\ref{conp}) and (\ref{conm}) as required for the invariance of these probe actions.  Therefore the $\nabla^{(\pm)}$-covariantly constant form bilinears $k^{\pm rs}$, $\pi^{\pm rs}$  and $\tau^{\pm rs}$  generate symmetries for the particle (\ref{pact}) and string (\ref{sact}) probe actions. These are given  by  the infinitesimal transformations
\bea
&&\delta x^M= \epsilon^{(\pm)}_{ rs}(k^{\pm rs})^M~,~~~ \delta x^M= \epsilon^{(\pm)}_{ rs} (\pi^{\pm rs})^M{}_{PQ} D_\pm x^P D_\pm x^Q~,~~~
\cr
&&\delta x^M= \epsilon^{(\pm)}_{ rs} (\tau^{\pm rs})^M{}_{N_1\dots N_4} D_\pm x^{N_1}\dots  D_\pm x^{N_4}~,
\label{iiainf}
\eea
where $\epsilon^{(\pm)}_{ rs}$  are the infinitesimal parameters.

Similarly after identifying $C$ with $H$  the spinning particle probes described by the action (\ref{1part})
are invariant under symmetries generated by the $\nabla^{(+)}$-covariantly constant forms  $k^{+ rs}$, $\pi^{+ rs}$  and $\tau^{+ rs}$.  The infinitesimal variations are given
 as in (\ref{iiainf}) after replacing the worldsheet superfields with the worldline ones and the superspace derivative $D_+$  with $D$. The $\nabla^{(-)}$-covariantly constant forms  $k^{- rs}$, $\pi^{- rs}$  and $\tau^{- rs}$ also generate symmetries but for the spinning particle probe with
action given in (\ref{1part}) but now with coupling $C$  identified with $-H$, $C=-H$.

The interpretation of the rest of the form bilinears satisfying the TCFH conditions (\ref{iaom})-(\ref{iiiatom}) as generators of symmetries of worldvolume probe actions
is not apparent. For generic common sector backgrounds, these bilinears do not generate symmetries for the probe actions we have considered here.  Nevertheless, they may generate symmetries for probes on some special backgrounds, as some terms in the TCFH may vanish and so the remaining TCFH conditions can be interpreted as invariance conditions of some worldvolume probe action.

\subsubsection{Hidden symmetries of probes on common sector IIA branes}

We have demonstrated that particle and string probes in common sector backgrounds exhibit a large number of symmetries generated by the $\nabla^{(\pm)}$-covariantly constant forms  $k^{\pm rs}$, $\pi^{\pm rs}$  and $\tau^{\pm rs}$. To present some examples, we shall explore the symmetries generated by the form bilinears of the fundamental string and NS5-brane. For this, we have to compute the form bilinears of these two backgrounds.

To begin, let us assume that  the worldsheet directions  of the fundamental string are along  $05$.  Then  the Killing spinors of the solution  can be written as $\epsilon=h^{-{1\over4}} \epsilon_0$, where $\epsilon_0$ is a constant spinor that satisfies the condition $\Gamma_0\Gamma_5\Gamma_{11}\epsilon_0=\pm \epsilon_0$ with the gamma matrices  in a frame basis\footnote{This will be the case for the conditions on the Killing spinors of all brane solutions that we shall investigate from now on.}. The metric of the solution  is given in (\ref{fstring}) after changing the worldvolume directions from $01$ to $05$ and taking $h$ to be any harmonic function on $\bR^8$, e.g. $h$ can be a multi-centred harmonic function as in (\ref{mhf}) for $p=1$. The choice of worldsheet directions we have made for the string above may be thought as unconventional.  However, it turns out that such a choice is aligned with the basis used in spinorial geometry  \cite{uggp} to construct realisations of Clifford algebras in terms of forms; for a review on spinorial geometry techniques see \cite{review}. We shall use spinorial geometry   to solve the condition on $\epsilon_0$ and so this  labelling of the coordinates is convenient.

 Indeed choosing the plus sign in the  condition on $\epsilon_0$ and using the realisation of spinors in terms of forms\footnote{In spinorial geometry the Dirac spinors of $\mathfrak{spin}(9,1)$ are identified with
 $\Lambda^*(\bC^5)$. The Gamma matrices are realised on $\Lambda^*(\bC^5)$ using the exterior multiplication and inner derivation operations with respect to a Hermitian basis $(e_1, \dots, e_5)$ in $\bC^5$. The Majorana spinors satisfy the reality condition $\Gamma_{6789} *\epsilon=\epsilon$. For more details see e.g. appendix B of \cite{review}.} write   $\epsilon_0=\eta+ e_5\wedge \lambda$, where $\eta$ and $\lambda$ are constant Majorana $\mathfrak{spin}(8)$ spinors. Then the condition $\Gamma_0\Gamma_5\Gamma_{11}\epsilon_0= \epsilon_0$ restricts  $\eta$ and $\lambda$ to be positive chirality Majorana-Weyl spinors of $\mathfrak{spin}(8)$, i.e. $\eta, \lambda \in \Delta_8^{+}\equiv \Lambda^{\mathrm{ev}}(\bR\langle e_1, e_2, e_3, e_4\rangle)$.
 Thus the most general solution of $\Gamma_0\Gamma_5\Gamma_{11}\epsilon_0= \epsilon_0$ is
 \bea
 \epsilon_0=\eta+ e_5\wedge \lambda~,
 \label{fsol}
 \eea
 where $\eta$ and $\lambda$ are positive chirality Majorana-Weyl spinors of $\mathfrak{spin}(8)$.

 Using (\ref{fsol}) one can easily express all the form bilinears of the fundamental string background in terms of the form bilinears of $\eta$ and $\lambda$. The explicit expressions have been collected in appendix \ref{apa}.  Using these one finds that

\bea
&&k^{+rs}= 2 h^{-{1\over2}}\langle \eta^r, \eta^s\rangle  (e^0-e^5)~,~~~ k^{-rs}=  h^{-{1\over2}} \langle \lambda^r, \lambda^s\rangle  (e^0+e^5)~,
\cr
&&
\pi^{+rs}=h^{-{1\over2}} \langle \eta^r, \Gamma_{ij}\eta^s\rangle   (e^0-e^5)\wedge e^i\wedge e^j~,~~~
\pi^{-rs}=  h^{-{1\over2}}\langle \lambda^r,\Gamma_{ij} \lambda^s\rangle  (e^0+e^5)\wedge e^i\wedge e^j~,
\cr
&&
\tau^{+rs}={2\over 4!}h^{-{1\over2}}\langle \eta^r,\Gamma_{ijk\ell} \eta^s\rangle   (e^0-e^5)\wedge  e^i\wedge e^j\wedge e^k\wedge e^\ell~,
\cr
&&
\tau^{-rs}= {2\over 4!} h^{-{1\over2}} \langle \lambda^r,\Gamma_{ijk\ell} \lambda^s\rangle  (e^0+e^5)\wedge  e^i\wedge e^j\wedge e^k\wedge e^\ell~,
\eea
where $(e^0, e^5, e^i)$ is a pseudo-orthonormal frame for the metric (\ref{fstring}), i.e. $g=-(e^0)^2+ (e^5)^2+\sum_i (e^i)^2$,  and $\langle\cdot, \cdot\rangle$ is the $\mathfrak{spin}(8)$-invariant (Hermitian) inner product on $\Delta_8^{+}$. Both $k^{\pm rs}$ are along the worldvolume directions and Killing. This implies that both $k$ and $\tilde k$ are Killing as well.  This is expected for $k$ but not for $\tilde k$. Nevertheless  $\tilde k$ is Killing because the fundamental string is a special background. Observe that the $\nabla^{(+)}$- ($\nabla^{(-)}$-) parallel form bilinears are left- (right-) handed from the string worldvolume perspective as indicated by their dependence on the worldsheet lightcone directions.

It remains to compute the bilinears of $\mathfrak{spin}(8)$  Majorana-Weyl spinors $\eta$ and $\lambda$.  These can be obtained using the decomposition of the product of two positive chirality Majorana-Weyl representations  $\Delta_8^{+}$ in terms of forms on $\bR^8$ as
\bea
&&\Delta_8^{+}\otimes \Delta_8^+= \Lambda^{0}(\bR^8)\oplus \Lambda^{2}(\bR^8)\oplus \Lambda^{4+}(\bR^8)~,
\label{repdec}
\eea
 where $\Lambda^{4+}(\bR^8)$ are the self-dual 4-forms on $\bR^8$. As $\eta$ and $\lambda$ are in $\Delta_8^{+}$ and  otherwise unrestricted, their bilinears span all 0-, 2- and self-dual 4-forms in $\bR^8$. As a consequence, the string probe  (\ref{sact}) and particle probe (\ref{pact}) actions  are invariant under $2^7$ independent symmetries.

Next let us turn to the symmetries of probes on the NS5-brane background. Choosing the worldvolume of the NS5-brane along the  $012567$ directions,  the Killing spinors $\epsilon=\epsilon_0$  of the background satisfy the condition
$\Gamma_{3489}\Gamma_{11}\epsilon_0=\pm\epsilon_0$, where $\epsilon_0$ is a constant Majorana spinor. The metric of the solution is given in (\ref{ns5}) after changing the worldvolume directions from $012345$ to $012567$ for similar reasons as those explained for the fundamental string above and after taking $h$ to be a harmonic function on $\bR^4$ as in (\ref{mhf}) for $p=5$.  Choosing the plus sign, the condition $\Gamma_{3489}\Gamma_{11}\epsilon_0=\epsilon_0$ can be solved using spinorial geometry. It is convenient to first solve this condition  for Dirac spinors and then impose the reality condition on $\epsilon$.  The solution can be expressed as
\bea
\epsilon=\eta^1+e_{34}\wedge \lambda^1+ e_3\wedge \eta^2+e_4\wedge \lambda^2~,
\label{ns5s}
\eea
where $\eta$ and $\lambda$ are positive chirality Weyl spinors of $\mathfrak{spin}(5,1)$, i.e. $\eta, \lambda\in \Delta^+_{(6)}\equiv  \Lambda^{\mathrm{ev}}(\bC\langle e_1,e_2, e_5\rangle)$.  Imposing the reality condition on $\epsilon$, $\Gamma_{6789}*\epsilon=\epsilon$, one finds that
\bea
\lambda^1=-\Gamma_{67} (\eta^1)^*~,~~~\lambda^2=-\Gamma_{67} (\eta^2)^*~.
\eea
So the Killing spinor $\epsilon$ is completely determined by  the (complex) positive chirality  $\mathfrak{spin}(5,1)$ spinors  $\eta^1$ and  $\eta^2$.

Using (\ref{ns5s}), one can easily compute all the form bilinears of the NS5-brane background and express them in terms of the form bilinears of $\eta^1$ and $\eta^2$. All these can be found in appendix \ref{apa}.

In particular the $\nabla^{(\pm)}$-covariantly constant spinor bilinears are
\bea
&&k^{+rs}=4 \mathrm{Re}\langle \eta^{1r}, \Gamma_a\eta^{1s}\rangle_D e^a~,~~~k^{-rs}=4\mathrm{Re}\langle \eta^{2r}, \Gamma_a\eta^{2s}\rangle_D ~e^a~,
\eea

\bea
&&\pi^{+rs}={2\over3}\mathrm{Re}\langle \eta^{1r},\Gamma_{abc} \eta^{1s}\rangle_D\,  e^a\wedge e^b\wedge e^c
-4\mathrm{Re}\langle \eta^{1r},\Gamma_{a} \lambda^{1s}\rangle_D\, (e^3\wedge e^4 -e^8\wedge e^9)\wedge e^a
\cr
&&\qquad\qquad
-4\mathrm{Im}\langle \eta^{1r},\Gamma_{a} \eta^{1s}\rangle_D\, (e^3\wedge e^8 +e^4\wedge e^9)\wedge e^a
\cr
&&\qquad\qquad
-4\mathrm{Im}\langle \eta^{1r},\Gamma_{a} \lambda^{1s}\rangle_D\, (e^3\wedge e^9 -e^4\wedge e^8)\wedge e^a~,
\eea

\bea
&&\pi^{-rs}={2\over3}\mathrm{Re}\langle \eta^{2r}, \Gamma_{abc} \eta^{2r}\rangle_D\,  e^a\wedge e^b\wedge e^c
+4\mathrm{Re}\langle \eta^{2r},\Gamma_{a} \lambda^{2s}\rangle_D\, (e^3\wedge e^4 +e^8\wedge e^9)\wedge e^a
\cr
&&\qquad\qquad
+4\mathrm{Im}\langle \eta^{2r},\Gamma_{a} \eta^{2s}\rangle_D\, (e^3\wedge e^8 -e^4\wedge e^9)\wedge e^a
\cr
&&\qquad\qquad
+4\mathrm{Im}\langle \eta^{2r},\Gamma_{a} \lambda^{2s}\rangle_D\, (e^3\wedge e^9 +e^4\wedge e^8)\wedge e^a~,
\eea

\bea
&&\tau^{+rs}= k^{+rs}\wedge e^3\wedge e^4\wedge e^8\wedge e^9
-{2\over3}\mathrm{Re}\langle \eta^{1r},\Gamma_{abc} \lambda^{1s}\rangle_D\, (e^3\wedge e^4 -e^8\wedge e^9)\wedge e^a\wedge e^b\wedge e^c
\cr
&&\qquad\qquad
-{2\over3}\mathrm{Im}\langle \eta^{1r},\Gamma_{abc} \eta^{1s}\rangle_D\, (e^3\wedge e^8 +e^4\wedge e^9)\wedge e^a\wedge e^b\wedge e^c
\cr
&&\qquad\qquad
-{2\over3}\mathrm{Im}\langle \eta^{1r},\Gamma_{abc} \lambda^{1s}\rangle_D\, (e^3\wedge e^9 -e^4\wedge e^8)\wedge e^a\wedge e^b\wedge e^c
\cr
&&\qquad\qquad
+{4\over5!} \mathrm{Re}\langle \eta^{1r}, \Gamma_{a_1\dots a_5}\eta^{1s}\rangle_D~e^{a_1}\wedge\dots\wedge e^{a_5}~,
\eea

\bea
&&\tau^{-rs}=- k^{-rs}\wedge e^3\wedge e^4\wedge e^8\wedge e^9
+{2\over3}\mathrm{Re}\langle \eta^{2r},\Gamma_{abc} \lambda^{2s}\rangle_D (e^3\wedge e^4 +e^8\wedge e^9)\wedge e^a\wedge e^b\wedge e^c
\cr
&&\qquad\qquad
+{2\over3}\mathrm{Im}\langle \eta^{2r},\Gamma_{abc} \eta^{2s}\rangle_D (e^3\wedge e^8 -e^4\wedge e^9)\wedge e^a\wedge e^b\wedge e^c
\cr
&&\qquad\qquad
+{2\over3}\mathrm{Im}\langle \eta^{2r},\Gamma_{abc} \lambda^{2s}\rangle_D (e^3\wedge e^9 +e^4\wedge e^8)\wedge e^a\wedge e^b\wedge e^c
\cr
&&\qquad\qquad
+{4\over5!} \mathrm{Re}\langle \eta^{2r}, \Gamma_{a_1\dots a_5}\eta^{2s}\rangle_D~e^{a_1}\wedge\dots\wedge e^{a_5}~,
\eea
where  $a,b,c=0,1,2,5,6,7$ are the worldvolume directions, $(e^a, e^3, e^4, e^8, e^9)$ is a pseudo-orthonormal frame for the metric (\ref{ns5}), $\langle\cdot,\cdot\rangle_D$ is the $\mathfrak{spin}(5,1)$ invariant Dirac inner product and $\epsilon_{3489}=1$. Both $k^{\pm rs}$ are along the worldvolume directions of the brane and  are Killing. This in turn implies that both $k$ and $\tilde k$ are Killing as well.   Again  $\tilde k$ is Killing  because the NS5-brane is a special background.  The 3- and 5-forms have mixed components along both worldvolume and transverse directions. Note that the anti-self-dual and self-dual 2-forms along the transverse directions contribute to $\nabla^{(+)}$ and $\nabla^{(-)}$ covariantly constant
forms, respectively.

 Therefore the  NS5-brane form bilinears have been expressed in terms of those of two positive chirality Weyl  $\mathfrak{spin}(5,1)$  spinors. The decomposition  of two positive chirality Weyl $\mathfrak{spin}(5,1)$ representations into forms on $\bC^6$ is given by
 \bea
 \otimes^2 \Delta^+_{(6)}= \Lambda^1(\bC^6)\oplus \Lambda^{3+} (\bC^6)
 \eea
  Therefore the string probe with action  (\ref{sact})  and
 particle probe with action  (\ref{pact}) are invariant under $2^5$ symmetries counted over the reals.  To see this, observe that from the decomposition above all 1- and self-dual 3-forms along the NS5-brane worldvolume are spanned by these spinors. So there are $6+10=2^4$ independent symmetries generated by the $\nabla^{(+)}$-covariantly constant forms and similarly for the $\nabla^{(-)}$-covariantly constant forms yielding $2^5$ in total.   These generate a symmetry algebra of W-type \cite{phgp}. For the remaining form bilinears in appendix \ref{apa}, there is not a straightforward way to relate them to symmetries of particle or string probe actions.

\subsection{IIB common sector}

\subsubsection{The TCFH and probe hidden symmetries}

The  TCFH of IIB common sector can be written as

\bea
 \nabla_M\, \phi^{rs}_{N_1\dots N_p} -\frac{p}{2}\,H_{M[N_1}{}^P\,\phi^{(3)rs}_{|P|\dots N_p]}=0 ~, ~~~\nabla_M\, \phi^{(3)rs}_{N_1\dots N_p} -\frac{p}{2}\,H_{M[N_1}{}^P\,\phi^{rs}_{|P|\dots N_p]}=0 ~,
\eea
for $\phi= k, \pi$ and $\tau$.
The rest of the TCFH is
\bea
 \nabla_M\, k^{(\alpha)rs}_P + \frac{i}{4}\,\varepsilon_{\alpha\beta}\,H_{M}{}^{N_1N_2}\pi^{(\beta)rs}_{PN_1N_2}
= 0 ~,
\eea
\bea
 \nabla_M\,\pi^{(\alpha)rs}_{P_1P_2P_3}  + \frac{i}{4}\,\varepsilon_{\alpha\beta}\,H_{MN_1N_2}\,\tau^{(\beta)rs}_{P_1P_2P_3}{}^{N_1N_2}
-\frac{3i}{2}\,\varepsilon_{\alpha\beta}\,H_{M[P_1P_2}\,k^{(\beta)rs}_{P_3]}
= 0~,
\eea

\bea
 &&
 \nabla_M\,\tau^{(\alpha)rs}_{P_1\dots P_5}
 +\frac{5i}{4}\,\varepsilon_{\alpha\beta}\,{}^{\star}H_{M[P_1\dots P_4}{}^{N_1N_2}\,\pi^{(\beta)rs}_{P_5]N_1N_2}
-5i\,\varepsilon_{\alpha\beta}\,H_{M[P_1P_2}\,\pi^{(\beta)rs}_{P_3P_4P_5]}
=
\cr
&&
\qquad
 +\frac{5i}{12}\varepsilon_{\alpha\beta}\,g_{M[P_1}\,{}^{\star}H_{P_2\dots P_5]}{}^{N_1N_2N_3}\,\pi^{(\beta)rs}_{N_1N_2N_3} -\frac{3i}{2}\,\varepsilon_{\alpha\beta}\,{}^{\star}H_{[P_1\dots P_5}{}^{N_1N_2}\,\pi^{(\beta)rs}_{M]N_1N_2} ~.
\eea
where $\alpha, \beta=1,2$ and $\epsilon_{12}=1$. As it has already been explained the TCFH is real after replacing the purely imaginary form bilinears $k^{(2)}, \pi^{(2)}$ and $\tau^{(2)}$ with $ik^{(2)}, i\pi^{(2)}$ and $i\tau^{(2)}$.

It is clear from the TCFH above, that the forms $k^{\pm}\defeq k\pm k^{(3)}$, $\pi^{\pm}\defeq\pi\pm \pi^{(3)}$ and $\tau^{\pm}\defeq\tau\pm\tau^{(3)}$ are covariantly constant with respect to the $\nabla^{(\pm)}$ connection defined in (\ref{nablapm}).  As a result these form bilinears generate symmetries in the worldvolume probe actions
given in (\ref{sact}) and (\ref{pact}). These are the form bilinears that have mostly been explored in the literature. The remaining form bilinears in the TCFH do not have such an apparent interpretation. Nevertheless they are part of the
geometric structure of the common sector backgrounds.

 A consequence of the TCFH above is that
the holonomy of the minimal connection ${\cal D}^{\cal F}$ of a generic common sector background is included in $\times^6 SO(9,1)\times^2 GL(256)$.
Note that the holonomy of the IIA common sector minimal connection is included in $\times^6 SO(9,1)\times^2 GL(255)$.  The difference is that the action of  ${\cal D}^{\cal F}$ on the IIA 0-form bilinears $\sigma$ and $\tilde \sigma$ is via a partial derivative and so the holonomy is trivial. However if instead we had  considered the maximal TCFH connections, see \cite{gptcfh},  of the IIA and IIB common sector both would have reduced holonomy contained in $\times^6 SO(9,1)\times^2 GL(256)$.

\subsubsection{Hidden symmetries of probes on common sector IIB branes}

As an example, we shall explicitly give the symmetries of string and particle probes on the IIB fundamental string and NS5-brane backgrounds. For this one has to calculate the form bilinears of these solutions. Starting with the fundamental string and
choosing the worldsheet  along the $05$ directions as in the IIA case, the Killing spinors of the background are $\epsilon=h^{-{1\over 4}} \epsilon_0$, where the constant spinor $\epsilon_0$, $\epsilon_0^t=(\epsilon^1_0, \epsilon_0^2)$, and the two components of $\epsilon_0$   satisfy the conditions
\bea
\Gamma_{05}\epsilon_0^1=\pm \epsilon^1_0~,~~~\Gamma_{05} \epsilon^2_0=\mp \epsilon^2_0~.
\label{iibfcon}
\eea
Both  $\epsilon_0^1$ and $\epsilon_0^2$  are Majorana-Weyl $\mathfrak{spin}(9,1)$ spinors.
The metric of the solution is described in (\ref{fstring}) after changing the worldsheet directions from $01$ to $05$ and  $h$ is  taken to be a general harmonic function on $\bR^8$, given in (\ref{mhf}) for  $p=1$.
 To solve the above condition, we shall again use spinorial geometry \cite{uggp}. In particular  choosing the plus sign in (\ref{iibfcon})   and writing $\epsilon_0=\eta+e_5\wedge \lambda$, where $\eta$ ($\lambda$) is a doublet of chiral (anti-chiral) Majorana-Weyl $\mathfrak{spin}(8)$ spinors, one finds that
\bea
\epsilon_0^1=\eta^1=\eta~,~~~\epsilon_0^2=e_5\wedge\lambda^2=e_5\wedge \lambda~,
\label{fiibsol}
\eea
i.e. the condition on the Killing spinor implies $\lambda^1=\eta^2=0$. One can use the solution (\ref{fiibsol}) to express the bilinears of the Killing spinors in terms of those of independent  $\mathfrak{spin}(8)$ spinors $\eta$ and $\lambda$.  The results can be found in appendix \ref{apa}.

In particular one finds that the $\nabla^{(\pm)}$-covariantly constant form bilinears can be expressed as
\bea
&&
k^{+rs}=2 h^{-{1\over2}}\langle \eta^r, \eta^s\rangle (e^0-e^5)~,~~k^{-rs}=2 h^{-{1\over2}} \langle \lambda^r, \lambda^s\rangle (e^0+e^5)~,
\cr
&&
\pi^{+rs}= h^{-{1\over2}} \langle\eta^r, \Gamma_{ij} \eta^s\rangle (e^0-e^5) \wedge e^i\wedge e^j~,~~~\pi^{-rs}= h^{-{1\over2}} \langle\lambda^r, \Gamma_{ij} \lambda^s\rangle (e^0+e^5) \wedge e^i\wedge e^j~,
\cr
&&
\tau^{+rs}={2\over4!} h^{-{1\over2}} \langle\eta^r, \Gamma_{i_1\dots i_4} \eta^s\rangle (e^0-e^5) \wedge e^{i_1}\wedge \dots \wedge e^{i_4}~,
\cr
&&
\tau^{-rs}={2\over4!} h^{-{1\over2}} \langle\lambda^r, \Gamma_{i_1\dots i_4} \lambda^s\rangle (e^0+e^5) \wedge e^{i_1}\wedge\dots \wedge e^{i_4}~,
\eea
where $(e^0, e^5, e^i)$ is a pseudo-orthonormal frame for the metric (\ref{fstring}) and $\langle\cdot, \cdot\rangle$ is the $\mathfrak{spin}(8)$ invariant inner product. As in the IIA case both $k^{\pm rs}$ are along the worldvolume directions and  Killing which in turn implies that $k$ and $k^{(3)}$ are Killing as well. The latter property is a special property of the IIB fundamental string solution.  In addition,  as in the IIA case,  the $\nabla^{(+)}-$ ($\nabla^{(-)}-$) parallel form bilinears are left- (right-) handed from the string worldvolume perspective as indicated by their dependence on the worldsheet lightcone directions.

It remains to find the form bilinears of the  $\mathfrak{spin}(8)$ spinors $\eta$ and $\lambda$. These can be identified from the decomposition
of the product of two chiral $\Delta^+_{(8)}$ and  two anti-chiral $\Delta^-_{(8)}$ Majorana-Weyl  representations of  $\mathfrak{spin}(8)$.  It is well known that
\bea
\Delta^\pm_{(8)}\otimes \Delta^\pm_{(8)}= \Lambda^0(\bR^8)\oplus \Lambda^2(\bR^8)\oplus \Lambda^{4\pm}(\bR^8)~.
\eea
Therefore these bilinears span all constant 0-, 2- and self-dual or anti-self-dual 4-forms on $\bR^8$.  As a result, the probe actions (\ref{sact}) and (\ref{pact})
admit $2^7$ independent symmetries generated by these forms.  Commutators of  symmetries generated by $\nabla^{(\pm)}$-covariantly constant forms have been examined  in \cite{phgp} and it was found that they are of W-type.
After some investigation it has been found that the remaining form bilinears do not generate symmetries in particle and string probe actions like
(\ref{sact}), (\ref{pact}) and  (\ref{1part}).

Next let us turn to investigate the form bilinears of the IIB NS5-brane. Choosing the worldvolume along the directions $051627$, the  Killing spinors  $\epsilon$ of the solution are constant, $\epsilon=\epsilon_0$, and satisfy the condition  $\Gamma_{3489}\epsilon^1=\pm \epsilon^1$ and $\Gamma_{3489}\epsilon^2=\mp \epsilon^2$, where
both $\epsilon^1$ and $\epsilon^2$ are Majorana-Weyl $\mathfrak{spin}(9,1)$ spinors. Choosing
the first sign, one can solve the above conditions using spinorial geometry \cite{uggp}. As in the IIA case, it is best to first  solve the condition for $\epsilon$ complex and then impose the reality condition.  The solution is
\bea
\epsilon^1=\eta^1+ e_{34}\wedge \lambda^1~,~~~\epsilon^2=e_3\wedge \eta^2+ e_4 \wedge \lambda^2~,
\label{iibsolns5}
\eea
where $\eta^1, \lambda^1$ are positive chirality $\mathfrak{spin}(5,1)$ spinors, i.e.  $\eta^1, \lambda^1\in \Lambda^{\mathrm{ev}}(\bC\langle e_1, e_2,  e_5\rangle)$,  and  $\eta^2,\lambda^2$ are negative chirality $\mathfrak{spin}(5,1)$ spinors, i.e. $\eta^2, \lambda^2\in \Lambda^{\mathrm{odd}}(\bC\langle e_1, e_2, e_5\rangle)$.
Moreover the reality condition on the $\epsilon^1$ and $\epsilon^2$ spinors implies that
\bea
\lambda^1=-\Gamma_{67} (\eta^1)^*~,~~~\lambda^2=-\Gamma_{67} (\eta^2)^*~.
\eea
Using (\ref{iibsolns5}), one can easily compute the form bilinears in terms of those of $\eta^1$ and $\eta^2$.  These can be found in appendix \ref{apa}.

Using  the expressions of the form bilinears in appendix \ref{apa}, one finds that the $\nabla^{(\pm)}$-covariant constant bilinears are
\bea
&&
k^{+rs}= 4 \mathrm{Re} \langle \eta^{1r}, \Gamma_a \eta^{1s}\rangle_D\, e^a~,~~~k^{-rs}=
4 \mathrm{Re} \langle \eta^{2r}, \Gamma_a \eta^{2s}\rangle_D\, e^a~,
\eea
\bea
&&\pi^{+rs}=- 4 \mathrm{Re} \langle \eta^{1r}, \Gamma_a \lambda^{1s}\rangle_D \, e^a \wedge (e^3\wedge e^4-e^8\wedge e^9)
-4 \mathrm{Im}\langle \eta^{1r}, \Gamma_a \eta^{1s}\rangle_D\, e^a\wedge (e^3\wedge e^8+ e^4\wedge e^9)
\cr
&&\qquad\qquad
- 4 \mathrm{Im} \langle \eta^{1r}, \Gamma_a \lambda^{1s}\rangle_D \, e^a \wedge (e^3\wedge e^9-e^4\wedge e^8)
\cr
&&\qquad\qquad
+{2\over3}\mathrm{Re} \langle \eta^{1r}, \Gamma_{abc} \eta^{1s}\rangle_D  e^a\wedge e^b \wedge e^c~,
\eea

\bea
&&\pi^{-rs}=4  \mathrm{Re} \langle \eta^{2r}, \Gamma_a \lambda^{2s}\rangle_D\, e^a \wedge (e^3\wedge e^4+e^8\wedge e^9)
+4 \mathrm{Im}\langle \eta^{2r}, \Gamma_a \eta^{2s}\rangle_D\, e^a\wedge (e^3\wedge e^8- e^4\wedge e^9)
\cr
&&\qquad\qquad
+4  \mathrm{Im} \langle \eta^{2r}, \Gamma_a \lambda^{2s}\rangle_D\, e^a \wedge (e^3\wedge e^9+e^4\wedge e^8)
\cr
&&\qquad\qquad
+{2\over3}  \mathrm{Re} \langle \eta^{2r}, \Gamma_{abc} \eta^{2s}\rangle_D  e^a\wedge e^b \wedge e^c~,
\eea

\bea
&&\tau^{+rs}= 4\mathrm{Re}\langle \eta^{1r}, \Gamma_a \eta^{1s}\rangle_D
\,e^a\wedge e^3\wedge e^4\wedge e^8\wedge e^9+{4\over5!} \mathrm{Re} \langle \eta^{1r}, \Gamma_{a_1\dots a_5} \eta^{1s}\rangle_D  e^{a_1}\wedge\dots \wedge e^{a_5}
\cr
&&\qquad\qquad
- {2\over3} \mathrm{Re} \langle \eta^{1r}, \Gamma_{abc} \lambda^{1s}\rangle_D \, e^a \wedge  e^b \wedge e^c \wedge(e^3\wedge e^4-e^8\wedge e^9)
\cr
&&\qquad\qquad
-{2\over3} \mathrm{Im}\langle \eta^{1r}, \Gamma_{abc} \eta^{1s}\rangle_D\, e^a\wedge  e^b \wedge e^c \wedge (e^3\wedge e^8+ e^4\wedge e^9)
\cr
&&\qquad\qquad
- {2\over3} \mathrm{Im} \langle \eta^{1r}, \Gamma_{abc} \lambda^{1s}\rangle_D \, e^a\wedge  e^b \wedge e^c  \wedge (e^3\wedge e^9-e^4\wedge e^8)~,
\eea

\bea
&&\tau^{-rs}= -4 \mathrm{Re}\langle \eta^{2r}, \Gamma_a \eta^{2s}\rangle_D
\,e^a\wedge e^3\wedge e^4\wedge e^8\wedge e^9+{4\over5!}  \mathrm{Re} \langle \eta^{2r}, \Gamma_{a_1\dots a_5} \eta^{2s}\rangle_D e^{a_1}\wedge\dots \wedge e^{a_5}
\cr
&&\qquad\qquad
+{2\over3}  \mathrm{Re} \langle \eta^{2r}, \Gamma_{abc} \lambda^{2s}\rangle_D\, e^a\wedge  e^b \wedge e^c  \wedge (e^3\wedge e^4+e^8\wedge e^9)
\cr
&&\qquad\qquad
+{2\over3} \mathrm{Im}\langle \eta^{2r}, \Gamma_{abc} \eta^{2s}\rangle_D\, e^a\wedge  e^b \wedge e^c \wedge (e^3\wedge e^8- e^4\wedge e^9)
\cr
&&\qquad\qquad
+{2\over3}  \mathrm{Im} \langle \eta^{2r}, \Gamma_{abc} \lambda^{2s}\rangle_D\, e^a\wedge  e^b \wedge e^c  \wedge (e^3\wedge e^9+e^4\wedge e^8)~,
\eea
where $(e^a, e^3, e^4, e^8, e^9)$ is a pseudo-orthonormal frame of the NS5-brane metric (\ref{ns5}) and $\langle\cdot,\cdot\rangle_D$ is the $\mathfrak{spin}(5,1)$ invariant Dirac inner product. Clearly   $k^{\pm}$ are Killing which implies that  both
$k$ and $k^{(3)}$ are Killing as well.  As in all previous common sector branes,  the latter generates an additional symmetry for particle and string actions on NS5-brane backgrounds.  The 3- and 5-form bilinears above have mixed components along both the worldvolume and transverse directions, and the anti-self-dual and self-dual 2-forms along the transverse directions contribute to $\nabla^{(+)}$- and $\nabla^{(-)}$- covariantly constant
forms, respectively.

We have expressed the $\nabla^{(\pm)}$-covariantly constant bilinears in terms of the bilinears of the  chiral and anti-chiral  $\mathfrak{spin}(5,1)$ spinors
$\eta^1$ and $\eta^2$, respectively. To determine those note that
\bea
 \otimes^2 \Delta^\pm_{(6)}= \Lambda^1(\bC^6)\oplus \Lambda^{3\pm} (\bC^6)~.
 \eea
Therefore these span all 1-forms and 3-forms on the worldvolume of the NS5-brane. In particular, they generate $2^5$ independent symmetries, counting over the real numbers, for the
spinning particle and string probe actions in (\ref{sact}) and (\ref{pact}). The algebra of these symmetries is of W-type \cite{phgp}.  An investigation reveals that the remaining bilinears do not generate symmetries for the  (\ref{sact}) and (\ref{pact}) probe actions.

\section{IIA D-branes}

There is no classification of IIA supersymmetric backgrounds.  So to give more examples for which the  TCFH can be interpreted as  invariance condition  for probe particle and string actions under symmetries generated by the form bilinears, we shall turn to some special solutions and in particular to the D-branes\footnote{A consequence of this investigation is that we shall find all form bilinears of the type II D-brane solutions.}.  It is convenient to organise the investigation  in electric-magnetic brane pairs as the non-vanishing fields  that appear in TCFH are the same.
The TCFH for each D-brane pair can be easily found from that of the IIA TCFH given in (\ref{iiatcfha})-(\ref{iiatcfhf}) and (\ref{iiatcfha1})-(\ref{iiatcfha4}) upon setting all the form field strengths to zero apart from those associated to the D-brane under investigation.

\subsection{D0- and D6-branes}

\subsubsection{D0-branes}

The Killing spinors of the D0-brane are given by $\epsilon= h^{-{1\over8}} \epsilon_0$, where $\epsilon_0$ is a constant spinor
restricted as $\Gamma_0\Gamma_{11}\epsilon_0=\pm \epsilon_0$, the worldline is along the 0-th direction and $h$ is a multi-centred harmonic function as in (\ref{mhf}) for $p=0$.  Choosing the plus sign and using spinorial geometry \cite{uggp}, one can solve this condition  by setting
\bea
\epsilon_0=\eta-e_5\wedge \Gamma_{11} \eta~,
\label{d0sol}
\eea
 where
$\eta\in \Lambda^*(\bR\langle e_1, \dots, e_4\rangle)$ and the reality condition is imposed by $\Gamma_{6789}*\eta=\eta$.  Using this, one can compute the form bilinears.  These are given in appendix \ref{apb}.

As expected $k$ is a Killing vector.  As a result $k$ generates a symmetry in all probe actions (\ref{sact}), (\ref{pact}) and (\ref{1part}) after setting the form couplings to zero. It also generates a symmetry in the probe action of \cite{gpeb} with the 2-form coupling; the D0-brane 2-form field strength $F=F_{0i}\, e^0\wedge e^i$ is invariant under the action of $k$. An investigation of the TCFH for the rest of the form bilinears using that $F_{0i}\not=0$ reveals that these  do not generate symmetries  for the probe actions we have been considering. Because of this we postpone a more detailed analysis
of the TCFH for later and in particular for the D6- and D2-branes.

\subsubsection{D6-brane}

Choosing the transverse directions of the D6-brane along $549$, the Killing spinor $\epsilon=h^{-{1\over8}} \epsilon_0$ satisfies the condition
\bea
\Gamma_{549} \Gamma_{11}\epsilon_0=\pm\epsilon_0~,
\label{d6pro}
\eea
where $\epsilon_0$ is a constant spinor and $h$ is a multi-centred harmonic function as in (\ref{mhf}) with $p=6$. To solve this condition with the plus sign using spinorial geometry,  set
\bea
\epsilon_0=\eta+e_4\wedge \lambda~,
\label{d6sol}
\eea
 where $\eta, \lambda\in \Lambda^*(\bC\langle e_1, e_2, e_3, e_5\rangle)$.  Then the  condition (\ref{d6pro}) gives
\bea
\Gamma_5 \Gamma_{11}\eta=-i\eta~,~~~\Gamma_5 \Gamma_{11}\lambda=i\lambda~.
\label{d6elcon}
\eea
One can  proceed to  expand $\eta$ and $\lambda$ as $\eta=\eta^1+e_5\wedge \eta^2$ and $\lambda=\lambda^1+e_5\wedge \lambda^2$ in which case the  conditions (\ref{d6elcon}) give  $\eta^2=i \Gamma_{11} \eta^1$ and $\lambda^2=-i \Gamma_{11}\lambda^1$, where $\eta^1,  \lambda^1 \in \Lambda^*(\bC\langle e_1, e_2, e_3\rangle)$
are the independent spinors. However if one proceeds in this way the form bilinears will not be manifestly worldvolume Lorentz covariant, as the $0$-th direction will be separated from the rest.  Because of this, we shall not solve (\ref{d6elcon})  and do the computation with $\eta$ and $\lambda$. After the computation of the form bilinears, one can substitute in the formulae the solution of (\ref{d6elcon}) in terms of $\eta^1$ and $\lambda^1$. However   this is not  necessary for the purpose of this paper.
It remains to impose the reality condition on $\epsilon_0$.  This gives $\eta=-i \Gamma_{678}*\lambda$ or equivalently $\lambda=-i \Gamma_{678}*\eta$.
The form bilinears are given in appendix \ref{apb}.

The TCFH for $k$ on  a background with  a 2-form field strength is
\begin{equation}
	\begin{split}
		\nabla_M k_N =  \frac{1}{8} e^\Phi F_{PQ}\tilde{\zeta}^{PQ}{}_{MN} +\frac{1}{4}e^\Phi F_{MN}\tilde{\sigma}~.
	\end{split}
\label{ftcfh1}
\end{equation}
As expected $k$ generates isometries  and so symmetries in all the probe actions (\ref{pact}), (\ref{sact}) and (\ref{1part})  with vanishing form couplings. It also generates a symmetry for the probe action of \cite{gpeb} with the 2-form coupling, as the D6-brane 2-form field strength $F=\frac{1}{2} F_{ij}\, e^i\wedge e^j$  is invariant under the action of $k$. In what follows we shall be mostly concerned with the symmetries generated by the form bilinears for the probe action (\ref{1part}). The invariance of this action imposes the weakest  conditions on the form bilinears amongst all  probe actions that we have been investigating.

Next consider the $\tilde k$ and $\omega$ bilinears on  a background with  a 2-form field strength. The
TCFH for these is
\begin{equation}
	\begin{split}
		\nabla_M \tilde{k}_N- \frac{1}{2} e^\Phi F_{MP}\omega^P{}_N =  \frac{1}{8}e^\Phi g_{MN} F_{PQ}\omega^{PQ} -\frac{1}{2}e^\Phi F_{[M|P|}\omega^P{}_{N]}~,
	\end{split}
\label{ftcfh2}
\end{equation}
\bea
\nabla_M \omega_{NR} + e^\Phi F_{M[N}\tilde{k}_{R]} & = & \frac{3}{4} e^\Phi F_{[MN}\tilde{k}_{R]} + \frac{1}{2} e^\Phi g_{M[N}F_{R]P}\tilde{k}^P
\cr
&& \qquad\qquad	
+\frac{1}{4 \cdot 5!} e^\Phi {}^\star{F}_{MNRP_1\dots P_5}\tau^{P_1\dots P_5}~.
\label{ftcfh3}
\eea
For $\tilde k$ to generate symmetries in probe action (\ref{1part}) with $C=0$, it must be a KY tensor. As for D6-branes $F_{ij}\not=0$, the term proportional to the spacetime metric in the first of the equations above must vanish.  This requires that $\omega_{ij}=0$. Then from the expressions of the form bilinears of D6-brane in appendix \ref{apb} and (\ref{d6elcon}), one concludes   that $\tilde k=0$.  Therefore $\tilde k$ does not generate symmetries for the probe action (\ref{1part}).

Similarly for $\omega$ to generate a symmetry for probe action (\ref{1part}) with $C=0$, one finds from the last TCFH above that $\tilde k=0$.  Then from the expressions for the D6-brane form bilinears in appendix \ref{apb}, this  implies that $\omega_{ij}=0$ or equivalently
\bea
{\D{\eta^r}{\Gamma_{11} \lambda^s}}=\mathrm{Im} \D{\eta^r}{\eta^s}=0~.
\eea
Then
\bea
\omega={1\over2} \omega_{ab}\, e^a\wedge e^b= h^{-\frac{1}{4}} \mathrm{Re}{\D{\eta^r}{\Gamma_{ab}\eta^s}}\, e^a \wedge e^b~,
\eea
 is a KY form and generates a (hidden) symmetry for the probe action (\ref{1part}) with $C=0$.  Note that there are Killing spinors for which $\omega\not=0$ even though $\omega_{ij}=0$. This can be verified using a decomposition similar to (\ref{repdec}) but now for $\mathfrak{spin}(6,1)$ spinors.

The TCFH for the bilinears $\tilde \omega$ and $\pi$ is
\begin{equation}
	\begin{split}
		\nabla_M \tilde\omega_{NR}  + \frac{1}{2}e^\Phi F_{MP} \pi^P{}_{NR}=  -\frac{1}{4}e^\Phi g_{M[N}F_{|PQ|}\pi^{PQ}{}_{R]} + \frac{3}{4} e^\Phi F_{[M|P|} \pi^P{}_{NR]}~,
	\end{split}
\label{ftcfh4}
\end{equation}

\begin{equation}
	\begin{split}
		\nabla_M \pi_{NRS} - &\frac{3}{2} e^\Phi F_{M[N} \tilde\omega_{RS]}=   \frac{1}{4 \cdot 4!} e^\Phi {}^\star{F}_{MNRSP_1 \dots P_4} \zeta^{P_1\dots P_4} \\
		&- \frac{3}{2} e^\Phi g_{M[N} F_{R|P|} \tilde \omega^P{}_{S]} - \frac{3}{2} e^\Phi F_{[MN} \tilde\omega_{RS]}~.
	\end{split}
\label{ftcfh5}
\end{equation}
For $\tilde \omega$ to be a KY form and so generate a symmetry in the probe action (\ref{1part}) with $C=0$, $\pi_{aij}=0$.  As it can be seen from the D6-brane bilinears in appendix \ref{apb} after using (\ref{d6elcon}), this implies that $\tilde \omega=0$ and so $\tilde \omega$ does not generate any symmetries. Turning to $\pi$, one finds that this is a KY tensor provided that $\tilde\omega=0$ which implies that $\pi_{aij}=0$ or equivalently
\bea
{\D{\eta^r}{\Gamma_a \Gamma_{5} \lambda^s}}=\mathrm{Im}{\D{\eta^r}{\Gamma_a \eta^s}}=0~.
\label{bibix}
\eea
 The remaining components of $\pi$,
\bea
\pi={1\over3!} \pi_{abc} e^a\wedge e^b\wedge e^c=\frac{1}{3} h^{-\frac{1}{4}} \mathrm{Re}{\D{\eta^r}{\Gamma_{abc} \eta^s}} e^a \wedge e^b \wedge e^c~,
\label{bibixx}
\eea
 generate a (hidden) symmetry for the probe action  (\ref{1part}) with $C=0$.

From now on to simplify  the analysis that  follows on the symmetries generated by TCFHs for all IIA  D-branes, we shall only mention the components of the form bilinears that are required to vanish in order for some others become KY forms.  In particular,     we shall not give the explicit
expressions for  the vanishing components of the form bilinears  and those of the   KY forms in terms of the Killing spinors as we have done in e.g. (\ref{bibix}) and (\ref{bibixx}), respectively.  These can be easily read from the expressions of the form bilinears of D-branes given in appendix \ref{apb}.

The TCFH for the bilinears $\zeta$ and $\tilde \pi$ is
\begin{equation}
	\begin{split}
		\nabla_M \tilde\pi_{NRS}-&\frac{1}{2} e^\Phi F_{MP} \zeta^P{}_{NRS}  =   \frac{3}{8} e^\Phi g_{M[N} F_{|PQ|} \zeta^{PQ}{}_{RS]} \\
		&- e^\Phi F_{[M|P|} \zeta^P{}_{NRS]} - \frac{3}{4}e^\Phi g_{M[N}F_{RS]} \sigma~,
	\end{split}
\label{ftcfh6}
\end{equation}

\begin{equation}
	\begin{split}
		\nabla_M \zeta_{N_1 \dots N_4}& +2 e^\Phi F_{M[N_1} \tilde\pi_{N_2 N_3 N_4]}=
		 - \frac{1}{4!} e^\Phi {}^\star{F}_{MN_1 \dots N_4 PQR} \pi^{PQR}
\\
		&+ 3 e^\Phi g_{M[N_1} F_{N_2|P|}\tilde\pi^P{}_{N_3N_4]} + \frac{5}{2} e^\Phi F_{[MN_1}\tilde\pi_{N_2N_3N_4]} ~  .
	\end{split}
\label{ftcfh7}
\end{equation}
A similar analysis to the one presented above reveals that $\tilde \pi$ does not generate symmetries in the probe actions we have been considering. While
for $\zeta$ to be a KY form, and so generate a (hidden) symmetry for the probe action (\ref{1part}) with $C=0$, one requires that $\tilde \pi=0$. This in turn implies that $\zeta_{abij}=0$ and that $\zeta={1\over 24} \zeta_{a_1\dots a_4} e^{a_1}\wedge\dots\wedge e^{a_4}$  is a KY form.

The TCFH for  $\tilde \zeta$ and $\tau$ is
\begin{equation}
	\begin{split}
		\nabla_M \tilde{\zeta}_{N_1 \dots N_4}+& \frac{1}{2} e^\Phi F_{MP}\tau^P{}_{N_1 \dots N_4} =
			-\frac{1}{2} e^\Phi g_{M[N_1}F_{|PQ|}\tau^{PQ}{}_{N_2N_3N_4]} + \frac{5}{8} e^\Phi F_{[M|P|}\tau^P{}_{N_1 \dots N_4]}  \\
			&+ 3e^\Phi g_{M[N_1}F_{N_2N_3}k_{N_4]}~,
	\end{split}
\label{ftcfh8}
\end{equation}
\begin{equation}
	\begin{split}
		\nabla_M \tau_{N_1 \dots N_5} &-\frac{5}{2}e^\Phi F_{M[N_1}\tilde{\zeta}_{N_2 \dots N_5]}= -\frac{1}{8} e^\Phi {}^\star{F}_{MN_1\dots N_5}{}^{PQ}\omega_{PQ}   \\
		&-5e^\Phi g_{M[N_1}F_{N_2|P|}\tilde{\zeta}^P{}_{N_3N_4N_5]} - \frac{15}{4} e^\Phi F_{[MN_1}\tilde{\zeta}_{N_2\dots N_5]}~.
\end{split}
\label{ftcfh9}
\end{equation}
For $\tilde \zeta$ to generate a symmetry, the above TCFH requires $k_a=0$ and $\tau_{abcij}=0$.  These imply that $\tilde \zeta=0$ and so this bilinear does not generate a symmetry.
It turns out that $\tau$ is a KY form provided that $\tilde\zeta_{abci}=0$.  As a result $\tau_{abcij}=0$. The remaining non-vanishing components of
$\tau$, $\tau={1\over 5!} \tau_{a_1\dots a_5} e^{a_1}\wedge\dots\wedge e^{a_5}$ generate a (hidden) symmetry of the probe action (\ref{1part}) with $C=0$.

It is clear from the TCFH in (\ref{ftcfh1})-(\ref{ftcfh3}), (\ref{ftcfh4}), (\ref{ftcfh5}), (\ref{ftcfh6}), (\ref{ftcfh7}), (\ref{ftcfh8}) and (\ref{ftcfh9})
that the holonomy of the minimal connection reduces for backgrounds with only a 2-form field strength. In particular, the (reduced) holonomy of the minimal connection reduces
to a subgroup of $O(9,1)\times GL(55, \bR)\times GL(165, \bR)\times GL(330, \bR)\times GL(462, \bR)$.
For completeness we state the TCFH on the scalar bilinears
\begin{equation}
	\nabla_M \tilde{\sigma} =  - \frac{1}{4} e^\Phi F_{MP}k^P~,~~ \nabla_M \sigma =  - \frac{1}{8} e^\Phi F_{PQ} \tilde\pi^{PQ}{}_M~.
\end{equation}
These give a trivial contribution to the holonomy of the minimal connection.

To summarise the results of this section, we have concluded as a consequence of the TCFH that there are Killing spinors such that  $k$, $\pi$, $\zeta$ and $\tau$, which have non-vanishing components only along the worldvolume directions
of the D6-brane, are KY forms.  Therefore they generate symmetries for the probe described by the action (\ref{1part}) with $C=0$ in a D6-brane background. This is the case for any multi-centred harmonic function  $h$ that the D6-brane solution  depends on.

\subsection{D2 and D4-branes}

\subsubsection{D2 brane}

Choosing the worldvolume directions of the D2-brane along $051$, the Killing spinors $\epsilon=h^{-{1\over8}} \epsilon_0$ of the solution satisfy the condition
\bea
\Gamma_{051}\epsilon_0=\pm\epsilon_0~,
\label{d2scon}
\eea
where $\epsilon_0$ is a constant spinor and $h$ is given in (\ref{mhf}) for $p=2$. To solve this condition  with the plus sign using spinorial geometry, set
\bea
\epsilon_0=\eta+ e_5\wedge \lambda~,
\label{d2solscon}
\eea
to find that the remaining restrictions on $\eta$ and $\lambda$ are
\bea
\Gamma_1\eta=\eta~,~~~\Gamma_1\lambda=\lambda~,
\eea
where $\eta, \lambda\in \Lambda^*(\bR \langle  e_1, e_2, e_3, e_4\rangle)$; the reality condition is imposed with $\Gamma_{6789}*\eta=\eta$ and $\Gamma_{6789}*\lambda=\lambda$. As in the D6-brane case, the remaining condition on $\eta$ and $\lambda$ can be solved by setting
$\eta=\eta^1+ e_1\wedge \eta^1$ and $\lambda=\lambda^1+e_1\wedge \lambda^1$, where  $\eta^1, \lambda^1\in \Lambda^*(\bR \langle e_2, e_3, e_4\rangle)$
 label the independent solutions of (\ref{d2scon}).
However, we shall perform the computation of the form bilinears using (\ref{d2solscon}) as otherwise their expression will not be manifestly covariant along the
transverse directions of the D2-brane, e.g. the 6-th direction will have to be treated separately from the rest. The form bilinears of the D2-brane can be found in appendix \ref{apb}.

D2-branes exhibit a non-vanishing 4-form field strength $G_{015i}\not=0$. As the probe actions we have been considering do not exhibit such a coupling,   the only remaining coupling is that of the spacetime metric. Therefore  for the form bilinears  to generate a symmetry,   they must be KY forms.
To investigate which of the form bilinears are KY, we shall organise the TCFH according to the domain  that the minimal connection acts on.
As expected the TCFH
\begin{equation}
	\begin{split}
		\nabla_M k_N = \frac{1}{4\cdot 4!}e^\Phi {}^\star{G}_{MNP_1\dots P_4}\tilde{\zeta}^{P_1 \dots P_4} + \frac{1}{8}e^\Phi G_{MNPQ}\omega^{PQ}~,
	\end{split}
\label{iiatcfhbx1}
\end{equation}
implies that $k$ is a Killing 1-form.  As a result it generates symmetries in all probe action (\ref{pact}), (\ref{sact}) and (\ref{1part})
after setting $b=C=0$.

Next observe that
\begin{equation}
	\begin{split}
	\nabla_M \tilde{k}_N - \frac{1}{12}e^\Phi G_{MPQR}\tilde{\zeta}^{PQR}{}_N= \frac{1}{4\cdot 4!}e^\Phi g_{MN} G_{P_1\dots P_4}\tilde{\zeta}^{P_1 \dots P_4}  - \frac{1}{12}e^\Phi G_{[M|PQR|}\tilde{\zeta}^{PQR}{}_{N]}~,
	\end{split}
\label{iiatcfhcx2}
\end{equation}

\bea
	   &&\nabla_M \tilde{\zeta}_{N_1 \dots N_4} -\frac{1}{2}e^\Phi {}^\star{G}_{M[N_1N_2|PQR|}\tau^{PQR}{}_{N_3N_4]}+ 2e^\Phi G_{M[N_1N_2N_3}\tilde{k}_{N_4]}
\cr
&&\qquad\qquad= \frac{1}{8}e^\Phi g_{M[N_1}{}^\star{G}_{N_2 N_3|P_1 \dots P_4|}\tau^{P_1 \dots P_4}{}_{N_4]}
\cr
			&&\qquad\qquad-\frac{5}{12}e^\Phi {}^\star{G}_{[MN_1N_2|PQR|}\tau^{PQR}{}_{N_3N_4]}  -\frac{1}{4}e^\Phi {}^\star{G}_{MN_1\dots N_4P}k^P
\cr
			&&\qquad\qquad+ \frac{5}{4}e^\Phi G_{[MN_1N_2N_3}\tilde{k}_{N_4]} + e^\Phi g_{M[N_1}G_{N_2N_3N_4]P}\tilde{k}^P ~.
\label{iiatcfhex3}
\eea

\bea
	&&
\nabla_M \tau_{N_1 \dots N_5}+\frac{5}{2}e^\Phi {}^\star{G}_{M[N_1N_2N_3|PQ|}\tilde{\zeta}^{PQ}{}_{N_4N_5]} +5e^\Phi G_{M[N_1N_2N_3}\omega_{N_4N_5]}
\cr
&&
\qquad\qquad=  \frac{15}{8}e^\Phi {}^\star{G}_{[MN_1N_2N_3|PQ|}\tilde{\zeta}^{PQ}{}_{N_4N_5]}
\cr
&&\qquad\qquad
+ \frac{1}{4}e^\Phi {}^\star{G}_{MN_1\dots N_5}\tilde{\sigma} + \frac{5}{6 }e^\Phi g_{M[N_1}{}^\star{G}_{N_2N_3N_4|PQR|}\tilde{\zeta}^{PQR}{}_{N_5]}  + \frac{15}{4} e^\Phi G_{[MN_1N_2N_3}\omega_{N_4N_5]}
\cr
&&\qquad\qquad
+ 5 e^\Phi g_{M[N_1}G_{N_2N_3N_4|P|}\omega^P{}_{N_5]} ~,
\label{iiatcfhfx4}
\eea

\bea
 &&\nabla_M \omega_{NR}  - \frac{1}{12} e^\Phi G_{MP_1P_2P_3}\tau^{P_1P_2P_3}{}_{NR} =    \frac{1}{2\cdot 4!}e^\Phi g_{M[N} G_{|P_1\dots P_4|}\tau^{P_1 \dots P_4}{}_{R]}
 \cr
 &&
 - \frac{1}{8}e^\Phi G_{[M|P_1 P_2 P_3|}\tau^{P_1 P_2 P_3}{}_{NR]}  - \frac{1}{4}e^\Phi G_{MNRP} k^P ~,
\label{iiatcfhdx5}
\eea
and so the minimal connection acts  on the domain of   $\tilde k$, $\tilde\zeta$, $\tau$ and $\omega$ form bilinears. Using that for D2-branes $G_{015i}\not=0$ and  the explicit expression for the form bilinears  in appendix \ref{apb}, one finds  that the TCFH implies that the form bilinears $\tilde k$, $\tilde\zeta$ and $\tau$ cannot  be KY tensors.  So these do not generate a symmetry in probe actions.
On the other hand for $\omega$ to be a KY tensor, the TCFH implies that $\tau_{abcij}=0$.  This in turn implies that $\omega_{ij}=0$. As a result $\omega={1\over2} \omega_{ab} e^a\wedge e^b$ is a KY form and generates a (hidden) symmetry in the probe action (\ref{1part}). The condition $\tau_{abcij}=0$ on the Killing spinors and the expression for $\omega_{ab}$ in terms of Killing spinors can be easily  read from the expressions of these form bilinears  in appendix \ref{apb}. There are Killing spinors such that $\tau_{abcij}=0$ and $\omega\not=0$.

The TCFH on the remaining form bilinears is
\bea
&&\nabla_M \pi_{NRS} - \frac{3}{4}e^\Phi G_{M[N|PQ|} \zeta^{PQ}{}_{RS]}=   - \frac{1}{4} e^\Phi G_{MNRS} \sigma
		-\frac{1}{8} e^\Phi {}^\star{G}_{MNRSPQ} \tilde\omega^{PQ}
\cr
&&\qquad\qquad
- \frac{1}{4} e^\Phi g_{M[N} G_{R|P_1P_2P_3|}\zeta^{P_1P_2P_3}{}_{S]} - \frac{3}{4}e^\Phi G_{[MN|PQ|}\zeta^{PQ}{}_{RS]} ~,
\label{iiatcfha2x1}
\eea

\bea
		&&
\nabla_M \zeta_{N_1 \dots N_4} + 3 e^\Phi G_{M[N_1N_2|P|}\pi^P{}_{N_3N_4]}- e^\Phi {}^\star{G}_{M[N_1N_2N_3|PQ|}\tilde\pi^{PQ}{}_{N_4]}
\cr
&& \qquad\qquad =  - \frac{1}{6} e^\Phi g_{M[N_1} {}^\star{G}_{N_2N_3N_4]PQR}\tilde\pi^{PQR}
- \frac{5}{8} e^\Phi {}^\star{G}_{[MN_1N_2N_3|PQ|}\tilde\pi^{PQ}{}_{N_4]}
\cr
&&
\qquad\qquad- \frac{3}{2} e^\Phi  g_{M[N_1} G_{N_2N_3|PQ|}\pi^{PQ}{}_{N_4]} + \frac{5}{2} e^\Phi G_{[MN_1N_2|P|} \pi^P{}_{N_3N_4]} ~,
\label{iiatcfha4x2}
\eea

\bea
&&
		\nabla_M \tilde\pi_{NRS} +\frac{3}{2} e^\Phi G_{M[NR|P|} \tilde\omega^P{}_{S]}
  + \frac{1}{4}e^\Phi {}^\star{G}_{M[NR|P_1P_2P_3|}\zeta^{P_1P_2P_3}{}_{S]}
   \cr
   &&\qquad\qquad =
   - \frac{3}{8} e^\Phi g_{M[N}G_{RS]PQ} \tilde \omega^{PQ}+ e^\Phi G_{[MNR|P|} \tilde\omega^P{}_{S]}  - \frac{1}{32} e^\Phi g_{M[N} {}^\star{G}_{RS]P_1 \dots P_4}\zeta^{P_1\dots P_4}
   \cr
   &&\qquad\qquad
    + \frac{1}{6} e^\Phi {}^\star{G}_{[MNR|P_1P_2P_3|}\zeta^{P_1P_2P_3}{}_{S]} ~,
\label{iiatcfha3x3}
\eea

\bea
		&&\nabla_M \tilde\omega_{NR}
 - \frac{1}{2} e^\Phi G_{M[N|PQ|} \tilde\pi^{PQ}{}_{R]}=  \frac{1}{4!} e^\Phi {}^\star{G}_{MNRP_1P_2P_3}\pi^{P_1P_2P_3}
 \cr
 &&
 \qquad\qquad- \frac{1}{12} e^\Phi g_{M[N}G_{R]P_1P_2P_3}\tilde\pi^{P_1P_2P_3}
 - \frac{3}{8}e^\Phi G_{[MN|PQ|} \tilde\pi^{PQ}{}_{R]} ~.
\label{iiatcfha1x4}
\eea
Requiring that these form bilinears must be KY tensors, the above TCFH together with the explicit expressions for the D2-brane form bilinears in \ref{apb} reveal that
$\zeta= \tilde \pi=\tilde \omega=0$. For $\pi$ to be a KY form, the TCFH implies that $ \zeta_{ijab}=0$ which in turn gives $\pi_{ija}=0$.  The remaining non-vanishing component of $\pi$, $\pi={1\over3!} \pi_{abc} e^a\wedge e^b\wedge e^c$, is a KY tensor and generates a (hidden) symmetry in the probe action (\ref{1part}) with $C=0$. Again the expression of the conditions $ \zeta_{ijab}=0$ and that of $\pi$ in terms of the Killing spinors can be found in appendix \ref{apb}.  There are Killing spinors such that $ \zeta_{ijab}=0$ and $ \pi\not=0$.

It is clear that the holonomy of the minimal connection of  the TCFH with only the 4-form field strength reduces.  In particular, the reduced holonomy is
included in $O(9,1)\times GL(517, \bR)\times GL(495,\bR)$. For completeness we give the TCFH on the scalars as
\bea
	\nabla_M \tilde{\sigma} =  - \frac{1}{4 \cdot 5!}{}^\star{G}_{MP_1\dots P_5}\tau^{P_1\dots P_5}~,~~\nabla_M \sigma = \frac{1}{4!} e^\Phi G_{MPQR} \pi^{PQR} ~,
\eea
which give a trivial contribution in the holonomy of the minimal connection.

To summarise the results of this section, we have shown that there are choices of Killing spinors such that  $\omega$ and $\pi$, with non-vanishing components only along the worldvolume directions of the D2-brane, are KY tensors.  Therefore these bilinears generate (hidden) symmetries for a probe described by the action (\ref{1part}) with $C=0$ on all D2-brane backgrounds, including those that depend on  a multi-centred harmonic function $h$.

\subsubsection{D4 brane}

Choosing  the transverse directions of the D4-brane as $23859$, the Killing spinors $\epsilon=h^{-{1\over8}} \epsilon_0$ of the solution satisfy the condition
\bea
\Gamma_{23849}\epsilon_0=\pm\epsilon_0~,
\label{d4ks}
\eea
where $\epsilon_0$ is a constant spinor and $h$ is a harmonic function as in (\ref{mhf}) for $p=4$. To solve this condition with the plus sign using spinorial geometry write
\bea
\epsilon_0=\eta^1+e_{34}\wedge \eta^2+ e_3\wedge \lambda^1+ e_4\wedge \lambda^2~,
\label{d4sol}
\eea
where $\eta, \lambda \in \Lambda^*\langle e_5, e_1, e_2\rangle$. Substituting this into (\ref{d4ks}), one finds that
\bea
\Gamma_2\eta^{1}=-\eta^{1}~,~~~\Gamma_2\lambda^{1}=-\lambda^{1}~,
\label{rd4ks}
\eea
and similarly for $\eta^2$ and $\lambda^2$.  The reality condition on $\epsilon$ implies that $\eta^1=\Gamma_{67}* \eta^2$ and $\lambda^1=\Gamma_{67}* \lambda^2$.
The remaining conditions (\ref{rd4ks}) can be solved by setting $\eta^1=\rho-e_2\wedge \rho$,  where $\rho\in \Lambda^*\langle e_5, e_1\rangle$, and similarly for the rest of the spinors.  However as for the D2-brane, we shall not do this as otherwise the expression for the form bilinears will not be manifestly covariant in the worldvolume directions because the 6-th
direction will have to be treated separately from the rest.  The form bilinears of the D4-brane can be expressed in terms of those of $\eta$ and $\lambda$ spinors.
Their expressions can be found in appendix \ref{apb}.

As in the D2-brane case, the form bilinears generate symmetries in the probe actions we have been considering provided that they are KY forms.  This condition requires that certain terms in the TCFH must vanish.  Using that for the D4-brane solution $G_{ijkl}\not=0$ and the explicit expression of the form bilinears  in appendix \ref{apb}, one finds after   a detailed analysis of the TCFH   that only $k$, $\tilde \zeta$, $\tau$, $\tilde \omega$ and $\pi$ can be KY tensors while the rest of the bilinears vanish.   In particular, as expected, $k$ is Killing
and so generates a symmetry for the probe actions we have been considering.

For $\tilde \zeta$ to be a KY tensor, the TCFH requires that $\tilde k=0$,  $\tau_{ija_1a_2a_3}=0$ and $\tau_{a_1\dots a_5}=0$.  These imply that $\tilde \zeta_{ija_1a_2}=0$.  The non-vanishing component of $\tilde \zeta$, $\tilde\zeta={1\over 4!} \tilde\zeta_{a_1\dots a_4} e^{a_1}\wedge \dots \wedge e^{a_4}$, generates a (hidden) symmetry for the probe action (\ref{1part}) with $C=0$.
Similarly for $\tau$ to be a KY form, the TCFH requires  that $\omega=0$ and $\tilde\zeta_{ijab}=0$. These imply that  $\tau={1\over 5!} \tau_{a_1\dots\tau_5} e^{a_1}\wedge \dots \wedge e^{a_5}$  is a KY form and  generates a (hidden) symmetry for the probe action (\ref{1part}) with $C=0$.

For $\tilde \omega$ to be a KY form the TCFH requires that $\tilde \pi_{ijk}=0$, which in turn implies that $\tilde \omega_{ij}=0$. The remaining component
of $\tilde \omega={1\over2} \tilde\omega_{ab} e^a\wedge e^b$ is a KY tensor and generates a symmetry for  probe action (\ref{1part}) with $C=0$.
Similarly for $\pi$ to be a KY form, the TCFH requires  that $\zeta_{aijk}=0$, which in turn gives $\pi_{aij}=0$. Then
$\pi={1\over3!} \pi_{abc} e^a\wedge e^b\wedge e^c$ is a KY form and generates a (hidden) symmetry for the probe action (\ref{1part}) with $C=0$.
In all the above cases, the explicit expressions for the vanishing conditions on some of the components of the form bilinears, as well as the expressions of KY forms in terms of the Killing spinors, can be easily  read from the results of appendix \ref{apb} and so they will not be repeated here.

To summarise the results of this section, there are Killing spinors such that  $k$, $\tilde \zeta$, $\tau$, $\tilde \omega$ and $\pi$ with non-vanishing components only along the worldvolume directions of D4-brane, are KY tensors. Therefore, they  generate (hidden) symmetries for the probe described by the action (\ref{1part}) with $C=0$
on any D4-brane background depending of a harmonic function $h$ as in (\ref{mhf}) for $p=4$.

\subsection {D8-brane}

To derive the TCFH on D8-brane type of backgrounds set all the IIA form fields strengths to zero apart from $S$. Then the IIA TCFH in section \ref{iiatcfhs} reduces to

\begin{equation}
	\nabla_M \tilde{\sigma} = \frac{1}{4}e^\Phi S \tilde{k}_M~,~~~
		\nabla_M k_N =  \frac{1}{4}e^\Phi S \omega_{MN} ~,~~~
		\nabla_M \tilde{k}_N =  \frac{1}{4}e^\Phi g_{MN}S\tilde{\sigma}~,
\end{equation}

\begin{equation}
	 \nabla_M \omega_{NR}  =   \frac{1}{2}e^\Phi S g_{M[N}k_{R]}~,~~~
	    \nabla_M \tilde{\zeta}_{N_1 \dots N_4} = - \frac{1}{4 \cdot 5!}e^\Phi {}^\star{S}_{MN_1\dots N_4P_1\dots P_5}\tau^{P_1\dots P_5} ~,
\end{equation}

\bea
		&&\nabla_M \tau_{N_1 \dots N_5} = \frac{1}{4\cdot 4!}e^\Phi {}^\star{S}_{MN_1\dots N_5 P_1 \dots P_4}\tilde{\zeta}^{P_1\dots P_4}~,~~
	 \nabla_M \sigma = 0 ~,~~
\cr
&&
		\nabla_M \tilde\omega_{NR} =  \frac{1}{4} e^\Phi S \tilde\pi_{MNR}~,
\eea

\bea
		&&\nabla_M \pi_{NRS}=  \frac{1}{4} e^\Phi S\zeta_{MNRS}~,~~
			\nabla_M \tilde\pi_{NRS}=  \frac{3}{4} e^\Phi S g_{M[N} \tilde\omega_{RS]}~,~~
\cr
&&
			\nabla_M \zeta_{N_1 \dots N_4} = e^\Phi S g_{M[N_1} \pi_{N_2N_3N_4]}~.
\eea
It is clear from this that $k$, $\tilde\zeta$, $\tau$, $\tilde\omega$ and $\pi$ are KY tensors and generate a (hidden) symmetry of the probe action (\ref{1part}) with $C=0$.
Note that all these form bilinears $k$, $\tilde\zeta$, $\tau$, $\tilde\omega$ and $\pi$    have components only along the worldvolume directions of the D8-brane.
Notice also that the (reduced) holonomy of the minimal TCFH connection is included in $SO(9,1)$.

To find an explicit expression of the form bilinears of D8-brane solution
choose the worldvolume directions along $012346789$.  The Killing spinors $\epsilon= h^{-{1\over8}} \epsilon_0$ of the solution satisfy the condition
$\Gamma_5\epsilon_0=\pm \epsilon_0$, where $\epsilon_0$ is a constant spinor and $h=1+\sum_\ell q_\ell |y-y_\ell|$. Taking the plus sign, this condition can be solved using spinorial geometry by setting
\bea
\epsilon_0=\eta+ e_5\wedge \eta~,
\label{d8sol}
\eea
where $\eta\in \Lambda^*(\bR\langle e_1, e_2, e_3, e_4\rangle)$ after imposing the reality condition $\Gamma_{6789} * \eta=\eta$.
Using the solution for $\epsilon_0$ above, one can easily compute the form bilinears of D8-brane in terms of those of $\eta$.  Their expressions can be found in appendix \ref{apb}. Imposing the condition that the remaining form bilinears $\tilde k$, $\zeta$, $\omega$ and $\tilde \pi$ must be KY forms, the TCFH together
with their explicit expressions in \ref{apb} imply that they should vanish. Therefore they do not generate symmetries for probe actions.

\section{TCFH and probe symmetries on  IIB D-branes}

As in the IIA, there is no classification of IIB supersymmetric backgrounds.  So we shall turn to IIB D-branes to give more examples of backgrounds for which the  TCFH can be interpreted as the condition for invariance  of  particle and string  probe actions under symmetries generated by the form bilinears.  The computation will again be  organised in D-brane electric-magnetic pairs.
The TCFH for each pair can be easily found from that of the IIB TCFH given in (\ref{iibtcfh1})-(\ref{iibtcfh6}) upon setting all the form field strengths to zero apart from those associated  to the D-brane under investigation.

\subsection{D1- and D5-branes}

\subsubsection{The TCFH of D1- and D5-branes}

To illustrate  the construction of symmetries for probes propagating on  D1- and  D5-brane backgrounds using the IIB TCFH, we shall present the D1- and D5-brane TCFH.  This is easily derived from  (\ref{iibtcfh1})-(\ref{iibtcfh6}) upon setting $G^{(1)}=G^{(5)}=0$.  After a re-arrangement of terms so that $e^{\Phi}\,G^{(3)}$ can be interpreted as torsion of a TCFH connection, one finds

\begin{align}
 &
 \nabla_M\, k^{rs}_P - \frac{1}{2}\,e^{\Phi}\,G^{(3)}_{MP}{}^N\, k^{(1)rs}_{N} = \frac{1}{12}\,e^{\Phi}\,G^{(3)N_1N_2N_3}\,\tau^{(1)rs}_{N_1N_2N_3MP} \nn
&
 \nabla_M\, k^{(i)rs}_P  -\frac{1}{2}\,\delta_{i1}\,e^{\Phi}\,G^{(3)}_{MP}{}^N\,k^{rs}_N +\frac{i}{2}\,\varepsilon_{1ij}\,e^{\Phi}\,G^{(3)N_1N_2}_M\,\pi^{(j)rs}_{PN_1N_2}
=
  \frac{1}{12}\,\delta_{i1}\,e^{\Phi}\,G^{(3)N_1N_2N_3}\,\tau^{rs}_{MPN_1N_2N_3} \nn
&
  \qquad+\frac{i}{12}\,\varepsilon_{1ij}\,e^{\Phi}\,g_{MP}\,G^{(3)N_1N_2N_3}\,\pi^{(j)rs}_{N_1N_2N_3}
  + \frac{i}{2}\,\varepsilon_{1ij}\,e^{\Phi}\,G^{(3)N_1N_2}{}_{[M}\,\pi^{(j)rs}_{P]N_1N_2}
  \nn
&
 \nabla_M\,\pi^{rs}_{P_1P_2P_3}  - 3\,e^{\Phi}\,G^{(3)}_{M[P_1}{}^N\,\pi^{(1)rs}_{P_2P_3]N}
= - \frac{1}{12}\,e^{\Phi}\,{}^{\star}G^{(7)}_{MP_1P_2P_3}{}^{N_1N_2N_3}\,\pi^{(1)rs}_{N_1N_2N_3}
\nn
&\qquad+\frac{3}{2}\,e^{\Phi}\,g_{M[P_1}
\,G^{(3)}_{P_2}{}^{N_1N_2}\,\pi^{(1)rs}_{P_3]N_1N_2} + 3\,e^{\Phi}\,G^{(3)}_{[P_1P_2}{}^N\,\pi^{(1)rs}_{P_3M]N}
\nn
&
\nabla_M\,\pi^{(i)rs}_{P_1P_2P_3}- 3\,\delta_{i1}\,e^{\Phi}\,G^{(3)}_{M[P_1}{}^N\,\pi^{rs}_{P_2P_3]N} +\frac{i}{2}\,\varepsilon_{1ij}\,e^{\Phi}\,G^{(3)}_{M}{}^{N_1N_2}\,\tau^{(j)rs}_{P_1P_2P_3N_1N_2}
\nn
&\qquad- 3i\,\varepsilon_{1ij}\,e^{\Phi}\,G^{(3)}_{M[P_1P_2}\,k^{(j)rs}_{P_3]}  =-  \frac{1}{12}\,\delta_{i1}\,e^{\Phi}\,{}^{\star}G^{(7)}_{MP_1P_2P_3}{}^{N_1N_2N_3}\,\pi^{rs}_{N_1N_2N_3} \nn
&
\qquad+
\frac{3}{2}\,\delta_{i1}\,e^{\Phi}\,g_{M[P_1}\,G^{(3)}_{P_2}{}^{N_1N_2}\,\pi^{rs}_{P_3]N_1N_2} + 3\,\delta_{i1}\,e^{\Phi}\,G^{(3)}_{[P_1P_2}{}^N\,\pi^{rs}_{P_3M]N} \nn
&
\qquad +\frac{i}{4}\,\varepsilon_{1ij}\,e^{\Phi}\,G^{(3)N_1N_2N_3}\,g_{M[P_1}\,\tau^{(j)rs}_{P_2P_3]N_1N_2N_3} - i\,\varepsilon_{1ij}\,G^{(3)}_{[P_1}{}^{N_1N_2}\,\tau^{(j)rs}_{P_2P_3M]N_1N_2} \nn
&
\qquad
-\frac{3i}{2}\,\varepsilon_{1ij}\,e^{\Phi}\,g_{M[P_1}\,G^{(3)}_{P_2P_3]}{}^N\,k^{(j)rs}_N + 2i\,\varepsilon_{1ij}\,e^{\Phi}\,G^{(3)}_{[P_1P_2P_3}\,k^{(j)rs}_{M]} ~, \nn
& \nabla_M\,\tau^{rs}_{P_1 \dots P_5} - 5\,e^{\Phi}\,G^{(3)}_{M[P_1}{}^N\,\tau^{(1)rs}_{P_2\dots P_5]N}
=  \frac{1}{2}\,e^{\Phi}\,{}^{\star}G^{(7)}_{MP_1\dots P_5}{}^N\,k^{(1)rs}_N
\nn
&\quad
+\frac{15}{2}\,e^{\Phi}\,G^{(3)}_{[P_1P_2}{}^N\,\tau^{(1)rs}_{P_3P_4P_5M]N} + 5\,e^{\Phi}\,g_{M[P_1}\,G^{(3)}_{P_2}{}^{N_1N_2}\,\tau^{(1)rs}_{P_3P_4P_5]N_1N_2} -10\,e^{\Phi}\,g_{M[P_1}\,G^{(3)}_{P_2P_3P_4}\,k^{(1)rs}_{P_5]}
\nn
&\nabla_M\,\tau^{(i)rs}_{P_1\dots P_5}  - 5\,\delta_{i1}\,e^{\Phi}\,G^{(3)}_{M[P_1}{}^N\,\tau^{rs}_{P_2 \dots P_5]N} +\frac{5i}{2}\,\varepsilon_{1ij}\,e^{\Phi}\,{}^{\star}G^{(7)}_{M[P_1 \dots P_4}{}^{N_1N_2}\,\pi^{(j)rs}_{P_5]N_1N_2}
\nn
&\qquad-10i\,\varepsilon_{1ij}\,e^{\Phi}\,G^{(3)}_{M[P_1P_2}\,\pi^{(j)rs}_{P_3P_4P_5]} = \frac{1}{2} \delta_{i1}\,e^{\Phi}\,{}^{\star}G^{(7)}_{MP_1\dots P_5}{}^N\,k^{rs}_N +5\,\delta_{i1}\,e^{\Phi}\,g_{M[P_1}\,G^{(3)}_{P_2}{}^{N_1N_2}\,\tau^{rs}_{P_3P_4P_5]N_1N_2} \nn
&\qquad+ \frac{15}{2}\delta_{i1}\,e^{\Phi}\,G^{(3)}_{[P_1P_2}{}^N\,\tau^{rs}_{P_3P_4P_5M]N} -10\delta_{i1}\,e^{\Phi}\,g_{M[P_1}\,G^{(3)}_{P_2P_3P_4}\,k^{rs}_{P_5]} \nn
&\qquad + 10i\,\varepsilon_{1ij}\,e^{\Phi}\,G^{(3)}_{[P_1P_2P_3}\,\pi^{(j)rs}_{P_4P_5M]} -15i\,\varepsilon_{1ij}\,e^{\Phi}\,g_{M[P_1}\,G^{(3)}_{P_2P_3}{}^N\,\pi^{(j)rs}_{P_4P_5]N} \nn
&\qquad + \frac{5i}{12}\,\varepsilon_{1ij}\,g_{M[P_1}\,{}^{\star}G^{(7)}_{P_2\dots P_5]}{}^{N_1N_2N_3}\,\pi^{(j)rs}_{N_1N_2N_3} -\frac{3i}{2}\,\varepsilon_{1ij}\,e^{\Phi}\,{}^{\star}G^{(7)}_{[P_1\dots P_5}{}^{N_1N_2}\,\pi^{(j)rs}_{M]N_1N_2}~.
\end{align}
Clearly the (reduced) holonomy of the minimal TCFH connection for generic backgrounds with only $G^{(3)}$ non-vanishing is included
in $\times^6 SO(9,1)\times ^2 GL(256)$.

The difficulties that one encounters when interpreting the TCFH  above as invariance conditions for a  particle probe
described by an action\footnote{A probe with action (\ref{1part}) is chosen because it gives the weakest invariance conditions on the couplings and on the forms that generate the symmetries.   }, like (\ref{1part}), for symmetries
generated by the form bilinears are twofold. One is that the TCFH connection  contains terms that involve double and higher contractions of indices between the
$G^{(3)}$  field strength and the form bilinears.  The other is that the right-hand side of the TCFH involves terms that contain the spacetime metric. Terms such as these  do not occur as invariance conditions for  actions like (\ref{1part}) under symmetries generated by spacetime forms, see (\ref{addcon}). The only option is to set both such terms to zero.  As $G^{(3)}$ is given for each solution,  this puts restrictions on the form bilinears and, in turn, on the choice of Killing spinors used to construct these bilinears.

\subsubsection{D1-brane}

To find the form bilinears of the D1-brane, choose the worldsheet  along the directions $05$. The Killing spinors of the solution are $\epsilon=h^{-{1\over8}} \epsilon_0$, where the constant spinor $\epsilon_0=(\epsilon_0^1, \epsilon^2_0)^t$ is a doublet of Majorana-Weyl $\mathfrak{spin}(9,1)$ spinors satisfying the additional condition
\bea
\Gamma_{05}\sigma_1 \epsilon_0 =\pm \epsilon_0~,
 \eea
 and $h$ is a harmonic function on $\bR^8$ as in (\ref{mhf}) for $p=1$. The metric of the D1-brane is given in (\ref{dbrane}) for $p=1$.  Choosing the plus sign in the condition above
the components of the doublet $\epsilon_0$ are restricted as $\Gamma_{05} \epsilon^1_0=\epsilon_0^2$ and $\Gamma_{05} \epsilon^2_0=\epsilon_0^1$. As in previous cases, these conditions are solved using spinorial geometry \cite{uggp}.  After a short computation, one finds that
\bea
\epsilon^1_0=\eta+ e_5\wedge \lambda~,~~~\epsilon^2_0=\eta-e_5\wedge \lambda~,
\label{d1ks}
\eea
where $\eta\in  \Delta^+_{(8)}=\Lambda^{\mathrm{ev}}(\bR\langle e_1, e_2, e_3, e_4\rangle)$ and $\lambda\in \Delta^+_{(8)}=\Lambda^{\mathrm{odd}}(\bR\langle e_1, e_2, e_3, e_4\rangle)$ are chiral and anti-chiral Majorana-Weyl $\mathfrak{spin}(8)$ spinors, respectively. The form bilinears of $\epsilon$ can be computed in terms of those of $\eta$ and $\lambda$.  The result can be found in appendix \ref{apb}.

For the D-string, the non-vanishing components of $G^{(3)}$ are proportional to $G^{(3)}_{05i}$. Using these, and the expression for the form bilinears in appendix \ref{apb}, one concludes from the TCFH that
\bea
\nabla_M\, k^{rs}_P - \frac{1}{2}\,e^{\Phi}\,G^{(3)}_{MP}{}^N\, k^{(1)rs}_{N} =0~,~~~\nabla_M\, k^{(1)rs}_P  -\frac{1}{2}\,e^{\Phi}\,G^{(3)}_{MP}{}^N\,k^{rs}_N=0~.
\eea
Therefore both $\tilde k^{\pm}=k\pm k^{(1)}$  are covariantly constant with respect to the connection $\nabla^{(\pm)}$, as in (\ref{nablapm}), but now with torsion $e^{\Phi}\,G^{(3)}$.  As $d(e^{\Phi}\,G^{(3)})=0$, $\tilde k^{\pm}$ generate symmetries for the probe actions (\ref{sact}) and (\ref{pact}), where the coupling $b$ is given by  $e^{\Phi}\,G^{(3)}=db$. Furthermore  $\tilde k^{+}$ ($\tilde k^{-}$) generates a symmetry for the probe action (\ref{1part}), where the coupling $C$ is $C=e^{\Phi}\,G^{(3)}$ ($C=-e^{\Phi}\,G^{(3)}$).  Note that both $\tilde k^{\pm}$ have components along the worldsheet directions of the D-string.

It can be shown that the remaining form bilinears do not generate symmetries for the probe actions
(\ref{sact}),  (\ref{pact}) and  (\ref{1part}).  The details of this analysis is similar to those explained for IIA D-branes and they will not be presented here.

\subsubsection{D5-brane}

Choosing the transverse directions of the D5-brane as $3489$, the condition on the Killing spinors $\epsilon=h^{-{1\over8}} \epsilon_0$ for the D5-branes is

\bea
\Gamma_{3489} \sigma_1\epsilon_0=\pm\epsilon_0~,
\eea
where $\epsilon_0=(\epsilon_0^1, \epsilon^2_0)^t$ is a doublet of constant Majorana-Weyl $\mathfrak{spin}(9,1)$ spinors and $h$ is a harmonic function as in (\ref{mhf}) for $p=5$. This condition with the plus sign can be solved using spinorial geometry to yield
\bea
\epsilon^1_0=\eta^1+e_{34}\wedge \lambda^1+e_3\wedge \eta^2+e_4\wedge \lambda^2~,~~~\epsilon^2_0=\eta^1+e_{34}\wedge \lambda^1-e_3\wedge \eta^2-e_4\wedge \lambda^2~,
\label{d5ks}
\eea
where $\eta^1, \lambda^1$ ($\eta^2, \lambda^2)$ are positive (negative) chirality spinors of $\mathfrak{spin}(5,1)$.  The reality condition on $\epsilon_0$ implies
that $\lambda^1=-\Gamma_{67}*\eta^1$ and $\lambda^2=-\Gamma_{67}*\eta^2$. Using this, one can calculate the form bilinears of the D5-brane solution. These  have been presented in appendix \ref{apb}.

As for D1-branes, let us define $\tilde k^{\pm }=k\pm k^{(1)}$.  The TCFH together with the expression of the form bilinears for this background in appendix \ref{apb} give
\bea
\nabla^{(\pm)}_M \tilde k_N^{\pm}=\nabla^{(\pm)}_{[M} \tilde k_{N]}^{\pm}~.
\eea
Therefore $\tilde k^{\pm}$ satisfy the KY equation with respect to the connection $\nabla^{(\pm)}$ as in (\ref{nablapm}) with torsion $\pm e^\Phi G^{(3)}$.
A consequence of this is that $\tilde k^{\pm}$ generate symmetries in the particle probe action (\ref{1part}) with 3-form coupling $\pm e^\Phi G^{(3)}$.  Note that the second condition in (\ref{addcon}) required for this is also satisfied as  $i_{\tilde k^{\pm}}d(e^\Phi G^{(3)})=0$ and $i_{\tilde k^{\pm}} G^{(3)}=0$ because $\tilde k^{\pm}$ have components only along the worldvolume directions of the D5-brane.  A similar investigation reveals  that $k^{(2)}$ and $k^{(3)}$ do not generate  symmetries for the probe actions we are considering.

Next define $\tilde \pi^{\pm}=\pi\pm \pi^{(1)}$. The TCFH can be re-organised as a KY equation with respect to a connection with skew-symmetric torsion provided
that the term proportional to the spacetime metric $g$ vanishes. For this the $\tilde \pi^{\pm}_{Mij}$ components of the 3-form bilinears should vanish. In particular, $\tilde \pi^{+}$ is a KY form with respect to $\nabla^{(+)}$ connection provided that
\bea
\langle \eta^{1r}, \Gamma_a \lambda^{1s}\rangle_D=\mathrm{Im} \langle \eta^{1r}, \Gamma_a \eta^{1s}\rangle_D=0~,
\eea
and similarly $\tilde \pi^{-}$ is a KY form with respect to $\nabla^{(-)}$ connection provided that
\bea
\langle \eta^{2r}, \Gamma_a \lambda^{2s}\rangle_D=\mathrm{Im} \langle \eta^{2r}, \Gamma_a \eta^{2s}\rangle_D=0~.
\eea
The remaining non-vanishing components of $\tilde \pi^{\pm}$ are
\begin{align}
&\tilde\pi^{+rs} = \frac{4}{3}\,h^{-1/4}\rep \left\langle \eta^{1r}, \Gamma_{abc} \eta^{1s} \right\rangle_D e^a \wedge e^b \wedge e^c~, \nn
&\tilde\pi^{-rs} = \frac{4}{3}\,h^{-1/4}\rep \left\langle \eta^{2r}, \Gamma_{abc} \eta^{2s} \right\rangle_D e^a \wedge e^b \wedge e^c~.
\end{align}
These generate (hidden) symmetries of the probe action (\ref{1part}) with 3-form coupling $\pm 2 e^\Phi G^{(3)}$, respectively.  Note that the second condition in (\ref{addcon}) required for the invariance of the action (\ref{1part}) generated by $\tilde\pi^{\pm }$ is also satisfied as $i_{\tilde \pi^{\pm}} d e^\Phi G^{(3)}=0$ and $i_{\tilde \pi^{\pm}} G^{(3)}=0$.   A similar investigation reveals that $\pi^{(2)}$ and $\pi^{(3)}$ do not generate symmetries for the probe actions we have been considering.
The same applies all four 5-form bilinears.

To summarise the results of this section, we have demonstrated that there are Killing spinors such that the form bilinears $\tilde k^{\pm}$ and $\tilde \pi^{\pm}$, with non-vanishing components only along the worldvolume directions of the D5-brane, are KY forms with respect to connections with skew-symmetric torsion proportional to $\pm e^\Phi G^{(3)}$ and $\pm 2 e^\Phi G^{(3)}$, respectively. It turns out that these forms $\tilde k^{\pm}$  ($\tilde \pi^{\pm}$) generate (hidden) symmetries for the probes described by action (\ref{1part}) with form coupling $C$
equal to $\pm e^\Phi G^{(3)}$ ( $\pm 2 e^\Phi G^{(3)}$).

\subsection{D3-brane}

Choosing the worldvolume directions of the D3-brane as 0549, the Killing spinors, $\epsilon=h^{-{1\over8}} \epsilon_0$, of this solution  satisfy the condition
\bea
\Gamma_{0549}\epsilon_0^1=\pm \epsilon^2_0~,
\eea
where $\epsilon_0=(\epsilon_0^1, \epsilon^2_0)^t$ is a  doublet of constant Majorana-Weyl spinors of $\mathfrak{spin}(9,1)$ and $h$ a harmonic function as in (\ref{mhf}) with $p=3$. This condition with the plus sign can be solved using spinorial geometry as
\bea
\epsilon^1_0= \eta^1+e_{45}\wedge \lambda^1+e_4\wedge \eta^2+e_5\wedge \lambda^2~,~~\epsilon_0^2=i \eta^1+i e_{45}\wedge \lambda^1-i e_4\wedge \eta^2-i e_5\wedge \lambda^2~,
\label{d3ks}
\eea
where $\eta^1, \lambda^1 \in \Lambda^{\mathrm{ev}}(\bC\langle e_1, e_2, e_3\rangle)$  ($\eta^2, \lambda^2 \in \Lambda^{\mathrm{odd}}(\bC\langle e_1, e_2, e_3\rangle)$) are positive  (negative) chirality Weyl spinors of $\mathfrak{spin}(6)$.  Furthermore the reality condition on $\epsilon$ implies that
\bea
\eta^2=-i \Gamma_{678}*\eta^1~,~~~\lambda^2=i \Gamma_{678} * \lambda^1~.
\eea
Using these, one can easily express the form bilinears of the D3-brane solution in terms of those of the $\eta$ and $\lambda$ $\mathfrak{spin}(6)$ spinors. The form bilinears can be found in appendix \ref{apb}.

As the probe actions (\ref{sact}), (\ref{pact}) and (\ref{1part}) do not exhibit a 5-form coupling,  the only coupling one should consider is that of the spacetime metric. For the form bilinears to generate a symmetry for the probe described by the action (\ref{1part}), they must be KY tensors.  To see whether this is the case,  let us begin with the 1-form bilinears $k$ and $k^{(2)}$.  The TCFH\footnote{We have replaced $k^{(2)}, \pi^{(2)}$ and $\tau^{(2)}$ with $ik^{(2)}, i\pi^{(2)}$ and $i\tau^{(2)}$ so that the TCFH for the D3-brane and later for the D7-brane to be manifestly real.}
gives
\bea
&& \nabla_M\, k^{rs}_P =  \frac{1}{12}\,e^{\Phi}\,G^{(5)}_{MP}{}^{N_1N_2N_3}\,\pi^{(2)rs}_{N_1N_2N_3}~,
\cr
&&
\nabla_M\, k^{(2)rs}_P
= -\frac{1}{12}\,e^{\Phi}\,G^{(5)}_{MP}{}^{N_1N_2N_3}\,\pi^{rs}_{N_1N_2N_3}~.
\eea
Clearly both are KY tensors and so generate symmetries for the probe action (\ref{1part}) with $C=0$.  Using that the components $G^{(5)}_{a_1\dots a_4 i}$ and  $G^{(5)}_{i_1\dots a_5}$ of the 5-form field strength of the D3-brane solution do not vanish and the expressions for the bilinears in appendix B, one can show that the remaining
two 1-form bilinears do not generate a symmetry for the probe actions we have been considering.

Next let us turn to the 3-form bilinears $\pi$ and $\pi^{(2)}$.  The TCFH on a D3-background reads
\bea
 \nabla_M\,\pi^{rs}_{P_1P_2P_3}
-\frac{1}{4}\,e^{\Phi}\,G^{(5)}_{M[P_1}{}^{N_1N_2N_3}\,\tau^{(2)rs}_{P_2P_3]N_1N_2N_3}
=  -\frac{1}{2}\,e^{\Phi}\,G^{(5)}_{MP_1P_2P_3}{}^N\,k^{(2)rs}_N~,
\eea
\bea
 \nabla_M\,\pi^{(2)rs}_{P_1P_2P_3} +\frac{1}{4}\,e^{\Phi}\,G^{(5)}_{M[P_1}{}^{N_1N_2N_3}\,\tau^{rs}_{P_2P_3]N_1N_2N_3}
=
\frac{1}{2}\,e^{\Phi}\,G^{(5)}_{MP_1P_2P_3}{}^N\,k^{rs}_N~.
\eea
For either $\pi$ or $\pi^{(2)}$ be KY forms, the connection term involving $G^{(5)}$ in the TCFH must vanish. For $\pi$, this requires that
$\tau^{(2)}_{ijabc}=0$ which in turn implies that
\begin{align}
&\rep\left\langle \eta^{1r}, \Gamma_{ij}\eta^{1s} \right\rangle = \rep\left\langle \lambda^{1r}, \Gamma_{ij} \lambda^{1s} \right\rangle =0~, \nn
&\rep\left\langle \eta^{1r}, \Gamma_{ij}\lambda^{1s} \right\rangle + \rep\left\langle \lambda^{1r}, \Gamma_{ij} \eta^{1s} \right\rangle = 0~, \nn
&\imp\left\langle \eta^{1r}, \Gamma_{ij}\lambda^{1s} \right\rangle - \imp\left\langle \lambda^{1r}, \Gamma_{ij}\eta^{1s} \right\rangle = 0~. \nn
\label{d3picon}
\end{align}
There are solutions to these conditions.  For example take $\eta^{1r}=\lambda^{1r}$ and $\eta^{1s}=\lambda^{1s}$. The decomposition of two Weyl representations
of $\mathfrak{spin}(6)$ is isomorphic to $\Lambda^0(\bC^6)\oplus \Lambda^2(\bC^6)$. Thus one can choose $\eta^{1r}$ and $\eta^{1s}$ such that the component of their tensor product in $\Lambda^2(\bC^6)$ vanishes. Imposing the above conditions,  the non-vanishing components of $\pi$ are
\begin{align}
\pi^{rs} = &-4 h^{-\frac{1}{4}}\imp\left\langle \eta^{1r}, \eta^{1s} \right\rangle(e^0-e^5)\wedge e^4 \wedge e^9 \nn
&+4 h^{-\frac{1}{4}} \imp\left\langle \lambda^{1r}, \lambda^{1s} \right\rangle(e^0+e^5)\wedge e^4 \wedge e^9 \nn
&+4 h^{-\frac{1}{4}} \left( \rep\left\langle \eta^{1r}, \lambda^{1s} \right\rangle - \rep\left\langle \lambda^{1r}, \eta^{1s} \right\rangle \right)e^0\wedge e^5 \wedge e^4 \nn
&+4 h^{-\frac{1}{4}} \left( \imp\left\langle \eta^{1r}, \lambda^{1s} \right\rangle + \imp\left\langle \lambda^{1r}, \eta^{1s} \right\rangle \right)e^0\wedge e^5 \wedge e^9~.
\end{align}

Similarly for $\pi^{(2)}$ to be a KY form, $\tau_{ijabc}=0$.  The conditions on the spinors are given as in (\ref{d3picon}) after replacing $\rep$ with $\imp$ and vice versa.    After imposing these conditions, the non-vanishing components of $\pi^{(2)}$ are
\begin{align}
\pi^{(2)rs} = &-4h^{-\frac{1}{4}} \rep\left\langle \eta^{1r}, \eta^{1s} \right\rangle(e^0-e^5)\wedge e^4 \wedge e^9 \nn
&+4 h^{-\frac{1}{4}} \rep\left\langle \lambda^{1r}, \lambda^{1s} \right\rangle(e^0+e^5)\wedge e^4 \wedge e^9 \nn
&-4 h^{-\frac{1}{4}} \left( \imp\left\langle \eta^{1r}, \lambda^{1s} \right\rangle - \imp\left\langle \lambda^{1r}, \eta^{1s} \right\rangle \right)e^0\wedge e^5 \wedge e^4 \nn
&+4 h^{-\frac{1}{4}} \left( \rep\left\langle \eta^{1r}, \lambda^{1s} \right\rangle + \rep\left\langle \lambda^{1r}, \eta^{1s} \right\rangle \right)e^0\wedge e^5 \wedge e^9~.
\end{align}
Next let us focus on the two remaining 3-form bilinears $\pi^{(1)}$ and $\pi^{(3)}$. It turns out that they do not generate symmetries for the probe action (\ref{1part}) that we are considering. In particular for  $\pi^{(1)}$  to be a KY form, the TCFH requires that  $\pi^{(3)}=0$. This in turn implies that $\pi^{(1)}=0$. To establish the latter
the Hodge duality properties of the transverse components of $\pi^{(3)}$ have to be used.

To find the conditions for $\tau$ and $\tau^{(2)}$ be KY forms, the TCFH for these bilinears on a D3-brane background is
\bea
 \nabla_M\,\tau^{rs}_{P_1 \dots P_5}
+10\,e^{\Phi}\,G^{(5)}_{M[P_1P_2P_3}{}^N\,\pi^{(2)rs}_{P_4P_5]N}
&=&
-5\,e^{\Phi}\,g_{M[P_1}\,G^{(5)}_{P_2P_3P_4}{}^{N_1N_2}\,\pi^{(2)rs}_{P_5]N_1N_2}
\cr
&&-\frac{15}{2}\,e^{\Phi}\,G^{(5)}_{[P_1\dots P_4}{}^N\,\pi^{(2)rs}_{P_5M]N} ~,
\eea
\bea
 \nabla_M\,\tau^{(2)rs}_{P_1\dots P_5}
-10\,e^{\Phi}\,G^{(5)}_{M[P_1P_2P_3}{}^N\,\pi^{rs}_{P_4P_5]N}
&=&  5\,e^{\Phi}\,g_{M[P_1}G^{(5)}_{P_2P_3P_4}{}^{N_1N_2}\,\pi^{rs}_{P_5]N_1N_2}
\cr
&&+\frac{15}{2}\,e^{\Phi}\,G^{(5)}_{[P_1\dots P_4}{}^N\,\pi^{rs}_{P_5M]N}~.
\eea
It turns out that for $\tau$ to be a KY tensor, $\pi^{(2)}=0$.  Using the chirality of $\eta^1$ and $\eta^2$ as $\mathfrak{spin}(6)$ spinors, one concludes that
$\tau=0$, and so there are no symmetries generated by this 5-form bilinear.  Similarly, $\tau^{(2)}$ does not generate any symmetries for the probe action (\ref{1part}) we have been considering.

Finally, let us turn to investigate the TCFH of $\tau^{(1)}$ and $\tau^{(3)}$ on a D3-brane background. One finds that
\bea
 \nabla_M\,\tau^{(1)rs}_{P_1\dots P_5}
 +5\,e^{\Phi}\,G^{(5)}_{M[P_1\dots P_4}\,k^{(3)rs}_{P_5]} -\frac{5}{2}\,e^{\Phi}\,G^{(5)}_{M[P_1P_2}{}^{N_1N_2}\,\tau^{(3)rs}_{P_3P_4P_5]N_1N_2} \nn
= \frac{5}{2}\,e^{\Phi}\,g_{M[P_1}\,G^{(5)}_{P_2\dots P_5]}{}^N\,k^{(3)rs}_N -3\,e^{\Phi}\,G^{(5)}_{[P_1\dots P_5}\,k^{(3)rs}_{M]}~, \nn
\eea

\bea
 \nabla_M\,\tau^{(3)rs}_{P_1\dots P_5}
- 5\,e^{\Phi}\,G^{(5)}_{M[P_1\dots P_4}\,k^{(1)rs}_{P_5]} +\frac{5}{2}\,e^{\Phi}\,G^{(5)}_{M[P_1P_2}{}^{N_1N_2}\,\tau^{(1)rs}_{P_3P_4P_5]N_1N_2} \nn
=- \frac{5}{2}\,e^{\Phi}\,g_{M[P_1}\,G^{(5)}_{P_2\dots P_5]}{}^N\,k^{(1)rs}_N +3\,e^{\Phi}\,G^{(5)}_{[P_1\dots P_5}\,k^{(1)rs}_{M]}~. \nn
\eea
Focusing on the former condition, $\tau^{(1)}$ is a KY tensor provided that $k^{(3)}=0$ and $\tau^{(3)}_{ijkab}=0$.  Using the chirality of $\eta^2$ and $\lambda^2$ as  $\mathfrak{spin}(6)$ spinors, one finds that  $\tau^{(1)}=0$. A similar calculation for $\tau^{(3)}$ reveals that $\tau^{(3)}=0$. These two forms
do not generate symmetries  for the probe action (\ref{1part}).

To summarise the results of this section, we have demonstrated that there are Killing spinors  such that the form bilinears $k$, $k^{(2)}$, $\pi$ and $\pi^{(2)}$
of the D3-brane background are KY forms and so generate (hidden) symmetries for the probe described by the action (\ref{1part}) with $C=0$. All these forms have components only along the worldvolume directions of the D3-brane.

\subsection{D7-brane}
Choosing the transverse directions of the D7-brane as $49$, the Killing spinor $\epsilon=h^{-{1\over8}} \epsilon_0$ of the solution satisfies the condition
\bea
\Gamma_{49}\epsilon_0^1=\pm\epsilon_0^2~,
\eea
where $\epsilon_0=(\epsilon_0^1, \epsilon_0^2)^t$ is a constant doublet of Majorana-Weyl $\mathfrak{spin}(9,1)$ spinors and $h=1+\sum_\ell q_\ell \log|y-y_\ell|$. This condition with the plus sign can be solved using spinorial geometry  as
\bea
\epsilon_0^1=\eta+e_4\wedge \lambda~,~~~\epsilon_0^2=i\eta-i e_4\wedge \lambda~,
\label{d7ks}
\eea
where $\eta$ ($\lambda$) is a positive, $\eta\in \Lambda^{\mathrm{ev}}(\bC\langle e_1, e_2, e_3, e_5\rangle)$,  (negative, $\lambda\in \Lambda^{\mathrm{odd}}(\bC\langle e_1, e_2, e_3, e_5\rangle)$,) chirality $\mathfrak{spin}(7,1)$ Weyl spinors. The reality condition on $\epsilon$ implies that
\bea
\lambda=-i \Gamma_{678}*\eta~.
\eea
Using the above expression for the Killing spinors, the form bilinears can be easily computed and can be found in appendix \ref{apb}.

The TCFH for the form bilinears $k$ and $k^{(2)}$ gives
\bea
\nabla_M\, k^{rs}_P =  \frac{1}{2}\, e^{\Phi}\,G^{(1)N}\,\pi^{(2)rs}_{NMP}~,~~~\nabla_M k^{(2)rs}_P = -\frac{1}{2}\,e^{\Phi}\,G^{(1)N}\,\pi^{rs}_{NMP}~.
\eea
As a result, they are both KY forms.  Therefore both generate symmetries for the probe action (\ref{1part}) with $C=0$.
It can be shown that the remaining two 1-form bilinears $k^{(1)}$ and $k^{(3)}$ do not generate symmetries for the probe actions we are considering.

Similarly the TCFH of $\pi$ and $\pi^{(2)}$ on D7-brane background reads
\bea
&&
\nabla_M\pi^{rs}_{P_1P_2P_3} = \frac{1}{2}\,e^{\Phi}\,G^{(1)N}\,\tau^{(2) rs}_{MP_1P_2P_3N} + 3\,e^{\Phi}\,g_{M[P_1}\,G^{(1)}_{P_2}\,k^{(2) rs}_{P_3]}~,
\cr
&&
\nabla_M\pi^{(2)rs}_{P_1P_2P_3} = -\frac{1}{2}\,e^{\Phi}\,G^{(1)N}\,\tau^{rs}_{MP_1P_2P_3N} - 3\,e^{\Phi}\,g_{M[P_1}\,G^{(1)}_{P_2}\,k^{rs}_{P_3]}~.
\eea
For these to be KY forms, it is required that the terms of the TCFH that explicitly contain  the spacetime metric must vanish.  As the form field strength for the D7-brane $G^{(1)}\not=0$, for $\pi$ this leads to the condition $k^{(2)}=0$, or equivalently,
\begin{equation}
\imp\left\langle \eta^r, \Gamma_a \eta^s \right\rangle_D = 0~.
\end{equation}
Therefore
\begin{align}
\pi^{rs} &= \frac{2}{3} h^{-\frac{1}{4}} \rep\left\langle \eta^r, \Gamma_{abc}\eta^s \right\rangle_D e^a\wedge e^b \wedge e^c~,
\end{align}
is a KY form and generates a (hidden) symmetry for the particle probe described by the action (\ref{1part})  with $C=0$.

Similarly the condition for $\pi^{(2)}$ to be a KY form is
\begin{equation}
\rep\left\langle \eta^r, \Gamma_a \eta^s \right\rangle_D = 0~.
\end{equation}
As a result
\begin{align}
\pi^{(2)rs} &= -\frac{2}{3} h^{-\frac{1}{4}} \imp\left\langle \eta^r, \Gamma_{abc}\eta^s \right\rangle_D e^a\wedge e^b \wedge e^c~,
\end{align}
is a KY form and generates a (hidden) symmetry for the particle probe described by the action (\ref{1part}) with $C=0$. The remaining two 3-form bilinears $\pi^{(1)}$ and $\pi^{(3)}$ do not generate symmetries for the probe action we have been considering.

It remains to investigate whether any of the 5-form bilinears generate symmetries for probe action (\ref{1part}). To begin consider $\tau$ and $\tau^{(2)}$.
The TCFH for these in a D7-background is
\bea
&&
\nabla_M\tau^{rs}_{P_1\dots P_5} =- \frac{1}{12}\,e^{\Phi}{}^{\star}G^{(9)}_{MP_1\dots P_5}{}^{N_1N_2N_3}\,\pi^{(2)rs}_{N_1N_2N_3} + 10\,e^{\Phi}\,g_{M[P_1}\,G^{(1)}_{P_2}\,\pi^{(2)rs}_{P_3P_4P_5]}~,
\cr
&&
\nabla_M\tau^{(2)rs}_{P_1\dots P_5} = \frac{1}{12}\,e^{\Phi}{}^{\star}G^{(9)}_{MP_1\dots P_5}{}^{N_1N_2N_3}\,\pi^{rs}_{N_1N_2N_3} - 10\,e^{\Phi}\,g_{M[P_1}\,G^{(1)}_{P_2}\,\pi^{rs}_{P_3P_4P_5]}~.
\eea
For $\tau$, the vanishing of the last term in the first TCFH that contains the metric  leads to the condition $\pi^{(2)}=0$.
As a result
\begin{equation}
\tau^{rs} = \frac{4}{5!} h^{-\frac{1}{4}} \rep\left\langle \eta^r, \Gamma_{a_1 \dots a_5} \eta^s \right\rangle_D e^{a_1}\wedge \dots \wedge e^{a_5}~,
\end{equation}
is a KY form and generates a (hidden) symmetry for the probe action (\ref{1part}) with $C=0$.

Similarly   $\tau^{(2)}$ is a KY form provided that $\pi=0$
and so
\begin{equation}
\tau^{(2)rs} = -\frac{4}{5!} h^{-\frac{1}{4}} \imp\left\langle \eta^r, \Gamma_{a_1 \dots a_5} \eta^s \right\rangle_D e^{a_1}\wedge \dots \wedge e^{a_5}~,
\end{equation}
is a KY form generates a (hidden) symmetry for the probe action (\ref{1part}) with $C=0$.  The remaining two 5-forms do not generate symmetries
for the probe actions we have been considering.

 In all the above cases, there are Killing spinors such that they satisfy the conditions required for the existence of non-vanishing KY forms.  This can be seen from the decomposition of products of two $\mathfrak{spin}(7,1)$ spinor representations in terms of forms as we have described in previous cases.

To summarise the results of this section, we have demonstrated that there are Killing spinors  such that the form bilinears $k$, $k^{(2)}$, $\pi$, $\pi^{(2)}$, $\tau$ and $\tau^{(2)}$
of the D7-brane background are KY forms and so generate (hidden) symmetries for the probe described by the action (\ref{1part}) with $C=0$. All these forms have components only along the worldvolume directions of the D7-brane.

\section{Concluding Remarks}

We have presented the TCFH of both IIA and IIB supergravities and demonstrated that the form bilinears satisfy a generalisation
of the CKY equation with respect to the minimal TCFH  connection in agreement with the general theorem in \cite{gptcfh}. Then prompted by the well-known result that KY forms generate (hidden) symmetries in spinning particle actions,  we explored the question
on whether the form bilinears of some known supergravity backgrounds, which include all type II branes, generate symmetries for various particle and string
probes propagating on these backgrounds.

We have also explored the complete integrability of geodesic flow on all type II brane backgrounds.  We have demonstrated that if the harmonic function
that the solutions depend on has at most one centre, i.e. they are spherically symmetric,  then the geodesic flow is completely integrable. We have explicitly given  all independent conserved
charges in involution. We have also presented the KS,  KY and CCKY  tensors of these brane backgrounds associated with their integrability structure.

Returning to the symmetries generated by the TCFH, supersymmetric type II common sector backgrounds admit form bilinears which are covariantly constant with respect to a connection
with skew-symmetric torsion given by the NS-NS 3-form field strength.  All these bilinears generate (hidden) symmetries for string and particle probe actions with 3-form couplings. The type II fundamental string and NS5-brane background form bilinears have explicitly been given. Common sector backgrounds admit additional form bilinears which satisfy a TCFH but they are not
covariantly constant with respect to a connection with skew-symmetric torsion. Although these forms are part of the geometric structure of common sector backgrounds, their geometric interpretation is less straightforward.

Moreover we found that there are Killing spinors in  all Dp-brane backgrounds, for $p\not=1,5$, such that the associated bilinears are KY forms and so generate (hidden) symmetries for spinning particle probes. All these form bilinears
have components only along the worldvolume directions of the Dp-branes.  A similar conclusion holds for the D1- and D5-brane solutions, only that in this case the form bilinears are KY forms with respect to a connection with skew-symmetric torsion
that is determined by the 3-form field strength of the backgrounds. These form bilinears generate (hidden) symmetries for particle probes described by the
action (\ref{1part}) with a non-vanishing 3-form coupling.  Again these form bilinears have non-vanishing components only along the worldvolume directions of the D-branes.

It is fruitful to compare the KY forms we have obtained from the TCFH with those that are needed to investigate the integrability of the geodesic flow
in type II brane backgrounds. TCFH KY forms exist for any choice of the harmonic function that the brane solutions depend on.  Moreover,  as we have mentioned, these KY forms have non-vanishing components only along the worldvolume directions of D-branes. It is clear from this that although they generate symmetries for particle probes propagating on D-brane backgrounds these symmetries are not necessarily connected to the integrability
properties of such dynamical systems. This is because it is not expected, for example, that the geodesic flow of brane solutions which depend on a  multi-centred harmonic function to be completely
integrable.  Indeed the KS and KY tensors we have found that are responsible for the integrability of the geodesic flow on spherically symmetric branes also have  components
along the transverse directions of these solutions. As the brane metrics have a non-trivial dependence on the transverse coordinates, this is essential for proving the integrability  of the geodesic flow.   Therefore one concludes that although the form bilinears of supersymmetric backgrounds can generate symmetries in string and
particle probes propagating in these backgrounds, they are not sufficient to prove the complete integrability of  probe dynamics.  Nevertheless the TCFH
KY tensors, when they exist, are associated with symmetries of probes propagating on brane backgrounds which are not necessarily spherically symmetric.

To find TCFH  KY tensors, we have imposed a rather stringent
set of conditions on the form bilinears. In particular in several D-brane backgrounds, we set all terms of the minimal TCFH connection that depend on a form field
strength to zero.  It is likely that such a restriction can be lifted and the only condition necessary for invariance of a probe action will be that the terms
in the TCFH which contain explicitly the metric should vanish. For this a new set a probe actions should be found that have couplings which depend on the form
field strengths of the supergravity theories and generalise (\ref{1part}) which exhibits only a 3-form coupling. We hope to report on such a development in the future.

\section*{Acknowledgments}

JP is supported by the EPSRC grant EP/R513064/1.

\newpage

\setcounter{section}{0}
\setcounter{subsection}{0}
\setcounter{equation}{0}

\begin{appendices}

\section{Common sector brane form bilinears}\label{apa}

\subsection{ Form bilinears of IIA branes}

\subsubsection{Fundamental String}

A direct computation using (\ref{fsol}) reveals that the form bilinears of  IIA fundamental string  are
\bea
&&\sigma^{rs}= h^{-{1\over2}}\big(-\langle \eta^r, \lambda^s\rangle+ \langle \lambda^r, \eta^s\rangle\big)~  ,~~~
\cr
&&k^{rs}=  h^{-{1\over2}}\langle \eta^r, \eta^s\rangle  (e^0-e^5)+  h^{-{1\over2}} \langle \lambda^r, \lambda^s\rangle  (e^0+e^5)~,
\cr
&&
\omega^{rs}=h^{-{1\over2}}\big(\langle \eta^r, \lambda^s\rangle+ \langle \lambda^r, \eta^s\rangle\big) e^0\wedge e^5+{1\over2} h^{-{1\over2}}
\big(-\langle \eta^r, \Gamma_{ij}\lambda^s\rangle+ \langle \lambda^r, \Gamma_{ij}\eta^s\rangle\big) e^i\wedge e^j~,
\cr
&&
\pi^{rs}={1\over2}h^{-{1\over2}} \langle \eta^r, \Gamma_{ij}\eta^s\rangle \rangle  (e^0-e^5)\wedge e^i\wedge e^j
+{1\over2}  h^{-{1\over2}}\langle \lambda^r,\Gamma_{ij} \lambda^s\rangle  (e^0+e^5)\wedge e^i\wedge e^j~,
\cr
&&
\zeta^{rs}={1\over2}h^{-{1\over2}}\big(\langle \eta^r, \Gamma_{ij}\lambda^s\rangle+ \langle \lambda^r, \Gamma_{ij}\eta^s\rangle\big) e^0\wedge e^5\wedge e^i\wedge e^j
\cr
&&\qquad\qquad +{1\over4!}h^{-{1\over2}}\big(-\langle \eta^r, \Gamma_{ijk\ell}\lambda^s\rangle+ \langle \lambda^r, \Gamma_{ijk\ell}\eta^s\rangle\big) e^i\wedge e^j\wedge e^k\wedge e^\ell ~,~~~
\cr
&&
\tau^{rs}={1\over 4!}h^{-{1\over2}}\langle \eta^r,\Gamma_{ijk\ell} \eta^s\rangle   (e^0-e^5)\wedge  e^i\wedge e^j\wedge e^k\wedge e^\ell
\cr
&&
\qquad\qquad+ {1\over 4!} h^{-{1\over2}} \langle \lambda^r,\Gamma_{ijk\ell} \lambda^s\rangle  (e^0+e^5)\wedge  e^i\wedge e^j\wedge e^k\wedge e^\ell~,
\eea
where $i,j,k,\ell=1, 2, 3, 4, 6, 7, 8, 9$ are the transverse directions of the string and $(e^0, e^5, e^i)$ is a pseudo-orthonormal frame of the
fundamental string metric (\ref{fstring}), i.e $g=-(e^0)^2+(e^5)^2+ \sum_i (e^i)^2$.
 The remaining form bilinears $\tilde\sigma$, $\tilde k$, $\tilde \omega$, $\tilde \pi$, $\tilde \zeta$ and $\tilde \tau$ can be obtained from the expressions above upon setting $\lambda^s$ to $-\lambda^s$.

\subsubsection{NS5-brane}

A direct computation using (\ref{ns5s}) reveals that the form bilinears of  NS5-brane  are
\bea
&&k^{rs}=2 \big(\mathrm{Re}\langle \eta^{1r}, \Gamma_a\eta^{1s}\rangle_D+\mathrm{Re}\langle \eta^{2r}, \Gamma_a\eta^{2s}\rangle_D\big)~e^a~,
\eea
\bea
&&
\omega^{rs}=2 \big(\mathrm{Re}\langle \eta^{1r}, \Gamma_{a}\eta^{2s}\rangle_D+\mathrm{Re}\langle \eta^{2r}, \Gamma_{a}\eta^{1s}\rangle_D\big)~e^a\wedge e^3
\cr
&&\qquad\qquad
 +2 \big(\mathrm{Re}\langle \eta^{1r}, \Gamma_{a}\lambda^{2s}\rangle_D- \mathrm{Re}\langle \eta^{2r}, \Gamma_{a}\lambda^{1s}\rangle_D\big)~e^a\wedge e^4
\cr
&&\qquad\qquad
+2 \big(\mathrm{Im}\langle \eta^{1r}, \Gamma_{a}\eta^{2s}\rangle_D-\mathrm{Im}\langle \eta^{2r}, \Gamma_{a}\eta^{1s}\rangle_D\big)~e^a\wedge e^8
\cr
&&\qquad\qquad
 +2 \big(\mathrm{Im}\langle \eta^{1r}, \Gamma_{a}\lambda^{2s}\rangle_D- \mathrm{Im}\langle \eta^{2r}, \Gamma_{a}\lambda^{1s}\rangle_D\big)~e^a\wedge e^9~,
 \eea

\bea
&&\pi^{rs}={1\over3}\big(\mathrm{Re}\langle \eta^{1r},\Gamma_{abc} \eta^{1s}\rangle_D+\mathrm{Re}\langle \eta^{2r}, \Gamma_{abc} \eta^{2r}\rangle_D \big) e^a\wedge e^b\wedge e^c
\cr
&& \qquad
-2\mathrm{Re}\langle \eta^{1r},\Gamma_{a} \lambda^{1s}\rangle_D (e^3\wedge e^4 -e^8\wedge e^9)\wedge e^a
\cr
&&\qquad
+2\mathrm{Re}\langle \eta^{2r},\Gamma_{a} \lambda^{2s}\rangle_D (e^3\wedge e^4 +e^8\wedge e^9)\wedge e^a
\cr
&&\qquad
-2\mathrm{Im}\langle \eta^{1r},\Gamma_{a} \eta^{1s}\rangle_D (e^3\wedge e^8 +e^4\wedge e^9)\wedge e^a
\cr
&&\qquad
+2\mathrm{Im}\langle \eta^{2r},\Gamma_{a} \eta^{2s}\rangle_D (e^3\wedge e^8 -e^4\wedge e^9)\wedge e^a
\cr
&&\qquad
-2\mathrm{Im}\langle \eta^{1r},\Gamma_{a} \lambda^{1s}\rangle_D (e^3\wedge e^9 -e^4\wedge e^8)\wedge e^a
\cr
&&\qquad
+2\mathrm{Im}\langle \eta^{2r},\Gamma_{a} \lambda^{2s}\rangle_D (e^3\wedge e^9 +e^4\wedge e^8)\wedge e^a~,
\eea

\bea
&&\zeta^{rs}= {1\over6} \tilde \omega^{rs}_{a\ell} \epsilon^\ell{}_{ijk} e^a\wedge e^i\wedge e^j\wedge e^k
\cr
&&\qquad\qquad
+{1\over3}  \big(\mathrm{Re}\langle \eta^{1r}, \Gamma_{abc}\eta^{2s}\rangle_D+\mathrm{Re}\langle \eta^{2r}, \Gamma_{a}\eta^{1s}\rangle_D\big)~e^a\wedge e^b\wedge e^c\wedge e^3
\cr
&&\qquad\qquad
 +{1\over3}  \big(\mathrm{Re}\langle \eta^{1r}, \Gamma_{abc}\lambda^{2s}\rangle_D- \mathrm{Re}\langle \eta^{2r}, \Gamma_{abc}\lambda^{1s}\rangle_D\big)~e^a\wedge e^b\wedge e^c\wedge e^4
\cr
&&\qquad\qquad
+{1\over3}  \big(\mathrm{Im}\langle \eta^{1r}, \Gamma_{abc}\eta^{2s}\rangle_D-\mathrm{Im}\langle \eta^{2r}, \Gamma_{abc}\eta^{1s}\rangle_D\big)~e^a\wedge e^b\wedge e^c\wedge e^8
\cr
&&\qquad\qquad
 +{1\over3}  \big(\mathrm{Im}\langle \eta^{1r}, \Gamma_{abc}\lambda^{2s}\rangle_D- \mathrm{Im}\langle \eta^{2r}, \Gamma_{abc}\lambda^{1s}\rangle_D\big)~e^a\wedge e^b\wedge e^c\wedge e^9~,
\eea

\bea
&&\tau^{rs}=\tilde k^{rs}\wedge e^3\wedge e^4\wedge e^8\wedge e^9
\cr &&\qquad
-{1\over3}\mathrm{Re}\langle \eta^{1r},\Gamma_{abc} \lambda^{1s}\rangle_D (e^3\wedge e^4 -e^8\wedge e^9)\wedge e^a\wedge e^b\wedge e^c
\cr
&&\qquad
+{1\over3}\mathrm{Re}\langle \eta^{2r},\Gamma_{abc} \lambda^{2s}\rangle_D (e^3\wedge e^4 +e^8\wedge e^9)\wedge e^a\wedge e^b\wedge e^c
\cr
&&\qquad
-{1\over3}\mathrm{Im}\langle \eta^{1r},\Gamma_{abc} \eta^{1s}\rangle_D (e^3\wedge e^8 +e^4\wedge e^9)\wedge e^a\wedge e^b\wedge e^c
\cr
&&\qquad
+{1\over3}\mathrm{Im}\langle \eta^{2r},\Gamma_{abc} \eta^{2s}\rangle_D (e^3\wedge e^8 -e^4\wedge e^9)\wedge e^a\wedge e^b\wedge e^c
\cr
&&\qquad
-{1\over3}\mathrm{Im}\langle \eta^{1r},\Gamma_{abc} \lambda^{1s}\rangle_D (e^3\wedge e^9 -e^4\wedge e^8)\wedge e^a\wedge e^b\wedge e^c
\cr
&&\qquad
+{1\over3}\mathrm{Im}\langle \eta^{2r},\Gamma_{abc} \lambda^{2s}\rangle_D (e^3\wedge e^9 +e^4\wedge e^8)\wedge e^a\wedge e^b\wedge e^c
\cr
&&\qquad
+{2\over5!} \big(\mathrm{Re}\langle \eta^{1r}, \Gamma_{a_1\dots a_5}\eta^{1s}\rangle_D+\mathrm{Re}\langle \eta^{2r}, \Gamma_{a_1\dots a_5}\eta^{2s}\rangle_D\big)~e^{a_1}\wedge\dots\wedge e^{a_5}~,
\eea
where  $a,b,c=0,1,2,5,6,7$ are the worldvolume directions, $\epsilon_{3489}=1$ and $(e^a, e^3, e^4, e^8, e^9)$ is a pseudo-orthonormal frame for the NS5-brane metric (\ref{ns5}). The remaining form bilinears $\tilde \sigma$, $\tilde k$, $\tilde \omega$, $\tilde \pi$, $\tilde \zeta$ and $\tilde \tau$ bilinears can be constructed from those above upon replacing both
$\eta^{2s}$  and $\lambda^{2s}$  with $-\eta^{2s}$ and $-\lambda^{2s}$, respectively.

\subsection{ Form bilinears IIB branes}

All the bilinears below are manifestly real. In particular we have replaced $k^{(2)}, \pi^{(2)}$ and $\tau^{(2)}$ with $ik^{(2)}, i\pi^{(2)}$ and $i\tau^{(2)}$, respectively.

\subsubsection{Fundamental string}

Choosing the worldvolume and transverse directions of the IIB fundamental string as in the IIA case,  a direct computation using (\ref{fiibsol}) reveals that the form bilinears of  IIB fundamental string  are
\bea
&&
k^{rs}= h^{-{1\over2}}\langle \eta^r, \eta^s\rangle (e^0-e^5)+ h^{-{1\over2}} \langle \lambda^r, \lambda^s\rangle (e^0+e^5)~,
\cr
&&
\pi^{rs}={1\over2} h^{-{1\over2}} \langle\eta^r, \Gamma_{ij} \eta^s\rangle (e^0-e^5) \wedge e^i\wedge e^j+{1\over2} h^{-{1\over2}} \langle\lambda^r, \Gamma_{ij} \lambda^s\rangle (e^0+e^5) \wedge e^i\wedge e^j~,
\cr
&&
\tau^{rs}={1\over4!} h^{-{1\over2}} \langle\eta^r, \Gamma_{i_1\dots i_4} \eta^s\rangle (e^0-e^5) \wedge e^{i_1}\wedge \dots \wedge e^{i_4}
\cr
&& \qquad\qquad
+{1\over4!} h^{-{1\over2}} \langle\lambda^r, \Gamma_{i_1\dots i_4} \lambda^s\rangle (e^0+e^5) \wedge e^{i_1}\wedge\dots \wedge e^{i_4}~,
\eea
where again $(e^0, e^5, e^i)$ is a pseudo-orthonormal frame of (\ref{fstring}).
The $k^{(3)}, \pi^{(3)}$ and $\tau^{(3)}$ bilinears can be obtained from those above upon replacing $\lambda^s$ with $-\lambda^s$.

For the remaining form bilinears a direct computation yields
\bea
&&k^{(1)}{}^{rs}=h^{-{1\over2}} \langle\eta^r, \Gamma_i \lambda^s\rangle\, e^i+h^{-{1\over2}} \langle\lambda^r, \Gamma_i \eta^s\rangle\, e^i~,
\eea
\bea
&&\pi^{(1)}{}^{rs}=-h^{-{1\over2}} \langle\eta^r, \Gamma_i \lambda^s\rangle e^0\wedge e^5\wedge e^i+{1\over3!}h^{-{1\over2}} \langle\eta^r, \Gamma_{ijk} \lambda^s\rangle e^i\wedge e^j\wedge e^k
\cr
&&\qquad\qquad
+h^{-{1\over2}} \langle\lambda^r, \Gamma_i \eta^s\rangle e^0\wedge e^5\wedge e^i+{1\over3!}h^{-{1\over2}} \langle\lambda^r, \Gamma_{ijk} \eta^s\rangle e^i\wedge e^j\wedge e^k~,
\eea
\bea
&&\tau^{(1)}{}^{rs}=-{1\over3!}h^{-{1\over2}} \langle\eta^r, \Gamma_{ijk} \lambda^s\rangle   e^0\wedge e^5\wedge  e^i\wedge e^j\wedge e^k
\cr && \qquad\qquad
+{1\over5!}h^{-{1\over2}} \langle\eta^r, \Gamma_{i_1\dots i_5} \lambda^s \rangle e^{i_1}\wedge\dots \wedge e^{i_5}
\cr
&&\qquad\qquad
+{1\over3!}h^{-{1\over2}} \langle\lambda^r, \Gamma_{ijk} \eta^s  \rangle e^0\wedge e^5\wedge  e^i\wedge e^j\wedge e^k
\cr && \qquad\qquad
+{1\over5!}h^{-{1\over2}} \langle\lambda^r, \Gamma_{i_1\dots i_5} \eta^s \rangle e^{i_1}\wedge\dots \wedge e^{i_5}~.
\eea
The $k^{(2)}, \pi^{(2)}$ and $\tau^{(2)}$ bilinears can be obtained from those above upon replacing  $\eta^s$ with  $-\eta^s$.

\subsubsection{NS5-brane}

Choosing the worldvolume and transverse directions  as in the IIA case above, a direct computation using (\ref{iibsolns5}) reveals that the form bilinears of  IIB NS5-brane  are
\bea
&&
k^{rs}= 2 \mathrm{Re} \langle \eta^{1r}, \Gamma_a \eta^{1s}\rangle_D\, e^a
+ 2 \mathrm{Re} \langle \eta^{2r}, \Gamma_a \eta^{2s}\rangle_D\, e^a~,
\eea
\bea
&&\pi^{rs}=- 2 \mathrm{Re} \langle \eta^{1r}, \Gamma_a \lambda^{1s}\rangle_D \, e^a \wedge (e^3\wedge e^4-e^8\wedge e^9)
\cr
&&\qquad\qquad
+2  \mathrm{Re} \langle \eta^{2r}, \Gamma_a \lambda^{2s}\rangle_D\, e^a \wedge (e^3\wedge e^4+e^8\wedge e^9)
\cr
&&\qquad\qquad
-2 \mathrm{Im}\langle \eta^{1r}, \Gamma_a \eta^{1s}\rangle_D\, e^a\wedge (e^3\wedge e^8+ e^4\wedge e^9)
\cr
&&\qquad\qquad
+2 \mathrm{Im}\langle \eta^{2r}, \Gamma_a \eta^{2s}\rangle_D\, e^a\wedge (e^3\wedge e^8- e^4\wedge e^9)
\cr
&&\qquad\qquad
- 2 \mathrm{Im} \langle \eta^{1r}, \Gamma_a \lambda^{1s}\rangle_D \, e^a \wedge (e^3\wedge e^9-e^4\wedge e^8)
\cr
&&\qquad\qquad
+2  \mathrm{Im} \langle \eta^{2r}, \Gamma_a \lambda^{2s}\rangle_D\, e^a \wedge (e^3\wedge e^9+e^4\wedge e^8)
\cr
&&\qquad\qquad
+{1\over3} \big(\mathrm{Re} \langle \eta^{1r}, \Gamma_{abc} \eta^{1s}\rangle_D + \mathrm{Re} \langle \eta^{2r}, \Gamma_{abc} \eta^{2s}\rangle_D\big) e^a\wedge e^b \wedge e^c~,
\eea
\bea
&&\tau^{rs}= 2\big(\mathrm{Re}\langle \eta^{1r}, \Gamma_a \eta^{1s}\rangle_D - \mathrm{Re}\langle \eta^{2r}, \Gamma_a \eta^{2s}\rangle_D\big)
\,e^a\wedge e^3\wedge e^4\wedge e^8\wedge e^9
\cr
&&\qquad\qquad
- {1\over3} \mathrm{Re} \langle \eta^{1r}, \Gamma_{abc} \lambda^{1s}\rangle_D \, e^a \wedge  e^b \wedge e^c \wedge(e^3\wedge e^4-e^8\wedge e^9)
\cr
&&\qquad\qquad
+{1\over3}  \mathrm{Re} \langle \eta^{2r}, \Gamma_{abc} \lambda^{2s}\rangle_D\, e^a\wedge  e^b \wedge e^c  \wedge (e^3\wedge e^4+e^8\wedge e^9)
\cr
&&\qquad\qquad
-{1\over3} \mathrm{Im}\langle \eta^{1r}, \Gamma_{abc} \eta^{1s}\rangle_D\, e^a\wedge  e^b \wedge e^c \wedge (e^3\wedge e^8+ e^4\wedge e^9)
\cr
&&\qquad\qquad
+{1\over3} \mathrm{Im}\langle \eta^{2r}, \Gamma_{abc} \eta^{2s}\rangle_D\, e^a\wedge  e^b \wedge e^c \wedge (e^3\wedge e^8- e^4\wedge e^9)
\cr
&&\qquad\qquad
- {1\over3} \mathrm{Im} \langle \eta^{1r}, \Gamma_{abc} \lambda^{1s}\rangle_D \, e^a\wedge  e^b \wedge e^c  \wedge (e^3\wedge e^9-e^4\wedge e^8)
\cr
&&\qquad\qquad
+{1\over3}  \mathrm{Im} \langle \eta^{2r}, \Gamma_{abc} \lambda^{2s}\rangle_D\, e^a\wedge  e^b \wedge e^c  \wedge (e^3\wedge e^9+e^4\wedge e^8)
\cr
&&\qquad\qquad
+{2\over5!} \big(\mathrm{Re} \langle \eta^{1r}, \Gamma_{a_1\dots a_5} \eta^{1s}\rangle_D + \mathrm{Re} \langle \eta^{2r}, \Gamma_{a_1\dots a_5} \eta^{2s}\rangle_D\big) e^{a_1}\wedge\dots \wedge e^{a_5}~,
\eea
where $(e^a, e^3, e^4, e^8, e^9)$ is a pseudo-orthonormal frame of the metric (\ref{ns5}).
The form bilinears $k^{(3)}$, $\pi^{(3)}$ and $\tau^{(3)}$ can be constructed from those above after changing the sign in front of the terms containing the inner products $\langle\eta^2, Q\eta^2\rangle_D$ and $\langle\eta^2, Q\lambda^2\rangle_D$ for all Clifford elements  $Q$.

The remaining bilinears can be obtained in a similar way to find
\bea
&&k^{(1)rs} = 2\big(\mathrm{Re}\langle \eta^{1r}, \eta^{2s} \rangle_D - \mathrm{Re}\langle \eta^{2r}, \eta^{1s} \rangle_D \big)\, e^3 + 2\big(\mathrm{Re}\langle \eta^{1r}, \lambda^{2s} \rangle_D + \mathrm{Re}\langle \eta^{2r}, \lambda^{1s} \rangle_D \big)\, e^4
\cr
&&\qquad\qquad
+2\big(\mathrm{Im}\langle \eta^{1r}, \eta^{2s} \rangle_D + \mathrm{Im}\langle \eta^{2r}, \eta^{1s} \rangle_D \big)\, e^8
\cr
&&\qquad\qquad
+2\big(\mathrm{Im}\langle \eta^{1r}, \lambda^{2s} \rangle_D + \mathrm{Im}\langle \eta^{2r}, \lambda^{1s} \rangle_D \big)\, e^9~,
\eea
\bea
&&\pi^{(1)}{}^{rs}=\Big[ \big(\mathrm{Re}\langle\eta^{1r},\Gamma_{ab} \eta^{2s}\rangle_D-\mathrm{Re}\langle\eta^{2r}, \Gamma_{ab}\eta^{1s}\rangle_D \big) \, e^3
+ \big(\mathrm{Re}\langle\eta^{1r},\Gamma_{ab} \lambda^{2s}\rangle_D+\mathrm{Re}\langle\eta^{2r}, \Gamma_{ab}\lambda^{1s}\rangle_D \big) \, e^4
\cr
&&\qquad\qquad
+ \big(\mathrm{Im}\langle\eta^{1r}, \Gamma_{ab}\eta^{2s}\rangle_D+\mathrm{Im}\langle\eta^{2r}, \Gamma_{ab}\eta^{1s}\rangle_D \big) \, e^8
\cr
&&\qquad\qquad
+ \big(\mathrm{Im}\langle\eta^{1r}, \Gamma_{ab}\lambda^{2s}\rangle_D+\mathrm{Im}\langle\eta^{2r}, \Gamma_{ab} \lambda^{1s}\rangle_D \big) \, e^9\Big ] \wedge e^a\wedge e^b
\cr
&&\qquad \qquad
+ 2 \big(-\mathrm{Re}\langle\eta^{1r}, \eta^{2s}\rangle_D-\mathrm{Re}\langle\eta^{2r}, \eta^{1s}\rangle_D \big) \, e^4\wedge e^ 8\wedge e^9
\cr
&&\qquad \qquad
- 2 \big(-\mathrm{Re}\langle\eta^{1r}, \lambda^{2s}\rangle_D+\mathrm{Re}\langle\eta^{2r}, \lambda^{1s}\rangle_D \big) \, e^3\wedge e^8\wedge e^9
\cr
&&\qquad\qquad
+ 2 \big(-\mathrm{Im}\langle\eta^{1r}, \eta^{2s}\rangle_D+\mathrm{Im}\langle\eta^{2r}, \eta^{1s}\rangle_D \big) \, e^3\wedge e^4\wedge e^9
\cr
&&\qquad\qquad
- 2 \big(-\mathrm{Im}\langle\eta^{1r}, \lambda^{2s}\rangle_D+\mathrm{Im}\langle\eta^{2r}, \lambda^{1s}\rangle_D \big) \, e^3\wedge e^4\wedge e^8~,
\eea
\bea
&&\tau^{(1)}{}^{rs}={1\over12}\Big[ \big(\mathrm{Re}\langle\eta^{1r},\Gamma_{a_1\dots a_4} \eta^{2s}\rangle_D-\mathrm{Re}\langle\eta^{2r}, \Gamma_{a_1\dots a_4}\eta^{1s}\rangle_D \big) \, e^3
\cr
&&\qquad\quad
+ \big(\mathrm{Re}\langle\eta^{1r},\Gamma_{a_1\dots a_4} \lambda^{2s}\rangle_D+\mathrm{Re}\langle\eta^{2r}, \Gamma_{a_1\dots a_4}\lambda^{1s}\rangle_D \big) \, e^4
\cr
&&\qquad\quad
+ \big(\mathrm{Im}\langle\eta^{1r}, \Gamma_{a_1\dots a_4}\eta^{2s}\rangle_D+\mathrm{Im}\langle\eta^{2r}, \Gamma_{a_1\dots a_4}\eta^{1s}\rangle_D \big) \, e^8
\cr
&&\qquad\quad
+ \big(\mathrm{Im}\langle\eta^{1r}, \Gamma_{a_1\dots a_4}\lambda^{2s}\rangle_D+\mathrm{Im}\langle\eta^{2r}, \Gamma_{a_1\dots a_4} \lambda^{1s}\rangle_D \big) \, e^9\Big ] \wedge e^{a_1}\wedge\dots\wedge  e^{a_4}
\cr
&&\qquad \quad
+ \big(-\mathrm{Re}\langle\eta^{1r},\Gamma_{ab} \eta^{2s}\rangle_D-\mathrm{Re}\langle\eta^{2r}, \Gamma_{ab}\eta^{1s}\rangle_D \big) \, \, e^a\wedge e^b\wedge e^4\wedge e^ 8\wedge e^9
\cr
&&\qquad \quad
- \big[\big(-\mathrm{Re}\langle\eta^{1r}, \Gamma_{ab}\lambda^{2s}\rangle_D+\mathrm{Re}\langle\eta^{2r}, \Gamma_{ab}\lambda^{1s}\rangle_D \big) \, \, e^a\wedge e^b\wedge e^3\wedge e^8\wedge e^9
\cr
&&\qquad\quad
+ \big(-\mathrm{Im}\langle\eta^{1r}, \Gamma_{ab}\eta^{2s}\rangle_D+\mathrm{Im}\langle\eta^{2r},\Gamma_{ab} \eta^{1s}\rangle_D \big) \, e^a\wedge e^b\wedge \, e^3\wedge e^4\wedge e^9
\cr
&&\qquad\quad
- \big(-\mathrm{Im}\langle\eta^{1r}, \Gamma_{ab} \lambda^{2s}\rangle_D+\mathrm{Im}\langle\eta^{2r}, \Gamma_{ab}\lambda^{1s}\rangle_D \big)\big]  \, e^a\wedge e^b\wedge e^3\wedge e^4\wedge e^8~,
\eea
where the form bilinears $k^{(2)}$, $\pi^{(2)}$ and $\tau^{(2)}$ can be obtained from those above upon changing the sign in front  of the components
containing  the bilinears $\langle \eta^2, Q \eta^1\rangle$, $\langle \eta^2, Q \lambda^1\rangle$ and $\langle \lambda^2, Q \eta^1\rangle$, where  $Q$ is a Clifford element.

\section{Form bilinears of D-branes} \label{apb}

\subsection{Form bilinears of IIA D-branes}

\subsubsection{D0-brane}

Using  the expression for the Killing spinors of the D0-brane (\ref{d0sol}), one finds that the  non-vanishing from bilinears of the solution are
\bea
\tilde \sigma^{rs}=-2 h^{-{1\over4}}\, \langle\eta^r, \eta^s\rangle~,~~~
k^{rs}=2 h^{-{1\over4}}\, \langle\eta^r, \eta^s\rangle\,  e^0~,
\eea
\bea
\tilde k^{rs}=-2 h^{-{1\over4}} \langle\eta^r, \Gamma_{11}\eta^s\rangle\,  e^5+ 2 h^{-{1\over4}} \langle\eta^r, \Gamma_i\eta^s\rangle\,  e^i~,
\eea
\bea
\omega^{rs}=-2 h^{-{1\over4}}\,  \langle\eta^r, \Gamma_{11} \eta^s\rangle\, e^0\wedge e^5+ 2 h^{-{1\over4}} \langle\eta^r, \Gamma_i\eta^s\rangle  e^0\wedge e^i~,
\eea

\bea
\tilde\omega^{rs}=-2 h^{-{1\over4}}\, \langle\eta^r, \Gamma_i \Gamma_{11} \eta^s\rangle\, e^5\wedge e^i- h^{-{1\over4}} \langle\eta^r, \Gamma_{ij}\eta^s\rangle  e^i\wedge e^j~,
\eea

\bea
&&\pi^{rs}=2 h^{-{1\over4}}\, \langle\eta^r, \Gamma_i \Gamma_{11} \eta^s\rangle\, e^0\wedge e^5\wedge e^i+ h^{-{1\over4}} \langle\eta^r, \Gamma_{ij}\eta^s\rangle  e^0\wedge e^i\wedge e^j~,
\eea

\bea
\tilde\pi^{rs}=
- h^{-{1\over4}} \langle\eta^r, \Gamma_{ij}\Gamma_{11}\eta^s\rangle\,  e^5\wedge e^i\wedge e^j
 +{1\over3} h^{-{1\over4}} \langle\eta^r, \Gamma_{ijk}\eta^s\rangle  e^i\wedge e^j\wedge e^k~,
\eea

\bea
\zeta^{rs}=-h^{-{1\over4}}\, \langle\eta^r, \Gamma_{ij} \Gamma_{11}  \eta^s\rangle\, e^0\wedge e^5\wedge e^i\wedge e^j
+{1\over3}
h^{-{1\over4}} \langle\eta^r, \Gamma_{ijk}\eta^s\rangle\,  e^0\wedge e^i\wedge e^j \wedge e^k~,
\eea

\bea
\tilde \zeta^{rs}=-{1\over3} h^{-{1\over4}}\, \langle\eta^r, \Gamma_{ijk} \Gamma_{11} \eta^s\rangle\, e^5\wedge e^i\wedge e^j\wedge e^k
-{2\over4!} h^{-{1\over4}}\,\langle\eta^r, \Gamma_{i_1\dots i_4}\eta^s\rangle e^{i_1}\wedge \dots \wedge e^{i_4}~,
\eea

\bea
&&\tau^{rs}={1\over3} h^{-{1\over4}}\, \langle\eta^r, \Gamma_{ijk} \Gamma_{11} \eta^s\rangle\, e^0\wedge e^5\wedge e^i\wedge e^j\wedge e^k
\cr
&&
\qquad
+ {2\over4!} h^{-{1\over4}} \langle\eta^r, \Gamma_{i_1\dots i_4}\eta^s\rangle\,  e^0\wedge e^{i_1}\wedge\dots \wedge  e^{i_4}~,
\eea
where $i,j,k=1,2,3,4,6,7,8,9$ and $(e^0, e^5, e^i)$ is a pseudo-orthonormal frame of the D0-brane metric (\ref{dbrane}) for $p=0$.

\subsubsection{D6-brane}

Using the  expression for the Killing spinors in (\ref{d6sol}), one can easily compute the non-vanishing form bilinears of   D6-brane  as follows
\begin{equation}
	\sigma^{rs} = 2 h^{-\frac{1}{4}} \mathrm{Re}{\D{\eta^r}{\eta^s}}~,~~~
	k^{rs} = 2 h^{-\frac{1}{4}} \mathrm{Re}{\D{\eta^r}{\Gamma_a \eta^s}}\, e^a~,
\end{equation}
\begin{equation}
	\tilde{k}^{rs} = -2 h^{-\frac{1}{4}} \mathrm{Re}{\D{\eta^r}{\Gamma_{11} \lambda^s}}\, e^4 + 2 h^{-\frac{1}{4}} \mathrm{Im} \D{\eta^r}{\eta^s}\, e^5 - 2 h^{-\frac{1}{4}} \mathrm{Im}{\D{\eta^r}{\Gamma_{11} \lambda^s}}\, e^9~,
\end{equation}
\begin{equation}
	\begin{aligned}
		\omega^{rs} &= -2 h^{-\frac{1}{4}} \mathrm{Re}{\D{\eta^r}{\Gamma_5 \lambda^s}} e^4 \wedge e^5 - 2 h^{-\frac{1}{4}} \mathrm{Im}{\D{\eta^r}{\eta^s}} e^4 \wedge e^9 \\
		+& 2 h^{-\frac{1}{4}} \mathrm{Im}{\D{\eta^r}{\Gamma_5 \lambda^s}} e^5 \wedge e^9 + h^{-\frac{1}{4}} \mathrm{Re}{\D{\eta^r}{\Gamma_{ab}\eta^s}} e^a \wedge e^b~,
	\end{aligned}
\end{equation}
\begin{equation}
	\begin{aligned}
		\tilde\omega^{rs} &= -2 h^{-\frac{1}{4}} \mathrm{Re}{\D{\eta^r}{\Gamma_a \Gamma_{11} \lambda^s}} e^a \wedge e^4 + 2 h^{-\frac{1}{4}} \mathrm{Im} {\D{\eta^r}{\Gamma_a \eta^s}} e^a \wedge e^5 \\
		-& 2 h^{-\frac{1}{4}} \mathrm{Im}{\D{\eta^r}{\Gamma_a \Gamma_{11} \lambda^s}} e^a \wedge e^9~,
	\end{aligned}
\end{equation}
\begin{equation}
	\begin{aligned}
		\pi^{rs} &= - 2 h^{-\frac{1}{4}} \mathrm{Re}{\D{\eta^r}{\Gamma_a \Gamma_{5} \lambda^s}} e^a \wedge e^4 \wedge e^5 - 2 h^{-\frac{1}{4}} \mathrm{Im}{\D{\eta^r}{\Gamma_a \eta^s}} e^a \wedge e^4 \wedge e^9 \\
		+& 2 h^{-\frac{1}{4}} \mathrm{Im}{\D{\eta^r}{\Gamma_a \Gamma_{5} \lambda^s}} e^a \wedge e^5 \wedge e^9 + \frac{1}{3} h^{-\frac{1}{4}} \mathrm{Re}{\D{\eta^r}{\Gamma_{abc} \eta^s}} e^a \wedge e^b \wedge e^c~,
	\end{aligned}
\end{equation}
\begin{equation}
	\begin{aligned}
		\tilde{\pi}^{rs} &= -2 h^{-\frac{1}{4}} \mathrm{Re}{\D{\eta^r}{\eta^s}} e^4 \wedge e^5 \wedge e^9 - h^{-\frac{1}{4}} \mathrm{Re}{\D{\eta^r}{\Gamma_{ab} \Gamma_{11} \lambda^s}} e^a \wedge e^b \wedge e^4 \\
		+& h^{-\frac{1}{4}} \mathrm{Im}{\D{\eta^r}{\Gamma_{ab} \eta^s}} e^a \wedge e^b \wedge e^5 - h^{-\frac{1}{4}} \mathrm{Im}{\D{\eta^r}{\Gamma_{ab} \Gamma_{11} \lambda^s}} e^a \wedge e^b \wedge e^9~,
	\end{aligned}
\end{equation}
\bea
		\zeta^{rs} &&= - h^{-\frac{1}{4}} \mathrm{Re}{\D{\eta^r}{\Gamma_{ab} \Gamma_5 \lambda^s}} e^a \wedge e^b \wedge e^4 \wedge e^5 - h^{-\frac{1}{4}} \mathrm{Im}{\D{\eta^r}{\Gamma_{ab} \eta^s}} e^a \wedge e^b \wedge e^4 \wedge e^9
\cr
		&&\qquad + h^{-\frac{1}{4}} \mathrm{Im}{\D{\eta^r}{\Gamma_{ab} \Gamma_5 \lambda^s}} e^a \wedge e^b \wedge e^5 \wedge e^9
\cr
&&\qquad+ \frac{1}{12} h^{-\frac{1}{4}} \mathrm{Re}{\D{\eta^r}{\Gamma_{abcd}\eta^s}} e^a \wedge e^b \wedge e^c \wedge e^d~,
	\eea
\bea
		\tilde{\zeta}^{rs} &&= -2 h^{-\frac{1}{4}} \mathrm{Re}{\D{\eta^r}{\Gamma_a \eta^s}} e^a \wedge e^4 \wedge e^5 \wedge e^9 - \frac{1}{3} h^{-\frac{1}{4}} \mathrm{Re}{\D{\eta^r}{\Gamma_{abc} \Gamma_{11} \lambda^s}} e^a \wedge e^b \wedge e^c \wedge e^4
\cr
		&&\qquad + \frac{1}{3} h^{-\frac{1}{4}} \mathrm{Im}{\D{\eta^r}{\Gamma_{abc} \eta^s}} e^a \wedge e^b \wedge e^c \wedge e^5
\cr
&&\qquad - \frac{1}{3} h^{-\frac{1}{4}} \mathrm{Im}{\D{\eta^r}{\Gamma_{abc} \Gamma_{11} \lambda^s}} e^a \wedge e^b \wedge e^c \wedge e^9~,
	\eea
\bea
		\tau^{rs} &&= - \frac{1}{3} h^{-\frac{1}{4}} \mathrm{Re}{\D{\eta^r}{\Gamma_{abc} \Gamma_5 \lambda^s}} e^a \wedge e^b \wedge e^c \wedge e^4 \wedge e^5 - \frac{1}{3} h^{-\frac{1}{4}} \mathrm{Im}{\D{\eta^r}{\Gamma_{abc} \eta^s}} e^a \wedge e^b \wedge e^c \wedge e^4 \wedge e^9
\cr
		&&\qquad + \frac{1}{3} h^{-\frac{1}{4}} \mathrm{Im}{\D{\eta^r}{\Gamma_{abc} \Gamma_5 \lambda^s}} e^a \wedge e^b \wedge e^c \wedge e^5 \wedge e^9
 \cr
 &&\qquad + \frac{2}{5!} h^{-\frac{1}{4}} \mathrm{Re}{\D{\eta^r}{\Gamma_{a_1 \dots a_5} \eta^s}} e^{a_1} \wedge \dots \wedge e^{a_5}~,
	\eea
where $a,b,c=0, 1,2,3, 6,7,8$ and $(e^a, e^5, e^4, e^9)$ is a pseudo-orthonormal frame of the metric (\ref{dbrane}) with $p=6$.

\subsubsection{D2-brane}

Using the Killing spinors (\ref{d2solscon}), one finds that the non-vanishing  form bilinears of the D2-brane solution  are as follows
\bea
\sigma^{rs}=h^{-\frac{1}{4}}(-\langle \eta^r, \lambda^s\rangle+\langle \lambda^r, \eta^s\rangle)~,~~\tilde k^{rs}=h^{-\frac{1}{4}}(-\langle \eta^r, \Gamma_i \Gamma_{11}\lambda^s\rangle+ \langle \lambda^r, \Gamma_i \Gamma_{11} \eta^s\rangle)\, e^i~,
\eea
\bea
k^{rs}&=& h^{-\frac{1}{4}}(\langle \eta^r, \eta^s\rangle+ \langle \lambda^r, \lambda^s\rangle)\, e^0+h^{-\frac{1}{4}}(-\langle \eta^r, \eta^s\rangle+ \langle \lambda^r, \lambda^s\rangle)\, e^5
\cr
&&\qquad
+h^{-\frac{1}{4}}(\langle \eta^r, \lambda^s\rangle+ \langle \lambda^r, \eta^s\rangle)\, e^1~,
\eea
\bea
\omega^{rs}&=&{h^{-\frac{1}{4}}\over2} (-\langle \eta^r, \Gamma_{ij} \lambda^s\rangle+\langle \lambda^r,\Gamma_{ij} \eta^s\rangle)\, e^i\wedge e^j+h^{-\frac{1}{4}}(\langle \eta^r, \lambda^s\rangle+ \langle \lambda^r, \eta^s\rangle)\, e^0\wedge e^5
\cr
&&
+h^{-\frac{1}{4}}(\langle \eta^r, \eta^s\rangle- \langle \lambda^r, \lambda^s\rangle)\, e^0\wedge e^1+h^{-\frac{1}{4}} (\langle \eta^r, \eta^s\rangle+ \langle \lambda^r, \lambda^s\rangle)\, e^1\wedge e^5~,
\eea
\bea
\tilde \omega^{rs}&=& -h^{-\frac{1}{4}}(\langle \eta^r, \Gamma_{i} \Gamma_{11}\eta^s\rangle+\langle \lambda^r,\Gamma_{i} \Gamma_{11}\lambda^s\rangle)\,e^i\wedge e^0
+ h^{-\frac{1}{4}} (\langle \eta^r, \Gamma_{i} \Gamma_{11}\eta^s\rangle-\langle \lambda^r,\Gamma_{i} \Gamma_{11}\lambda^s\rangle)\, e^i\wedge e^5
\cr
&&
-h^{-\frac{1}{4}}
(\langle \eta^r, \Gamma_i \Gamma_{11} \lambda^s\rangle+ \langle \lambda^r, \Gamma_i \Gamma_{11}\eta^s\rangle)\, e^i\wedge e^1~,
\eea
\bea
\pi^{rs}&&=h^{-\frac{1}{4}}(-\langle \eta^r,  \lambda^s\rangle+ \langle \lambda^r, \eta^s\rangle)\,e^0\wedge e^5\wedge e^1
\cr
&&
\qquad
+{h^{-\frac{1}{4}}\over2}\Big(
 (\langle \eta^r,\Gamma_{ij} \eta^s\rangle+ \langle \lambda^r, \Gamma_{ij}\lambda^s\rangle)\, e^0+(-\langle \eta^r, \Gamma_{ij}\eta^s\rangle+ \langle \lambda^r,\Gamma_{ij} \lambda^s\rangle)\, e^5
 \cr
&&
\qquad
+(\langle \eta^r, \Gamma_{ij}\lambda^s\rangle+ \langle \lambda^r, \Gamma_{ij}\eta^s\rangle)\,  e^1 \Big)\,\wedge e^i\wedge e^j~,
\eea
\bea
\tilde \pi^{rs}&=&h^{-\frac{1}{4}}(\langle \lambda^r, \Gamma_i \Gamma_{11} \eta^s\rangle+ \langle \eta^r, \Gamma_i \Gamma_{11}\lambda^s\rangle)\, e^0\wedge e^5\wedge e^i
\cr
&&
\qquad
+ h^{-\frac{1}{4}}
(\langle \eta^r, \Gamma_i \Gamma_{11} \eta^s\rangle- \langle \lambda^r, \Gamma_i \Gamma_{11}\lambda^s\rangle)\, e^0\wedge e^1\wedge e^i
\cr
&&
-h^{-\frac{1}{4}}(\langle \eta^r, \Gamma_i \Gamma_{11} \eta^s\rangle+ \langle \lambda^r, \Gamma_i \Gamma_{11}\lambda^s\rangle)\, e^5\wedge e^1\wedge e^i
\cr
&&
\cr
&&
+{h^{-\frac{1}{4}}\over3!}(-\langle \eta^r, \Gamma_{ijk} \Gamma_{11}\lambda^s\rangle+ \langle \lambda^r, \Gamma_{ijk} \Gamma_{11} \eta^s\rangle)\, e^i\wedge e^j\wedge e^k~,
\eea
\bea
\zeta^{rs}&=&{h^{-\frac{1}{4}}\over4!}(-\langle \eta^r, \Gamma_{i_1\dots i_4} \lambda^s\rangle+\langle \lambda^r, \Gamma_{i_1\dots i_4}\eta^s\rangle)\,
e^{i_1}\wedge\dots \wedge e^{i_4}
\cr
&&
+{h^{-\frac{1}{4}}\over2} \Big((\langle \eta^r, \Gamma_{ij}\lambda^s\rangle+ \langle \lambda^r,\Gamma_{ij} \eta^s\rangle)\, e^0\wedge e^5
+(\langle \eta^r, \Gamma_{ij}\eta^s\rangle- \langle \lambda^r, \Gamma_{ij}\lambda^s\rangle)\, e^0\wedge e^1
\cr
&&
+ (\langle \eta^r, \Gamma_{ij}\eta^s\rangle+ \langle \lambda^r, \Gamma_{ij}\lambda^s\rangle)\, e^1\wedge e^5\Big)\wedge e^i\wedge e^j~,
\eea
\bea
\tilde \zeta^{rs}&=& {h^{-\frac{1}{4}}\over3!} \Big(-(\langle \eta^r, \Gamma_{ijk} \Gamma_{11}\eta^s\rangle+\langle \lambda^r,\Gamma_{ijk} \Gamma_{11}\lambda^s\rangle)\,e^i\wedge e^j\wedge e^k\wedge e^0
\cr
&&
\qquad
+ (\langle \eta^r, \Gamma_{ijk} \Gamma_{11}\eta^s\rangle-\langle \lambda^r,\Gamma_{ijk} \Gamma_{11}\lambda^s\rangle)\, e^i\wedge e^j\wedge e^k\wedge e^5
\cr
&&
-
(\langle \eta^r, \Gamma_{ijk} \Gamma_{11} \lambda^s\rangle+ \langle \lambda^r, \Gamma_{ijk} \Gamma_{11}\eta^s\rangle)\Big)\, e^i\wedge e^j\wedge e^k\wedge e^1
\cr
&&
+h^{-\frac{1}{4}}(-\langle \eta^r, \Gamma_i \Gamma_{11} \lambda^s\rangle+ \langle \lambda^r, \Gamma_i \Gamma_{11}\eta^s\rangle)\, e^0\wedge e^5\wedge e^1\wedge  e^i~,
\eea
\bea
\tau^{rs}&=&{h^{-\frac{1}{4}}\over2}(-\langle \eta^r, \Gamma_{ij} \lambda^s\rangle+ \langle \lambda^r, \Gamma_{ij}\eta^s\rangle)\,e^0\wedge e^5\wedge e^1\wedge e^i\wedge e^j
\cr
&&
\qquad
+{h^{-\frac{1}{4}}\over 4!}\Big((\langle \eta^r, \Gamma_{i_1\dots i_4} \eta^s\rangle
 + \langle \lambda^r, \Gamma_{i_1\dots i_4}\lambda^s\rangle)\, e^0
 \cr
&&
\qquad
+(-\langle \eta^r, \Gamma_{i_1\dots i_4}\eta^s\rangle+ \langle \lambda^r, \Gamma_{i_1\dots i_4} \lambda^s\rangle)\, e^5
\cr
&&
\qquad
+(\langle \eta^r, \Gamma_{i_1\dots i_4}\lambda^s\rangle+ \langle \lambda^r, \Gamma_{i_1\dots i_4}\eta^s\rangle)\, e^1\Big)\wedge e^{i_1}\wedge \dots \wedge e^{i_4}~,
\eea
where $i,j,k=2,3,4,6,7,8,9 $ and $(e^0, e^5, e^1, e^i)$ is a pseudo-orthonormal frame of the metric (\ref{dbrane}) with $p=2$.

\subsubsection{D4-brane}

Using the Killing spinors (\ref{d4sol}), one finds that the non-vanishing form bilinears of the D4-brane solution are as follows
\bea
&&
\tilde\sigma^{rs}=2h^{-\frac{1}{4}} \mathrm{Re} \langle \eta^{1r}, \Gamma_{11} \eta^{1s}\rangle_D+2h^{-\frac{1}{4}} \mathrm{Re} \langle \lambda^{1r}, \Gamma_{11} \lambda^{1s}\rangle_D~,~~
\eea
\bea
k^{rs}=2 h^{-\frac{1}{4}} ( \mathrm{Re} \langle \eta^{1r}, \Gamma_{a} \eta^{1s}\rangle_D+ \mathrm{Re} \langle \lambda^{1r}, \Gamma_{a} \lambda^{1s}\rangle_D) \, e^a~,
\eea
\bea
\tilde k^{rs}&=&2 h^{-\frac{1}{4}}(\mathrm{Re} \langle \eta^{1r}, \Gamma_{11} \eta^{1s}\rangle_D- \mathrm{Re} \langle \lambda^{1r}, \Gamma_{11} \lambda^{1s}\rangle_D)\, e^2
\cr
&&
+2h^{-\frac{1}{4}} (-\mathrm{Re} \langle \eta^{1r}, \Gamma_{11} \lambda^{2s}\rangle_D+ \mathrm{Re} \langle \lambda^{1r}, \Gamma_{11} \eta^{2s}\rangle_D)\, e^4
\cr
&&
-2 h^{-\frac{1}{4}}(\mathrm{Re} \langle \eta^{1r}, \Gamma_{11} \lambda^{1s}\rangle_D+ \mathrm{Re} \langle \lambda^{1r}, \Gamma_{11} \eta^{1s}\rangle_D)\, e^3
\cr
&&
+2 h^{-\frac{1}{4}}(-\mathrm{Im} \langle \eta^{1r}, \Gamma_{11} \lambda^{1s}\rangle_D+ \mathrm{Im} \langle \lambda^{1r}, \Gamma_{11} \eta^{1s}\rangle_D)\, e^8
\cr
&&
+2h^{-\frac{1}{4}} (-\mathrm{Im} \langle \eta^{1r}, \Gamma_{11} \lambda^{2s}\rangle_D+ \mathrm{Im} \langle \lambda^{1r}, \Gamma_{11} \eta^{2s}\rangle_D)\, e^9~,
\eea
\bea
\omega^{rs}&=&2 h^{-\frac{1}{4}}(-\mathrm{Re} \langle \eta^{1r}, \Gamma_a \eta^{1s}\rangle_D+ \mathrm{Re} \langle \lambda^{1r}, \Gamma_a \lambda^{1s}\rangle_D)\, e^a\wedge e^2
\cr
&&
+2 h^{-\frac{1}{4}}(\mathrm{Re} \langle \eta^{1r}, \Gamma_a \lambda^{1s}\rangle_D+ \mathrm{Re} \langle \lambda^{1r}, \Gamma_a \eta^{1s}\rangle_D)\,e^a\wedge e^3
\cr
&&
+2h^{-\frac{1}{4}} (\mathrm{Re} \langle \eta^{1r}, \Gamma_a \lambda^{2s}\rangle_D- \mathrm{Re} \langle \lambda^{1r}, \Gamma_a \eta^{2s}\rangle_D)\, e^a\wedge e^4
\cr
&&
+2h^{-\frac{1}{4}} (\mathrm{Im} \langle \eta^{1r}, \Gamma_a \lambda^{1s}\rangle_D- \mathrm{Im} \langle \lambda^{1r}, \Gamma_a \eta^{1s}\rangle_D)\, e^a\wedge e^8
\cr
&&
+2h^{-\frac{1}{4}} (\mathrm{Im} \langle \eta^{1r}, \Gamma_a \lambda^{2s}\rangle_D- \mathrm{Im} \langle \lambda^{1r}, \Gamma_a \eta^{2s}\rangle_D)\, e^a\wedge e^9~,
\eea
\bea
\tilde \omega^{rs}&=&h^{-\frac{1}{4}}(\mathrm{Re} \langle \eta^{1r}, \Gamma_{ab} \Gamma_{11} \eta^{1s}\rangle_D+ \mathrm{Re} \langle \lambda^{1r}, \Gamma_{ab} \Gamma_{11} \lambda^{1s}\rangle_D)\, e^a\wedge e^b
\cr
&&
-2h^{-\frac{1}{4}}(\mathrm{Re} \langle \eta^{1r},  \Gamma_{11} \lambda^{1s}\rangle_D- \mathrm{Re} \langle \lambda^{1r},  \Gamma_{11} \eta^{1s}\rangle_D)\, e^2\wedge e^3
\cr
&&
+2h^{-\frac{1}{4}}(-\mathrm{Re} \langle \eta^{1r},  \Gamma_{11} \lambda^{2s}\rangle_D- \mathrm{Re} \langle \lambda^{1r},  \Gamma_{11} \eta^{2s}\rangle_D)\, e^2\wedge e^4
\cr
&&
-2h^{-\frac{1}{4}}(\mathrm{Im} \langle \eta^{1r},  \Gamma_{11} \lambda^{1s}\rangle_D+ \mathrm{Im} \langle \lambda^{1r},  \Gamma_{11} \eta^{1s}\rangle_D)\, e^2\wedge e^8
\cr
&&
-2h^{-\frac{1}{4}}(\mathrm{Im} \langle \eta^{1r},  \Gamma_{11} \lambda^{2s}\rangle_D+ \mathrm{Im} \langle \lambda^{1r},  \Gamma_{11} \eta^{2s}\rangle_D)\, e^2\wedge e^9
\cr
&&
+2h^{-\frac{1}{4}}(-\mathrm{Re} \langle \eta^{1r},  \Gamma_{11} \eta^{2s}\rangle_D+ \mathrm{Re} \langle \lambda^{1r},  \Gamma_{11} \lambda^{2s}\rangle_D)\, e^3\wedge e^4
\cr
&&
+2h^{-\frac{1}{4}}(-\mathrm{Im} \langle \eta^{1r},  \Gamma_{11} \eta^{1s}\rangle_D+ \mathrm{Im} \langle \lambda^{1r},  \Gamma_{11} \lambda^{1s}\rangle_D)\, e^3\wedge e^8
\cr
&&
+2h^{-\frac{1}{4}}(-\mathrm{Im} \langle \eta^{1r},  \Gamma_{11} \eta^{2s}\rangle_D+ \mathrm{Im} \langle \lambda^{1r},  \Gamma_{11} \lambda^{2s}\rangle_D)\, e^3\wedge e^9
\cr
&&
+2h^{-\frac{1}{4}}(\mathrm{Im} \langle \eta^{1r},  \Gamma_{11} \eta^{2s}\rangle_D+ \mathrm{Im} \langle \lambda^{1r},  \Gamma_{11} \lambda^{2s}\rangle_D)\, e^4\wedge e^8
\cr
&&
-2h^{-\frac{1}{4}}(\mathrm{Im} \langle \eta^{1r},  \Gamma_{11} \eta^{1s}\rangle_D+ \mathrm{Im} \langle \lambda^{1r},  \Gamma_{11} \lambda^{1s}\rangle_D)\, e^4\wedge e^9
\cr
&&
+2h^{-\frac{1}{4}}(\mathrm{Re} \langle \eta^{1r},  \Gamma_{11} \eta^{2s}\rangle_D+ \mathrm{Re} \langle \lambda^{1r},  \Gamma_{11} \lambda^{2s}\rangle_D)\, e^8\wedge e^9~,
\eea
\bea
\pi^{rs}&=&{h^{-\frac{1}{4}}\over3} (\mathrm{Re} \langle \eta^{1r},  \Gamma_{abc} \eta^{1s}\rangle_D+ \mathrm{Re} \langle \lambda^{1r},  \Gamma_{abc} \lambda^{1s}\rangle_D)\, e^a\wedge e^b\wedge e^c
\cr
&&
-2h^{-\frac{1}{4}}(\mathrm{Re} \langle \eta^{1r},  \Gamma_a \lambda^{1s}\rangle_D- \mathrm{Re} \langle \lambda^{1r},  \Gamma_a\eta^{1s}\rangle_D)\, e^2\wedge e^3\wedge e^a
\cr
&&
+2h^{-\frac{1}{4}}(-\mathrm{Re} \langle \eta^{1r},  \Gamma_a \lambda^{2s}\rangle_D- \mathrm{Re} \langle \lambda^{1r},  \Gamma_a \eta^{2s}\rangle_D)\, e^2\wedge e^4\wedge e^a
\cr
&&
-2h^{-\frac{1}{4}}(\mathrm{Im} \langle \eta^{1r},  \Gamma_a \lambda^{1s}\rangle_D+ \mathrm{Im} \langle \lambda^{1r},  \Gamma_a \eta^{1s}\rangle_D)\, e^2\wedge e^8\wedge e^a
\cr
&&
-2h^{-\frac{1}{4}}(\mathrm{Im} \langle \eta^{1r},  \Gamma_a \lambda^{2s}\rangle_D+ \mathrm{Im} \langle \lambda^{1r},  \Gamma_a \eta^{2s}\rangle_D)\, e^2\wedge e^9\wedge e^a
\cr
&&
+2h^{-\frac{1}{4}}(-\mathrm{Re} \langle \eta^{1r},  \Gamma_a \eta^{2s}\rangle_D+ \mathrm{Re} \langle \lambda^{1r},  \Gamma_a \lambda^{2s}\rangle_D)\, e^3\wedge e^4\wedge e^a
\cr
&&
+2h^{-\frac{1}{4}}(-\mathrm{Im} \langle \eta^{1r},  \Gamma_a \eta^{1s}\rangle_D+ \mathrm{Im} \langle \lambda^{1r},  \Gamma_a \lambda^{1s}\rangle_D)\, e^3\wedge e^8\wedge e^a
\cr
&&
+2h^{-\frac{1}{4}}(-\mathrm{Im} \langle \eta^{1r},  \Gamma_a \eta^{2s}\rangle_D+ \mathrm{Im} \langle \lambda^{1r},  \Gamma_a \lambda^{2s}\rangle_D)\, e^3\wedge e^9\wedge e^a
\cr
&&
+2h^{-\frac{1}{4}}(\mathrm{Im} \langle \eta^{1r},  \Gamma_a \eta^{2s}\rangle_D+ \mathrm{Im} \langle \lambda^{1r},  \Gamma_a \lambda^{2s}\rangle_D)\, e^4\wedge e^8\wedge e^a
\cr
&&
-2h^{-\frac{1}{4}}(\mathrm{Im} \langle \eta^{1r},  \Gamma_a \eta^{1s}\rangle_D+ \mathrm{Im} \langle \lambda^{1r},  \Gamma_a \lambda^{1s}\rangle_D)\, e^4\wedge e^9\wedge e^a
\cr
&&
+2h^{-\frac{1}{4}}(\mathrm{Re} \langle \eta^{1r},  \Gamma_a \eta^{2s}\rangle_D+ \mathrm{Re} \langle \lambda^{1r},  \Gamma_a \lambda^{2s}\rangle_D)\, e^8\wedge e^9\wedge e^a
\eea
\bea
\tilde\pi^{rs}&=&h^{-\frac{1}{4}}\Big( (\mathrm{Re} \langle \eta^{1r}, \Gamma_{ab} \Gamma_{11} \eta^{1s}\rangle_D- \mathrm{Re} \langle \lambda^{1r}, \Gamma_{ab}\Gamma_{11} \lambda^{1s}\rangle_D)\, e^2
\cr
&&
+ (-\mathrm{Re} \langle \eta^{1r},\Gamma_{ab} \Gamma_{11} \lambda^{2s}\rangle_D+ \mathrm{Re} \langle \lambda^{1r}, \Gamma_{ab}\Gamma_{11} \eta^{2s}\rangle_D)\, e^4
\cr
&&
- (\mathrm{Re} \langle \eta^{1r}, \Gamma_{ab}\Gamma_{11} \lambda^{1s}\rangle_D+ \mathrm{Re} \langle \lambda^{1r}, \Gamma_{ab}\Gamma_{11} \eta^{1s}\rangle_D)\, e^3
\cr
&&
+ (-\mathrm{Im} \langle \eta^{1r}, \Gamma_{ab}\Gamma_{11} \lambda^{1s}\rangle_D+ \mathrm{Im} \langle \lambda^{1r}, \Gamma_{ab} \Gamma_{11} \eta^{1s}\rangle_D)\, e^8
\cr
&&
+ (-\mathrm{Im} \langle \eta^{1r}, \Gamma_{ab}\Gamma_{11} \lambda^{2s}\rangle_D+ \mathrm{Im} \langle \lambda^{1r}, \Gamma_{ab}\Gamma_{11} \eta^{2s}\rangle_D)\, e^9\Big) \wedge e^a\wedge e^b
\cr
&&
+{1\over 2\cdot 3!} \epsilon_{ijk}{}^{mn} \tilde \omega^{rs}_{mn}\, e^i\wedge e^i\wedge e^k~,
\eea
\bea
\zeta^{rs}&=&{h^{-\frac{1}{4}}\over3} (-\mathrm{Re} \langle \eta^{1r}, \Gamma_{abc} \eta^{1s}\rangle_D+ \mathrm{Re} \langle \lambda^{1r}, \Gamma_{abc} \lambda^{1s}\rangle_D)\, e^a\wedge e^b\wedge e^c\wedge e^2
\cr
&&
+{h^{-\frac{1}{4}}\over3} (\mathrm{Re} \langle \eta^{1r}, \Gamma_{abc} \lambda^{1s}\rangle_D+ \mathrm{Re} \langle \lambda^{1r},  \Gamma_{abc} \eta^{1s}\rangle_D)\,e^a\wedge e^b\wedge e^c\wedge e^3
\cr
&&
+{h^{-\frac{1}{4}}\over3} (\mathrm{Re} \langle \eta^{1r}, \Gamma_{abc} \lambda^{2s}\rangle_D- \mathrm{Re} \langle \lambda^{1r}, \Gamma_{abc} \eta^{2s}\rangle_D)\, e^a\wedge e^b\wedge e^c\wedge e^4
\cr
&&
+{h^{-\frac{1}{4}}\over3} (\mathrm{Im} \langle \eta^{1r}, \Gamma_{abc} \lambda^{1s}\rangle_D- \mathrm{Im} \langle \lambda^{1r}, \Gamma_{abc} \eta^{1s}\rangle_D)\, e^a\wedge e^b\wedge e^c\wedge e^8
\cr
&&
+{h^{-\frac{1}{4}}\over3} (\mathrm{Im} \langle \eta^{1r}, \Gamma_{abc} \lambda^{2s}\rangle_D- \mathrm{Im} \langle \lambda^{1r}, \Gamma_{abc} \eta^{2s}\rangle_D)\, e^a\wedge e^b\wedge e^c\wedge e^9
\cr
&&
-{1\over12}  \pi^{rs}_{amn} \epsilon_{ijk}{}^{mn} e^a\wedge e^i\wedge e^j\wedge e^k~,
\eea
\bea
\tilde\zeta^{rs}&=&{2 h^{-\frac{1}{4}}\over 4!} (\mathrm{Re} \langle \eta^{1r}, \Gamma_{a_1\dots a_4} \Gamma_{11} \eta^{1s}\rangle_D+ \mathrm{Re} \langle \lambda^{1r},\Gamma_{a_1\dots a_4} \Gamma_{11} \lambda^{1s}\rangle_D) \, e^{a_1}\wedge \dots\wedge e^{a_4}
\cr
&&
-h^{-\frac{1}{4}}(\mathrm{Re} \langle \eta^{1r}, \Gamma_{ab} \Gamma_{11} \lambda^{1s}\rangle_D- \mathrm{Re} \langle \lambda^{1r}, \Gamma_{ab} \Gamma_{11} \eta^{1s}\rangle_D)\, e^a\wedge e^b\wedge e^2\wedge e^3
\cr
&&
+h^{-\frac{1}{4}}(-\mathrm{Re} \langle \eta^{1r}, \Gamma_{ab} \Gamma_{11} \lambda^{2s}\rangle_D- \mathrm{Re} \langle \lambda^{1r}, \Gamma_{ab} \Gamma_{11} \eta^{2s}\rangle_D)\, e^a\wedge e^b\wedge e^2\wedge e^4
\cr
&&
-h^{-\frac{1}{4}}(\mathrm{Im} \langle \eta^{1r}, \Gamma_{ab} \Gamma_{11} \lambda^{1s}\rangle_D+ \mathrm{Im} \langle \lambda^{1r}, \Gamma_{ab} \Gamma_{11} \eta^{1s}\rangle_D)\,  e^a\wedge e^b\wedge e^2\wedge e^8
\cr
&&
-h^{-\frac{1}{4}}(\mathrm{Im} \langle \eta^{1r}, \Gamma_{ab} \Gamma_{11} \lambda^{2s}\rangle_D+ \mathrm{Im} \langle \lambda^{1r}, \Gamma_{ab} \Gamma_{11} \eta^{2s}\rangle_D)\, e^a\wedge e^b\wedge e^2\wedge e^9
\cr
&&
+h^{-\frac{1}{4}}(-\mathrm{Re} \langle \eta^{1r}, \Gamma_{ab} \Gamma_{11} \eta^{2s}\rangle_D+ \mathrm{Re} \langle \lambda^{1r}, \Gamma_{ab} \Gamma_{11} \lambda^{2s}\rangle_D)\, e^a\wedge e^b\wedge e^3\wedge e^4
\cr
&&
+h^{-\frac{1}{4}}(-\mathrm{Im} \langle \eta^{1r}, \Gamma_{ab} \Gamma_{11} \eta^{1s}\rangle_D+ \mathrm{Im} \langle \lambda^{1r}, \Gamma_{ab} \Gamma_{11} \lambda^{1s}\rangle_D)\, e^a\wedge e^b\wedge e^3\wedge e^8
\cr
&&
+h^{-\frac{1}{4}}(-\mathrm{Im} \langle \eta^{1r}, \Gamma_{ab} \Gamma_{11} \eta^{2s}\rangle_D+ \mathrm{Im} \langle \lambda^{1r}, \Gamma_{ab} \Gamma_{11} \lambda^{2s}\rangle_D)\, e^a\wedge e^b\wedge e^3\wedge e^9
\cr
&&
+h^{-\frac{1}{4}}(\mathrm{Im} \langle \eta^{1r}, \Gamma_{ab} \Gamma_{11} \eta^{2s}\rangle_D+ \mathrm{Im} \langle \lambda^{1r}, \Gamma_{ab} \Gamma_{11} \lambda^{2s}\rangle_D)\, e^a\wedge e^b\wedge e^4\wedge e^8
\cr
&&
-h^{-\frac{1}{4}}(\mathrm{Im} \langle \eta^{1r}, \Gamma_{ab} \Gamma_{11} \eta^{1s}\rangle_D+ \mathrm{Im} \langle \lambda^{1r}, \Gamma_{ab} \Gamma_{11} \lambda^{1s}\rangle_D)\, e^a\wedge e^b\wedge e^4\wedge e^9
\cr
&&
+h^{-\frac{1}{4}}(\mathrm{Re} \langle \eta^{1r}, \Gamma_{ab} \Gamma_{11} \eta^{2s}\rangle_D+ \mathrm{Re} \langle \lambda^{1r}, \Gamma_{ab} \Gamma_{11} \lambda^{2s}\rangle_D)\, e^a\wedge e^b\wedge e^8\wedge e^9
\cr
&&
-{1\over 4!} \epsilon_{i_1\dots i_4}{}^j \tilde k^{rs}_j\, e^{i_1}\wedge\dots\wedge e^{i_4}~,
\eea
\bea
\tau^{rs}&=& {2h^{-\frac{1}{4}}\over 5!}( \mathrm{Re} \langle \eta^{1r}, \Gamma_{a_1\dots a_5} \eta^{1s}\rangle_D+ \mathrm{Re} \langle \lambda^{1r}, \Gamma_{a_1\dots a_5} \lambda^{1s}\rangle_D) \, e^{a_1}\wedge\dots\wedge e^{a_5}
\cr
&&
-{h^{-\frac{1}{4}}\over3}(\mathrm{Re} \langle \eta^{1r},  \Gamma_{abc} \lambda^{1s}\rangle_D- \mathrm{Re} \langle \lambda^{1r},  \Gamma_{abc}\eta^{1s}\rangle_D)\, e^2\wedge e^3\wedge e^a \wedge e^b\wedge e^c
\cr
&&
+{h^{-\frac{1}{4}}\over3}(-\mathrm{Re} \langle \eta^{1r},  \Gamma_{abc} \lambda^{2s}\rangle_D- \mathrm{Re} \langle \lambda^{1r},  \Gamma_{abc}\eta^{2s}\rangle_D)\, e^2\wedge e^4\wedge e^a\wedge e^b\wedge e^c
\cr
&&
-{h^{-\frac{1}{4}}\over3}(\mathrm{Im} \langle \eta^{1r},  \Gamma_{abc} \lambda^{1s}\rangle_D+ \mathrm{Im} \langle \lambda^{1r},  \Gamma_{abc} \eta^{1s}\rangle_D)\, e^2\wedge e^8\wedge e^a\wedge e^b\wedge e^c
\cr
&&
-{h^{-\frac{1}{4}}\over3}(\mathrm{Im} \langle \eta^{1r},  \Gamma_{abc} \lambda^{2s}\rangle_D+ \mathrm{Im} \langle \lambda^{1r},  \Gamma_{abc} \eta^{2s}\rangle_D)\, e^2\wedge e^9\wedge e^a\wedge e^b\wedge e^c
\cr
&&
+{h^{-\frac{1}{4}}\over3}(-\mathrm{Re} \langle \eta^{1r},  \Gamma_{abc} \eta^{2s}\rangle_D+ \mathrm{Re} \langle \lambda^{1r},  \Gamma_{abc} \lambda^{2s}\rangle_D)\, e^3\wedge e^4\wedge e^a\wedge e^b\wedge e^c
\cr
&&
+{h^{-\frac{1}{4}}\over3}(-\mathrm{Im} \langle \eta^{1r}, \Gamma_{abc} \eta^{1s}\rangle_D+ \mathrm{Im} \langle \lambda^{1r},  \Gamma_{abc} \lambda^{1s}\rangle_D)\, e^3\wedge e^8\wedge e^a\wedge e^b\wedge e^c
\cr
&&
+{h^{-\frac{1}{4}}\over3}(-\mathrm{Im} \langle \eta^{1r},  \Gamma_{abc} \eta^{2s}\rangle_D+ \mathrm{Im} \langle \lambda^{1r},  \Gamma_{abc} \lambda^{2s}\rangle_D)\, e^3\wedge e^9\wedge e^a\wedge e^b\wedge e^c
\cr
&&
+{h^{-\frac{1}{4}}\over3}(\mathrm{Im} \langle \eta^{1r},  \Gamma_{abc} \eta^{2s}\rangle_D+ \mathrm{Im} \langle \lambda^{1r},  \Gamma_{abc} \lambda^{2s}\rangle_D)\, e^4\wedge e^8\wedge e^a\wedge e^b\wedge e^c
\cr
&&
-{h^{-\frac{1}{4}}\over3}(\mathrm{Im} \langle \eta^{1r},  \Gamma_{abc} \eta^{1s}\rangle_D+ \mathrm{Im} \langle \lambda^{1r},  \Gamma_{abc} \lambda^{1s}\rangle_D)\, e^4\wedge e^9\wedge e^a\wedge e^b\wedge e^c
\cr
&&
+{h^{-\frac{1}{4}}\over3}(\mathrm{Re} \langle \eta^{1r}, \Gamma_{abc} \eta^{2s}\rangle_D+ \mathrm{Re} \langle \lambda^{1r},  \Gamma_{abc} \lambda^{2s}\rangle_D)\, e^8\wedge e^9\wedge e^a\wedge e^b\wedge e^c
\cr
&&
+{1\over4!} \epsilon_{i_1\dots i_4}{}^j \omega_{aj} e^a\wedge e^{i_1}\wedge \dots\wedge e^{i_4}~,
\eea
where $\epsilon_{23849}=1$,  $a,b,c=0, 5, 1, 6, 7$, $i,j,k=2,3,4,8,9$  and $(e^a, e^i)$  is a pseudo-orthonormal frame of the metric (\ref{dbrane}) for $p=4$.

\subsubsection{D8-brane}

Using the Killing spinors (\ref{d8sol}), the non-vanishing form bilinears of D8-brane are as follows
\bea
&&\tilde \sigma^{rs}= 2 h^{-\frac{1}{4}} \langle \eta^r, \Gamma_{11} \eta^s\rangle~,~~~k^{rs}=2 h^{-\frac{1}{4}} \langle \eta^r,  \eta^s\rangle\, e^0+2 h^{-\frac{1}{4}} \langle \eta^r, \Gamma_{a'} \eta^s\rangle\, e^{a'}~,~~~
\cr
&&\tilde k^{rs} =-2h^{-\frac{1}{4}} \langle \eta^r, \Gamma_{11} \eta^s\rangle\, e^5~,~~~\omega^{rs}=2 h^{-\frac{1}{4}}\langle \eta^r,  \eta^s\rangle~e^0\wedge e^5+2 h^{-\frac{1}{4}}\langle \eta^r, \Gamma_{a'} \eta^s\rangle~e^{a'}\wedge e^5~,~~~
\cr
&&
\tilde \omega^{rs}=2 h^{-\frac{1}{4}} \langle \eta^r, \Gamma_{a'} \Gamma_{11} \eta^s\rangle~e^0\wedge e^{a'}+h^{-\frac{1}{4}} \langle \eta^r, \Gamma_{a'b'} \Gamma_{11} \eta^s\rangle~e^{a'}\wedge e^{b'}~,
\cr
&&
\pi^{rs}=h^{-\frac{1}{4}} \langle \eta^r, \Gamma_{b'c'} \eta^s\rangle~e^0\wedge e^{b'}\wedge e^{c'}+{h^{-\frac{1}{4}}\over3} \langle \eta^r, \Gamma_{a'b'c'} \eta^s\rangle~e^{a'}\wedge e^{b'}\wedge e^{c'}~,~~~
\cr
&&
\tilde \pi^{rs}=-2h^{-\frac{1}{4}}\langle \eta^r, \Gamma_{a'}\Gamma_{11} \eta^s\rangle~e^{0}\wedge e^{a'}\wedge e^5-h^{-\frac{1}{4}}\langle \eta^r, \Gamma_{ab}\Gamma_{11} \eta^s\rangle~e^a\wedge e^b\wedge e^5~,
\cr
&&
\zeta^{rs}={h^{-\frac{1}{4}}} \langle \eta^r, \Gamma_{b'c'} \eta^s\rangle~e^0\wedge e^{b'}\wedge e^{c'}\wedge e^5+{h^{-\frac{1}{4}}\over3} \langle \eta^r, \Gamma_{a'b'c'} \eta^s\rangle~e^{a'}\wedge e^{b'}\wedge e^{c'}\wedge e^5~,~~~
\cr
&&
\tilde \zeta^{rs}={h^{-\frac{1}{4}}\over3} \langle \eta^r, \Gamma_{a'b'c'} \Gamma_{11} \eta^s\rangle~e^{0}\wedge e^{a'}\wedge e^{b'} \wedge e^{c'}+{h^{-\frac{1}{4}}\over12} \langle \eta^r, \Gamma_{a'_1\dots a'_4} \Gamma_{11} \eta^s\rangle~e^{a'_1}\wedge \dots \wedge e^{a'_4}~,
\cr
&&
\tau^{rs}={h^{-\frac{1}{4}}\over 12} \langle \eta^r, \Gamma_{a'_1\dots a'_4} \eta^s\rangle~e^0\wedge e^{a'_1}\wedge \dots\wedge e^{a'_4}+{2h^{-\frac{1}{4}}\over 5!} \langle \eta^r, \Gamma_{a'_1\dots a'_5} \eta^s\rangle~e^{a'_1}\wedge \dots\wedge e^{a'_5}~,
\eea
where $a',b',c'=1,6,2,7,3,8,4,9$ and $(e^a, e^5)$ is a pseudo-orthonormal frame of the D8-brane metric  (\ref{dbrane}) for $p=8$.

\subsection{Form bilinears of IIB D-branes}

As for the common sector IIB branes, all the bilinears below are manifestly real. In particular, $k^{(2)}, \pi^{(2)}$ and $\tau^{(2)}$ have been replaced with $ik^{(2)}, i\pi^{(2)}$ and $i\tau^{(2)}$, respectively.

\subsubsection{D-string}

Using the Killing spinors (\ref{d1ks}), one can easily compute
the form bilinears of the D-string background to find
\bea
k^{rs}=2 h^{-{1\over4}}(\langle \eta^r, \eta^s\rangle+\langle \lambda^r, \lambda^s\rangle) e^0+ 2  h^{-{1\over4}} (-\langle \eta^r, \eta^s\rangle+\langle \lambda^r, \lambda^s\rangle) e^5~,
\eea

\bea
k^{(3)}{}^{rs}= 2 h^{-{1\over4}} (\langle \eta^r, \Gamma_i\lambda^s\rangle+\langle \lambda^r, \Gamma_i\eta^s\rangle) e^i~,
\eea
\bea
k^{(1)}{}^{rs}=2 h^{-{1\over4}}(\langle \eta^r, \eta^s\rangle-\langle \lambda^r, \lambda^s\rangle) e^0- 2  h^{-{1\over4}} (\langle \eta^r, \eta^s\rangle+\langle \lambda^r, \lambda^s\rangle) e^5~,
\eea

\bea
k^{(2)}{}^{rs}=2 h^{-{1\over4}} (-\langle \eta^r, \Gamma_i\lambda^s\rangle+\langle \lambda^r, \Gamma_i\eta^s\rangle) e^i~,
\eea

\bea
&&\pi^{rs}=h^{-{1\over4}} (\langle \eta^r, \Gamma_{ij}\eta^s\rangle+\langle \lambda^r, \Gamma_{ij}\lambda^s\rangle) e^0\wedge e^i\wedge e^j
\cr
&&
\qquad
+h^{-{1\over4}} (-\langle \eta^r, \Gamma_{ij}\eta^s\rangle+\langle \lambda^r, \Gamma_{ij}\lambda^s\rangle) e^5\wedge e^i\wedge e^j~,
\eea

\bea
&&\pi^{(3)}{}^{rs}=2h^{-{1\over4}} (-\langle \eta^r, \Gamma_{i}\lambda^s\rangle+\langle \lambda^r, \Gamma_{i}\eta^s\rangle) e^0\wedge e^5\wedge e^i
\cr
&&
\qquad
+{1\over3}h^{-{1\over4}} (\langle \eta^r, \Gamma_{ijk}\lambda^s\rangle+\langle \lambda^r, \Gamma_{ijk}\eta^s\rangle) e^i\wedge e^j\wedge e^k~,
\eea

\bea
&&\pi^{(1)}{}^{rs}=h^{-{1\over4}} (\langle \eta^r, \Gamma_{ij}\eta^s\rangle-\langle \lambda^r, \Gamma_{ij}\lambda^s\rangle) e^0\wedge e^i\wedge e^j
\cr
&&
\qquad
-h^{-{1\over4}} (\langle \eta^r, \Gamma_{ij}\eta^s\rangle+\langle \lambda^r, \Gamma_{ij}\lambda^s\rangle) e^5\wedge e^i\wedge e^j~,
\eea

\bea
&&\pi^{(2)}{}^{rs}=2h^{-{1\over4}} (\langle \eta^r, \Gamma_{i}\lambda^s\rangle+\langle \lambda^r, \Gamma_{i}\eta^s\rangle) e^0\wedge e^5\wedge e^i
\cr
&&
\qquad
+{1\over3}h^{-{1\over4}} (-\langle \eta^r, \Gamma_{ijk}\lambda^s\rangle+\langle \lambda^r, \Gamma_{ijk}\eta^s\rangle) e^i\wedge e^j\wedge e^k~,
\eea

\bea
&&\tau^{rs}={2\over 4!}h^{-{1\over4}} (\langle \eta^r, \Gamma_{i_1\dots i_4}\eta^s\rangle+\langle \lambda^r, \Gamma_{i_1\dots i_4}\lambda^s\rangle) e^0\wedge e^{i_1}\wedge \dots \wedge e^{i_4}
\cr
&&
\qquad
+{2\over 4!} h^{-{1\over4}} (-\langle \eta^r, \Gamma_{i_1\dots i_4}\eta^s\rangle+\langle \lambda^r, \Gamma_{i_1\dots i_4}\lambda^s\rangle) e^5\wedge e^{i_1}\wedge \dots \wedge e^{i_4}~,
\eea

\bea
&&\tau^{(3)}{}^{rs}={1\over3}h^{-{1\over4}} (-\langle \eta^r, \Gamma_{ijk}\lambda^s\rangle+\langle \lambda^r, \Gamma_{ijk}\eta^s\rangle) e^0\wedge e^5\wedge e^i\wedge e^j\wedge e^k
\cr
&&
\qquad
+{2\over5!}h^{-{1\over4}} (\langle \eta^r, \Gamma_{i_1\dots i_5}\lambda^s\rangle+\langle \lambda^r, \Gamma_{i_1\dots i_5}\eta^s\rangle) e^{i_1}\wedge \dots\wedge e^{i_5}~,
\eea

\bea
&&\tau^{(1)}{}^{rs}={2\over4!}h^{-{1\over4}} (\langle \eta^r, \Gamma_{i_1\dots i_4}\eta^s\rangle-\langle \lambda^r, \Gamma_{i_1\dots i_4}\lambda^s\rangle) e^0\wedge e^{i_1}\wedge \dots \wedge e^{i_4}
\cr
&&
\qquad
-{2\over4!}h^{-{1\over4}} (\langle \eta^r, \Gamma_{i_1\dots i_4}\eta^s\rangle+\langle \lambda^r, \Gamma_{i_1\dots i_4}\lambda^s\rangle) e^5\wedge e^{i_1}\wedge \dots \wedge e^{i_4}~,
\eea

\bea
&&\tau^{(2)}{}^{rs}={1\over3}h^{-{1\over4}} (\langle \eta^r, \Gamma_{ijk}\lambda^s\rangle+\langle \lambda^r, \Gamma_{ijk}\eta^s\rangle) e^0\wedge e^5\wedge e^i\wedge e^j\wedge e^k
\cr
&&
\qquad
+{2\over5!}h^{-{1\over4}} (-\langle \eta^r, \Gamma_{i_1\dots i_5}\lambda^s\rangle+\langle \lambda^r, \Gamma_{i_1\dots i_5}\eta^s\rangle) e^{i_1}\wedge \dots\wedge e^{i_5}~,
\eea
where $i,j,k=1,6,2,7,3,8,4,9$ and $(e^0, e^5, e^i)$ is a pseudo-orthonormal frame of the D-string metric (\ref{dbrane}) for $p=1$.

\subsubsection{D5-brane}

Using (\ref{d5ks}), one can find that the form bilinears of the D5-brane background are
\begin{equation}
k^{rs} = 4h^{-1/4} \left( \rep\left\langle \eta^{1r}, \Gamma_a \eta^{1s} \right\rangle_D + \rep\left\langle \eta^{2r}, \Gamma_a \eta^{2s} \right\rangle_D \right)e^a~,
\end{equation}
\bea
\pi^{rs}& =& 4h^{-1/4} \Big(- \rep \left\langle \eta^{1r}, \Gamma_a \lambda^{1s} \right\rangle_D e^a\wedge(e^3\wedge e^4 - e^8\wedge e^9)
\cr \qquad &&
+\rep \left\langle \eta^{2r}, \Gamma_a \lambda^{2s} \right\rangle_D e^a\wedge (e^3\wedge e^4 + e^8\wedge e^9)
\cr \qquad &&
-\imp \left\langle \eta^{1r}, \Gamma_a \eta^{1s} \right\rangle_D e^a\wedge (e^3\wedge e^8 + e^4\wedge e^9)
\cr \qquad &&
+\imp \left\langle \eta^{2r}, \Gamma_a \eta^{2s} \right\rangle_D e^a\wedge (e^3\wedge e^8 - e^4\wedge e^9)
\cr \qquad &&
-\imp \left\langle \eta^{1r}, \Gamma_a \lambda^{1s} \right\rangle_D e^a\wedge (e^3\wedge e^9 - e^4\wedge e^8)
\cr \qquad &&
+\imp \left\langle \eta^{2r}, \Gamma_a \lambda^{2s} \right\rangle_D e^a\wedge (e^3\wedge e^9 + e^4\wedge e^8)\Big)
\cr \qquad &&
+\frac{2}{3}\,h^{-1/4}\left( \rep\left\langle \eta^{1r}, \Gamma_{abc} \eta^{1s} \right\rangle_D + \rep\left\langle \eta^{2r}, \Gamma_{abc} \eta^{2s} \right\rangle_D \right)e^a\wedge e^b \wedge e^c~,
\eea

\begin{align}
\tau^{rs} = &-4h^{-1/4}\left( -\rep\left\langle \eta^{1r}, \Gamma_a \eta^{1s} \right\rangle_D + \rep\left\langle \eta^{2r}, \Gamma_a \eta^{2s} \right\rangle_D \right)e^a\wedge e^3 \wedge e^4 \wedge e^8 \wedge e^9 \nn
&+\frac{2}{3}\,h^{-1/4}\Big(-\rep \left\langle \eta^{1r}, \Gamma_{abc} \lambda^{1s} \right\rangle_D (e^3\wedge e^4 - e^8\wedge e^9)  \nn
&+\rep \left\langle \eta^{2r}, \Gamma_{abc} \lambda^{2s} \right\rangle_D (e^3\wedge e^4 + e^8\wedge e^9) - \imp \left\langle \eta^{1r}, \Gamma_{abc} \eta^{1s} \right\rangle_D (e^3\wedge e^8 + e^4\wedge e^9)  \nn
&+\imp \left\langle \eta^{2r}, \Gamma_{abc} \eta^{2s} \right\rangle_D (e^3\wedge e^8 - e^4\wedge e^9)  -\imp \left\langle \eta^{1r}, \Gamma_{abc} \lambda^{1s} \right\rangle_D (e^3\wedge e^9 - e^4\wedge e^8)  \nn
&+\imp \left\langle \eta^{2r}, \Gamma_{abc} \lambda^{2s} \right\rangle_D (e^3\wedge e^9 + e^4\wedge e^8) \Big) \wedge e^a \wedge e^b \wedge e^c \nn
&+\frac{4}{5!}\,h^{-1/4} \left( \rep \left\langle \eta^{1r}, \Gamma_{a_1 \dots a_5} \eta^{1s} \right\rangle_D + \rep \left\langle \eta^{2r}, \Gamma_{a_1 \dots a_5} \eta^{2s} \right\rangle_D \right)e^{a_1}\wedge \dots \wedge e^{a_5}~,
\end{align}
where $a,b,c=0,5,1,6,2,7$ and $(e^a, e^3, e^4, e^8, e^9)$ is a pseudo-orthonormal frame of the D5-brane metric (\ref{dbrane}) for $p=5$.
The formula for the form bilinears $k^{(1)}, \pi^{(1)}$ and $\tau^{(1)}$ can be obtained from that of  $k$, $\pi$ and $\tau$ by adding a minus sign in the $\eta^2-\eta^2$ and $\eta^2-\lambda^2$ terms.

The rest of the form  bilinears are
\begin{align}
k^{(3)rs} = 4h^{-1/4}&(\left( \rep\left\langle \eta^{1r}, \eta^{2s} \right\rangle_D - \rep\left\langle \eta^{2r}, \eta^{1s} \right\rangle_D \right)e^3 \nn
&+\left( \rep\left\langle \eta^{1r}, \lambda^{2s} \right\rangle_D + \rep\left\langle \eta^{2r}, \lambda^{1s} \right\rangle_D \right)e^4 \nn
&+\left( \imp\left\langle \eta^{1r}, \eta^{2s} \right\rangle_D + \imp\left\langle \eta^{2r}, \eta^{1s} \right\rangle_D \right)e^8 \nn
&+\left( \imp\left\langle \eta^{1r}, \lambda^{2s} \right\rangle_D + \imp\left\langle \eta^{2r}, \lambda^{1s} \right\rangle_D \right))e^9 ~,
\end{align}

\begin{align}
\pi^{(3)rs} = 2h^{-1/4}&(\left( \rep \left\langle \eta^{1r}, \Gamma_{ab} \eta^{2s} \right\rangle_D -\rep \left\langle \eta^{2r}, \Gamma_{ab} \eta^{1s} \right\rangle_D \right)e^3\wedge e^a \wedge e^b \nn
&+\left( \rep \left\langle \eta^{1r}, \Gamma_{ab} \lambda^{2s} \right\rangle_D +\rep \left\langle \eta^{2r}, \Gamma_{ab} \lambda^{1s} \right\rangle_D \right)e^4\wedge e^a \wedge e^b \nn
&+\left( \imp \left\langle \eta^{1r}, \Gamma_{ab} \eta^{2s} \right\rangle_D +\imp \left\langle \eta^{2r}, \Gamma_{ab} \eta^{1s} \right\rangle_D \right)e^8\wedge e^a \wedge e^b \nn
&+\left( \imp \left\langle \eta^{1r}, \Gamma_{ab} \lambda^{2s} \right\rangle_D +\imp \left\langle \eta^{2r}, \Gamma_{ab} \lambda^{1s} \right\rangle_D \right)e^9\wedge e^a \wedge e^b \nn
&+2\left( -\rep \left\langle \eta^{1r}, \eta^{2s} \right\rangle_D - \rep \left\langle \eta^{2r}, \eta^{1s} \right\rangle_D \right)e^4\wedge e^8 \wedge e^9 \nn
&-2\left( -\rep \left\langle \eta^{1r}, \lambda^{2s} \right\rangle_D + \rep \left\langle \eta^{2r}, \lambda^{1s} \right\rangle_D \right)e^3\wedge e^8 \wedge e^9 \nn
&-2\left( \imp \left\langle \eta^{1r}, \eta^{2s} \right\rangle_D - \imp \left\langle \eta^{2r}, \eta^{1s} \right\rangle_D \right)e^3\wedge e^4 \wedge e^9 \nn
&+2\left( \imp \left\langle \eta^{1r}, \lambda^{2s} \right\rangle_D - \imp \left\langle \eta^{2r}, \lambda^{1s} \right\rangle_D \right))e^3\wedge e^4 \wedge e^8 ~, \nn
\end{align}

\bea
&&\tau^{(3)rs} = -\frac{1}{6}\,h^{-1/4} \Big( \left( \rep \left\langle \eta^{1r}, \Gamma_{a_1\dots a_4} \eta^{2s} \right\rangle_D -\rep \left\langle \eta^{2r}, \Gamma_{a_1\dots a_4} \eta^{1s} \right\rangle_D \right)e^3  
\cr
&&\qquad
+ \left( \rep \left\langle \eta^{1r}, \Gamma_{a_1\dots a_4} \lambda^{2s} \right\rangle_D +\rep \left\langle \eta^{2r}, \Gamma_{a_1\dots a_4} \lambda^{1s} \right\rangle_D \right)e^4 
\cr
&&\qquad
+\left( \imp \left\langle \eta^{1r}, \Gamma_{a_1\dots a_4} \eta^{2s} \right\rangle_D +\imp \left\langle \eta^{2r}, \Gamma_{a_1\dots a_4} \eta^{1s} \right\rangle_D \right)e^8 
 \cr
&&\qquad
+ \left( \imp \left\langle \eta^{1r}, \Gamma_{a_1\dots a_4} \lambda^{2s} \right\rangle_D +\imp \left\langle \eta^{2r}, \Gamma_{a_1\dots a_4} \lambda^{1s} \right\rangle_D \right)e^9\Big)\wedge e^{a_1}\wedge \dots \wedge e^{a_4} 
\cr
&&\qquad
+2h^{-1/4}\Big( \left( -\rep \left\langle \eta^{1r}, \Gamma_{ab} \eta^{2s} \right\rangle_D - \rep \left\langle \eta^{2r}, \Gamma_{ab} \eta^{1s} \right\rangle_D \right) e^4 \wedge e^8 \wedge e^9 
\cr
&&\qquad
-\left( -\rep \left\langle \eta^{1r}, \Gamma_{ab} \lambda^{2s} \right\rangle_D + \rep \left\langle \eta^{2r}, \Gamma_{ab} \lambda^{1s} \right\rangle_D \right) e^3 \wedge e^8 \wedge e^9
\cr
&&\qquad
-\left( \imp \left\langle \eta^{1r}, \Gamma_{ab} \eta^{2s} \right\rangle_D - \imp \left\langle \eta^{2r}, \Gamma_{ab} \eta^{1s} \right\rangle_D \right) e^3 \wedge e^4 \wedge e^9
\cr
&&\qquad
+\left( \imp \left\langle \eta^{1r}, \Gamma_{ab} \lambda^{2s} \right\rangle_D - \imp \left\langle \eta^{2r}, \Gamma_{ab} \lambda^{1s} \right\rangle_D \right) e^3 \wedge e^4 \wedge e^8\Big)\wedge e^a\wedge e^b~.
\eea
The formula for the form bilinears $k^{(2)}, \pi^{(2)}$ and $\tau^{(2)}$ can be obtained from that of $k^{(3)}, \pi^{(3)}$ and $\tau^{(3)}$  by changing the  sign in front of the $\eta^1-\eta^2$ and $\eta^1-\lambda^2$ terms.

\subsubsection{D3-brane}

Using  for the Killing spinors (\ref{d3ks}), one finds  that the form bilinears of the D3-brane solution are as follows
\begin{align}
k^{rs} &= 4 h^{-{1\over4}}\Big(\rep \left\langle \eta^{1r}, \eta^{1s} \right\rangle(e^0-e^5) + \rep\left\langle \lambda^{1r}, \lambda^{1s} \right\rangle(e^0+e^5) \nn
&-\left( \rep\left\langle \eta^{1r}, \lambda^{1s} \right\rangle + \rep\left\langle \lambda^{1r}, \eta^{1s} \right\rangle \right)e^4 \nn
&-\left( \imp\left\langle \eta^{1r}, \lambda^{1s} \right\rangle - \imp\left\langle \lambda^{1r}, \eta^{1s} \right\rangle \right)e^9\Big)~,
\end{align}
\begin{align}
k^{(1)rs} = 4  h^{-{1\over4}} \left( \imp\left\langle \eta^{1r}, \Gamma_i \lambda^{2s} \right\rangle + \imp\left\langle \lambda^{1r}, \Gamma_i \eta^{2s} \right\rangle \right)e^i ~,
\end{align}

\begin{align}
\pi^{rs} &= h^{-{1\over4}}\Big(2\rep \left\langle \eta^{1r}, \Gamma_{ij} \eta^{1s} \right\rangle (e^0 - e^5) \wedge e^i \wedge e^j \nn
&+2\rep \left\langle \lambda^{1r}, \Gamma_{ij} \lambda^{1s} \right\rangle(e^0 + e^5) \wedge e^i \wedge e^j \nn
&-2\left( \rep \left\langle \eta^{1r}, \Gamma_{ij} \lambda^{1s} \right\rangle + \rep\left\langle \lambda^{1r}, \Gamma_{ij} \eta^{1s} \right\rangle \right)e^4 \wedge e^i \wedge e^j \nn
&-2\left( \imp \left\langle \eta^{1r}, \Gamma_{ij} \lambda^{1s} \right\rangle - \imp\left\langle \lambda^{1r}, \Gamma_{ij} \eta^{1s} \right\rangle \right)e^9 \wedge e^i \wedge e^j \nn
&-4\imp \left\langle \eta^{1r}, \eta^{1s} \right\rangle(e^0-e^5)\wedge e^4 \wedge e^9 \nn
&+4\imp \left\langle \lambda^{1r}, \lambda^{1s} \right\rangle(e^0+e^5)\wedge e^4 \wedge e^9 \nn
&+4 \left( \rep\left\langle \eta^{1r}, \lambda^{1s} \right\rangle - \rep\left\langle \lambda^{1r}, \eta^{1s} \right\rangle \right)e^0 \wedge e^5 \wedge e^4 \nn
&+4\left( \imp\left\langle \eta^{1r}, \lambda^{1s}\right\rangle + \imp\left\langle \lambda^{1r}, \eta^{1s} \right\rangle \right)e^0 \wedge e^5 \wedge e^9\Big)~,
\end{align}
\begin{align}
\pi^{(1)rs} &= h^{-{1\over4}}\Big(-4\imp \left\langle \eta^{1r}, \Gamma_i \eta^{2s} \right\rangle (e^0-e^5)\wedge e^4 \wedge e^i \nn
&-4\imp\left\langle \lambda^{1r}, \Gamma_i \lambda^{2s} \right\rangle(e^0+e^5)\wedge e^4 \wedge e^i \nn
&+4\rep\left\langle \eta^{1r}, \Gamma_i \eta^{2s} \right\rangle(e^0-e^5)\wedge e^9 \wedge e^i \nn
&-4\rep\left\langle \lambda^{1r}, \Gamma_i \lambda^{2s} \right\rangle(e^0+e^5) \wedge e^9 \wedge e^i \nn
&-4\left(\imp\left\langle \eta^{1r}, \Gamma_i \lambda^{2s} \right\rangle - \imp\left\langle \lambda^{1r}, \Gamma_i \eta^{2s} \right\rangle \right)e^0 \wedge e^5 \wedge e^i \nn
&+4\left(\rep\left\langle \eta^{1r}, \Gamma_i \lambda^{2s} \right\rangle - \rep\left\langle \lambda^{1r}, \Gamma_i \eta^{2s} \right\rangle \right)e^4 \wedge e^9 \wedge e^i \nn
&+\frac{2}{3}\left(\imp\left\langle \eta^{1r}, \Gamma_{ijk} \lambda^{2s} \right\rangle + \imp\left\langle \lambda^{1r}, \Gamma_{ijk}\eta^{2s} \right\rangle \right)e^i \wedge e^j \wedge e^k\Big)~,
\end{align}
\begin{align}
\tau^{rs} &= \frac{h^{-{1\over4}}}{6} \Big( \rep \left\langle \eta^{1r}, \Gamma_{i_1\dots i_4} \eta^{1s} \right\rangle (e^0 - e^5) +\rep \left\langle \lambda^{1r}, \Gamma_{i_1\dots i_4} \lambda^{1s} \right\rangle(e^0 + e^5) 
 \nn
&-\left( \rep \left\langle \eta^{1r}, \Gamma_{i_1\dots i_4} \lambda^{1s} \right\rangle + \rep\left\langle \lambda^{1r}, \Gamma_{i_1\dots i_4} \eta^{1s} \right\rangle \right)e^4  \nn
&-\left( \imp \left\langle \eta^{1r}, \Gamma_{i_1\dots i_4} \lambda^{1s} \right\rangle - \imp\left\langle \lambda^{1r}, \Gamma_{i_1\dots i_4} \eta^{1s} \right\rangle \right)e^9 \Big)\wedge e^{i_1} \wedge\dots \wedge e^{i_4} \nn
&-2 h^{-{1\over4}} \imp \left\langle \eta^{1r},\Gamma_{ij} \eta^{1s} \right\rangle(e^0-e^5)\wedge e^4 \wedge e^9\wedge e^i \wedge e^j \nn
&+2 h^{-{1\over4}}\imp \left\langle \lambda^{1r}, \Gamma_{ij}\lambda^{1s} \right\rangle(e^0+e^5)\wedge e^4 \wedge e^9\wedge e^i \wedge e^j \nn
&+2 h^{-{1\over4}} \left( \rep\left\langle \eta^{1r}, \Gamma_{ij}\lambda^{1s} \right\rangle - \rep\left\langle \lambda^{1r},\Gamma_{ij} \eta^{1s} \right\rangle \right)e^0 \wedge e^5 \wedge e^4\wedge e^i \wedge e^j \nn
&+2 h^{-{1\over4}}\left( \imp\left\langle \eta^{1r}, \Gamma_{ij}\lambda^{1s}\right\rangle + \imp\left\langle \lambda^{1r}, \Gamma_{ij}\eta^{1s} \right\rangle \right)e^0 \wedge e^5 \wedge e^9\wedge e^i \wedge e^j~,
\end{align}
\begin{align}
\tau^{(1)rs} &= -4 h^{-{1\over4}} \left( \rep\left\langle \eta^{1r}, \Gamma_i \lambda^{2s} \right\rangle + \rep\left\langle \lambda^{1r}, \Gamma_i \eta^{2s} \right\rangle \right)e^0\wedge e^5 \wedge e^4 \wedge e^9 \wedge e^i \nn
&
\frac{2 h^{-{1\over4}}}{3}\Big(-\imp \left\langle \eta^{1r}, \Gamma_{ijk} \eta^{2s} \right\rangle (e^0-e^5)\wedge e^4  -\imp\left\langle \lambda^{1r}, \Gamma_{ijk} \lambda^{2s} \right\rangle(e^0+e^5)\wedge e^4  \nn
&+\rep\left\langle \eta^{1r}, \Gamma_{ijk} \eta^{2s} \right\rangle(e^0-e^5)\wedge e^9  -\rep\left\langle \lambda^{1r}, \Gamma_{ijk} \lambda^{2s} \right\rangle(e^0+e^5) \wedge e^9  \nn
&-\left(\imp\left\langle \eta^{1r}, \Gamma_{ijk} \lambda^{2s} \right\rangle - \imp\left\langle \lambda^{1r}, \Gamma_{ijk} \eta^{2s} \right\rangle \right)e^0 \wedge e^5  \nn
&+\left(\rep\left\langle \eta^{1r}, \Gamma_{ijk} \lambda^{2s} \right\rangle - \rep\left\langle \lambda^{1r}, \Gamma_{ijk} \eta^{2s} \right\rangle \right)e^4 \wedge e^9 \Big)\wedge e^i\wedge e^j \wedge e^k \nn
&+\frac{4h^{-{1\over4}}}{5!}\left(\imp\left\langle \eta^{1r}, \Gamma_{i_1 \dots i_5} \lambda^{2s} \right\rangle + \imp\left\langle \lambda^{1r}, \Gamma_{i_1\dots i_5}\eta^{2s} \right\rangle \right)e^{i_1} \wedge \dots \wedge e^{i_5}~,
\end{align}
where $i,j,k=1,6,2,7,3,8$ and $(e^0, e^5, e^4, e^9)$ is a pseudo-orthonormal frame of the D3-brane metric (\ref{dbrane}) for $p=3$.
The $k^{(2)}, \pi^{(2)}$, and $\tau^{(2)}$  form bilinears can be obtained from $k, \pi$, and $\tau$, and the  $k^{(3)}, \pi^{(3)}$, and $\tau^{(3)}$ form bilinears can be obtained from $k^{(1)}, \pi^{(1)}$, and $\tau^{(1)}$ after replacing $\rep$ and $\imp$ with $\imp$ and $-\rep$, respectively.

\subsubsection{D7-brane}

Using the Killing spinors (\ref{d7ks}), one can compute the form bilinears of the D7-brane to find

\begin{equation}
k^{rs} = 4 h^{-{1\over4}}\rep \left\langle \eta^r, \Gamma_a \eta^s \right\rangle_D e^a~,
\end{equation}
\begin{equation}
\pi^{rs} = -4 h^{-{1\over4}}\imp \left\langle \eta^r, \Gamma_a \eta^s \right\rangle_D e^a \wedge e^4 \wedge e^9 + \frac{2}{3} h^{-{1\over4}}\rep \left\langle \eta^r, \Gamma_{abc} \eta^s \right\rangle_D e^a \wedge e^b \wedge e^c ~,
\end{equation}
\bea
&&\tau^{rs} = -\frac{2}{3}h^{-{1\over4}}\imp \left\langle \eta^r, \Gamma_{abc} \eta^s \right\rangle_D e^a \wedge e^b \wedge e^c \wedge e^4 \wedge e^9
 \cr
 &&\qquad\qquad + \frac{4}{5!} h^{-{1\over4}}\rep \left\langle \eta^r, \Gamma_{a_1 \dots a_5} \eta^s \right\rangle_D e^{a_1} \wedge \dots \wedge e^{a_5} ~,
\eea
where $a,b,c=0,5,1,6,2,7,3,8$ and $(e^a, e^4, e^9)$ is a pseudo-orthonormal frame of the metric (\ref{dbrane}) for $p=7$.
The form bilinears $k^{(2)}, \pi^{(2)}$ and $\tau^{(2)}$ can be obtained from $k$, $\pi$ and $\tau$ after replacing $\rep$ and $\imp$ with $\imp$ and $-\rep$, respectively,  in all the above expressions.

The rest bilinears are given by
\begin{equation}
k^{(3)rs} = 4h^{-{1\over4}}\rep\left\langle \eta^r, \lambda^s \right\rangle_De^4 + 4 h^{-{1\over4}}\imp \left\langle \eta^r, \lambda^s \right\rangle_D e^9~,
\end{equation}
\begin{equation}
\pi^{(3)rs} = 2h^{-{1\over4}}\rep \left\langle \eta^r, \Gamma_{ab} \lambda^s \right\rangle_D e^a \wedge e^b \wedge e^4 +2h^{-{1\over4}} \imp \left\langle \eta^r, \Gamma_{ab} \lambda^s \right\rangle_D e^a \wedge e^b \wedge e^9~,
\end{equation}
\bea
&&\tau^{(3)rs} = \frac{1}{6} h^{-{1\over4}}\rep \left\langle \eta^r, \Gamma_{a_1 \dots a_4} \lambda^s \right\rangle_D e^{a_1} \wedge \dots \wedge e^{a_4} \wedge e^4 \cr
&& \qquad \qquad
+\frac{1}{6} h^{-{1\over4}}\imp \left\langle \eta^r, \Gamma_{a_1 \dots a_4} \lambda^s \right\rangle_D e^{a_1} \wedge \dots \wedge e^{a_4} \wedge e^9~.
\eea
Again the  bilinears $k^{(1)}, \pi^{(1)}$ and $\tau^{(1)}$ can be derived from $k^{(3)}, \pi^{(3)}$ and $\tau^{(3)}$  after replacing $\rep$ and $\imp$ with $\imp$ and $-\rep$, respectively, in all three  expressions above.

\end{appendices}


\begin{thebibliography}{99}









  \bibitem{carter-b}
  B.~Carter,
  ``Global structure of the Kerr family of gravitational fields,''
  Phys.\ Rev.\  {\bf 174} (1968) 1559.


  \bibitem{penrose}
R. Penrose, Ann. N.Y. Acad. Sci. 224 (1973) 125.


\bibitem{floyd}
  R. Floyd, The dynamics of Kerr fields,. Ph. D. Thesis, London (1973).



  \bibitem{chandrasekhar}
  S.~Chandrasekhar,
  ``The Solution Of Dirac's Equation In Kerr Geometry,''
  Proc.\ Roy.\ Soc.\ Lond.\  A {\bf 349} (1976) 571.

   \bibitem{carter-a}
  B.~Carter,
  ``Killing Tensor Quantum Numbers And Conserved Currents In Curved Space,''
  Phys.\ Rev.\  D {\bf 16} (1977) 3395.

  \bibitem{carter-c}
  B.~Carter and R.~G.~Mclenaghan,
  ``Generalized Total Angular Momentum Operator For The Dirac Equation In
  Curved Space-Time,''
  Phys.\ Rev.\  D {\bf 19} (1979) 1093.

  \bibitem{page}
  P.~Krtous, D.~Kubiznak, D.~N.~Page and V.~P.~Frolov,
  ``Killing-Yano tensors, rank-2 Killing tensors, and conserved quantities in
  higher dimensions,''
  JHEP {\bf 0702} (2007) 004
  [arXiv:hep-th/0612029].

   \bibitem{sfetsos}
  F.~De Jonghe, K.~Peeters and K.~Sfetsos,
  ``Killing-Yano supersymmetry in string theory,''
  Class.\ Quant.\ Grav.\  {\bf 14} (1997) 35
  [arXiv:hep-th/9607203].

\bibitem{lun}
  Y.~Chervonyi and O.~Lunin,
  ``Killing(-Yano) Tensors in String Theory,''
  JHEP {\bf 1509} (2015) 182
  doi:10.1007/JHEP09(2015)182
  [arXiv:1505.06154 [hep-th]].





  \bibitem{revky}
  M.~Cariglia,
  ``Hidden Symmetries of Dynamics in Classical and Quantum Physics,''
  Rev.\ Mod.\ Phys.\  {\bf 86} (2014) 1283
  doi:10.1103/RevModPhys.86.1283
  [arXiv:1411.1262 [math-ph]].


\bibitem{frolov}
V.~Frolov, P.~Krtous and D.~Kubiznak,
``Black holes, hidden symmetries, and complete integrability,''
Living Rev. Rel. \textbf{20} (2017) no.1, 6
doi:10.1007/s41114-017-0009-9
[arXiv:1705.05482 [gr-qc]].





  \bibitem{gibbons}
  G.~W.~Gibbons, R.~H.~Rietdijk and J.~W.~van Holten,
  ``SUSY in the sky,''
  Nucl.\ Phys.\  B {\bf 404} (1993) 42
  [arXiv:hep-th/9303112].


\bibitem{bvh}
L.~Brink, P.~Di Vecchia and P.~S.~Howe,
``A Lagrangian Formulation of the Classical and Quantum Dynamics of Spinning Particles,''
Nucl. Phys. B \textbf{118} (1977), 76-94
doi:10.1016/0550-3213(77)90364-9



\bibitem{gptcfh}
G.~Papadopoulos,
``Twisted form hierarchies, Killing-Yano equations and supersymmetric backgrounds,''
JHEP \textbf{07} (2020), 025
doi:10.1007/JHEP07(2020)025
[arXiv:2001.07423 [hep-th]].


\bibitem{jggp}
J.~Gutowski and G.~Papadopoulos,
``Eigenvalue estimates for multi-form modified Dirac operators,''
J. Geom. Phys. \textbf{160} (2021), 103954
doi:10.1016/j.geomphys.2020.103954
[arXiv:1911.02281 [math.DG]].

\bibitem{gpeb}
G.~Papadopoulos and E.~P\'erez-Bola\~nos,
``Symmetries, spinning particles and the TCFH of $D=4,5$ minimal supergravities,''
Phys. Lett. B \textbf{819} (2021), 136441
doi:10.1016/j.physletb.2021.136441
[arXiv:2101.10709 [hep-th]].

\bibitem{hulltown}
C.~M.~Hull and P.~K.~Townsend,
``Unity of superstring dualities,''
Nucl. Phys. B \textbf{438} (1995), 109-137
doi:10.1016/0550-3213(94)00559-W
[arXiv:hep-th/9410167 [hep-th]].


\bibitem{town}
P.~K.~Townsend,
``D-branes from M-branes,''
Phys. Lett. B \textbf{373} (1996), 68-75
doi:10.1016/0370-2693(96)00104-9
[arXiv:hep-th/9512062 [hep-th]].


\bibitem{funstring}
A.~Dabholkar, G.~W.~Gibbons, J.~A.~Harvey and F.~Ruiz Ruiz,
``Superstrings and Solitons,''
Nucl. Phys. B \textbf{340} (1990), 33-55
doi:10.1016/0550-3213(90)90157-9


\bibitem{ns5}
C.~G.~Callan, Jr., J.~A.~Harvey and A.~Strominger,
``Worldbrane actions for string solitons,''
Nucl. Phys. B \textbf{367} (1991), 60-82
doi:10.1016/0550-3213(91)90041-U


\bibitem{callan}
C.~G.~Callan, Jr., J.~A.~Harvey and A.~Strominger,
``Supersymmetric string solitons,''
[arXiv:hep-th/9112030 [hep-th]].


\bibitem{d5}
M.~J.~Duff and J.~X.~Lu,
``Elementary five-brane solutions of D = 10 supergravity,''
Nucl. Phys. B \textbf{354} (1991), 141-153
doi:10.1016/0550-3213(91)90180-6

\bibitem{d5hs}
G.~T.~Horowitz and A.~Strominger,
``Black strings and P-branes,''
Nucl. Phys. B \textbf{360} (1991), 197-209
doi:10.1016/0550-3213(91)90440-9



\bibitem{d3}
M.~J.~Duff and J.~X.~Lu,
``The Selfdual type IIB superthreebrane,''
Phys. Lett. B \textbf{273} (1991), 409-414
doi:10.1016/0370-2693(91)90290-7

\bibitem{d7}
G.~W.~Gibbons, M.~B.~Green and M.~J.~Perry,
``Instantons and seven-branes in type IIB superstring theory,''
Phys. Lett. B \textbf{370} (1996), 37-44
doi:10.1016/0370-2693(95)01565-5
[arXiv:hep-th/9511080 [hep-th]].



\bibitem{d8a}
J.~Polchinski and E.~Witten,
``Evidence for heterotic - type I string duality,''
Nucl. Phys. B \textbf{460} (1996), 525-540
doi:10.1016/0550-3213(95)00614-1
[arXiv:hep-th/9510169 [hep-th]].



\bibitem{d8}
E.~Bergshoeff, M.~de Roo, M.~B.~Green, G.~Papadopoulos and P.~K.~Townsend,
``Duality of type II 7 branes and 8 branes,''
Nucl. Phys. B \textbf{470} (1996), 113-135
doi:10.1016/0550-3213(96)00171-X
[arXiv:hep-th/9601150 [hep-th]].


\bibitem{romans}
L. J. Romans, “Massive N=2a Supergravity in Ten-Dimensions,” Phys. Lett. B \textbf{169} (1986)
374.


\bibitem{eric}
E. A. Bergshoeff, M. de Roo, S. F. Kerstan, T. Ortin and F. Riccioni, “IIA ten-forms and
the gauge algebras of maximal supergravity theories,” JHEP \textbf{0607} (2006) 018
[hep-th/0602280]

\bibitem{gpdt}
G.~Papadopoulos and D.~Tsimpis,
``The Holonomy of IIB supercovariant connection,''
Class. Quant. Grav. \textbf{20} (2003), L253
doi:10.1088/0264-9381/20/20/103
[arXiv:hep-th/0307127 [hep-th]].








\bibitem{hull}
C.~Hull,
``Holonomy and symmetry in M theory,''
[arXiv:hep-th/0305039 [hep-th]].



\bibitem{duff}
M.~J.~Duff and J.~T.~Liu,
``Hidden space-time symmetries and generalized holonomy in M theory,''
Nucl. Phys. B \textbf{674} (2003), 217-230
doi:10.1016/j.nuclphysb.2003.09.019
[arXiv:hep-th/0303140 [hep-th]].




\bibitem{gpdtx}
G.~Papadopoulos and D.~Tsimpis,
``The Holonomy of the supercovariant connection and Killing spinors,''
JHEP \textbf{07} (2003), 018
doi:10.1088/1126-6708/2003/07/018
[arXiv:hep-th/0306117 [hep-th]].


\bibitem{gpug}
U.~Gran, G.~Papadopoulos and C.~von Schultz,
``Supersymmetric geometries of IIA supergravity I,''
JHEP \textbf{05} (2014), 024
doi:10.1007/JHEP05(2014)024
[arXiv:1401.6900 [hep-th]].


U.~Gran, G.~Papadopoulos and C.~von Schultz,
``Supersymmetric geometries of IIA supergravity II,''
JHEP \textbf{12} (2015), 113
doi:10.1007/JHEP12(2015)113
[arXiv:1508.05006 [hep-th]].

U.~Gran, G.~Papadopoulos and C.~von Schultz,
``Supersymmetric geometries of IIA supergravity III,''
JHEP \textbf{06} (2016), 045
doi:10.1007/JHEP06(2016)045
[arXiv:1602.07934 [hep-th]].


\bibitem{js}
J.~H.~Schwarz,
``Covariant Field Equations of Chiral N=2 D=10 Supergravity,''
Nucl. Phys. B \textbf{226} (1983), 269
doi:10.1016/0550-3213(83)90192-X



\bibitem{Bergshoeff:2005ac}
E.~A.~Bergshoeff, M.~de Roo, S.~F.~Kerstan and F.~Riccioni,
``IIB supergravity revisited,''
JHEP \textbf{08} (2005), 098
doi:10.1088/1126-6708/2005/08/098
[arXiv:hep-th/0506013 [hep-th]].




\bibitem{ugjggp}
U.~Gran, J.~Gutowski and G.~Papadopoulos,
``The Spinorial geometry of supersymmetric IIb backgrounds,''
Class. Quant. Grav. \textbf{22} (2005), 2453-2492
doi:10.1088/0264-9381/22/12/010
[arXiv:hep-th/0501177 [hep-th]].



U.~Gran, J.~Gutowski and G.~Papadopoulos,
``The G(2) spinorial geometry of supersymmetric IIB backgrounds,''
Class. Quant. Grav. \textbf{23} (2006), 143-206
doi:10.1088/0264-9381/23/1/009
[arXiv:hep-th/0505074 [hep-th]].




\bibitem{gggpks}
G.~W.~Gibbons, G.~Papadopoulos and K.~S.~Stelle,
``HKT and OKT geometries on soliton black hole moduli spaces,''
Nucl. Phys. B \textbf{508} (1997), 623-658
doi:10.1016/S0550-3213(97)00599-3
[arXiv:hep-th/9706207 [hep-th]].



\bibitem{kubiznak}
  D.~Kubiznak, H.~Kunduri and Y.~Yasui,
  ``Generalized Killing-Yano equations in D=5 gauged supergravity,''
   Nucl.\ Phys.\  B {\bf 678} (2009) 240
  [arXiv:hep-th/0905.0722].


  \bibitem{houri1}
  T.~Houri, D.~Kubiznak, C.~M.~Warnick and Y.~Yasui,
  ``Generalized hidden symmetries and the Kerr-Sen black hole,''
  JHEP {\bf 1007} (2010) 055
  [arXiv:1004.1032 [hep-th]].



\bibitem{houri2}
  T.~Houri, D.~Kubiznak, C.~Warnick and Y.~Yasui,
  ``Symmetries of the Dirac operator with skew-symmetric torsion,''
  Class.\ Quant.\ Grav.\  {\bf 27} (2010) 185019
  [arXiv:1002.3616 [hep-th]].



\bibitem{kygp}
G.~Papadopoulos,
``Killing-Yano Equations with Torsion, Worldline Actions and G-Structures,''
Class. Quant. Grav. \textbf{29} (2012), 115008
doi:10.1088/0264-9381/29/11/115008
[arXiv:1111.6744 [hep-th]].


\bibitem{howe2}
P.~S.~Howe and U.~Lindstr\"om,
``Some remarks on (super)-conformal Killing-Yano tensors,''
JHEP \textbf{11} (2018), 049
doi:10.1007/JHEP11(2018)049
[arXiv:1808.00583 [hep-th]].


\bibitem{Ggp}
G.~Papadopoulos,
``Killing-Yano equations and G-structures,''
Class. Quant. Grav. \textbf{25} (2008), 105016
doi:10.1088/0264-9381/25/10/105016
[arXiv:0712.0542 [hep-th]].




\bibitem{sat}
O.~P.~Santillan,
``Hidden symmetries and supergravity solutions,''
J. Math. Phys. \textbf{53} (2012), 043509
doi:10.1063/1.3698087
[arXiv:1108.0149 [hep-th]].


\bibitem{thimm}
A.~Thimm, ``Integrable geodesic flows on homogeneous spaces,'' Ergod. Th. $\&$ Dynam. Sys. (1981), {\bf 1}, 495.


















\bibitem{phgp}
P.~S.~Howe and G.~Papadopoulos,
``Holonomy groups and W symmetries,''
Commun. Math. Phys. \textbf{151} (1993), 467-480
doi:10.1007/BF02097022
[arXiv:hep-th/9202036 [hep-th]].


P.~S.~Howe, G.~Papadopoulos and V.~Stojevic,
``Covariantly constant forms on torsionful geometries from world-sheet and spacetime perspectives,''
JHEP \textbf{09} (2010), 100
doi:10.1007/JHEP09(2010)100
[arXiv:1004.2824 [hep-th]].

\bibitem{zumino}
B.~Zumino,
``Supersymmetry and Kahler Manifolds,''
Phys. Lett. B \textbf{87} (1979), 203
doi:10.1016/0370-2693(79)90964-X


\bibitem{lag}
L.~Alvarez-Gaume and D.~Z.~Freedman,
``Geometrical Structure and Ultraviolet Finiteness in the Supersymmetric Sigma Model,''
Commun. Math. Phys. \textbf{80} (1981), 443
doi:10.1007/BF01208280


\bibitem{ghull}
S.~J.~Gates, Jr., C.~M.~Hull and M.~Rocek,
``Twisted Multiplets and New Supersymmetric Nonlinear Sigma Models,''
Nucl. Phys. B \textbf{248} (1984), 157-186
doi:10.1016/0550-3213(84)90592-3



\bibitem{howesierra}
P.~S.~Howe and G.~Sierra,
``Two-Dimensional Supersymmetric Nonlinear Sigma Models with Torsion,''
Phys. Lett. B \textbf{148} (1984), 451-455
doi:10.1016/0370-2693(84)90736-6



\bibitem{howegp2x}
P.~S.~Howe and G.~Papadopoulos,
``Ultraviolet Behavior of Two-dimensional Supersymmetric Nonlinear $\sigma$ Models,''
Nucl. Phys. B \textbf{289} (1987), 264-276
doi:10.1016/0550-3213(87)90380-4


P.~S.~Howe and G.~Papadopoulos,
``Further Remarks on the Geometry of Two-dimensional Nonlinear $\sigma$ Models,''
Class. Quant. Grav. \textbf{5} (1988), 1647-1661
doi:10.1088/0264-9381/5/12/014


\bibitem{hetgpug}
U.~Gran, P.~Lohrmann and G.~Papadopoulos,
``The Spinorial geometry of supersymmetric heterotic string backgrounds,''
JHEP \textbf{02} (2006), 063
doi:10.1088/1126-6708/2006/02/063
[arXiv:hep-th/0510176 [hep-th]].


U.~Gran, G.~Papadopoulos, D.~Roest and P.~Sloane,
``Geometry of all supersymmetric type I backgrounds,''
JHEP \textbf{08} (2007), 074
doi:10.1088/1126-6708/2007/08/074
[arXiv:hep-th/0703143 [hep-th]].



\bibitem{colesgp}
R.~A.~Coles and G.~Papadopoulos,
``The Geometry of the one-dimensional supersymmetric nonlinear sigma models,''
Class. Quant. Grav. \textbf{7} (1990), 427-438
doi:10.1088/0264-9381/7/3/016



\bibitem{uggp}
J.~Gillard, U.~Gran and G.~Papadopoulos,
``The Spinorial geometry of supersymmetric backgrounds,''
Class. Quant. Grav. \textbf{22} (2005), 1033-1076
doi:10.1088/0264-9381/22/6/009
[arXiv:hep-th/0410155 [hep-th]].


\bibitem{review}
U.~Gran, J.~Gutowski and G.~Papadopoulos,
``Classification, geometry and applications of supersymmetric backgrounds,''
Phys. Rept. \textbf{794} (2019), 1-87
doi:10.1016/j.physrep.2018.11.005
[arXiv:1808.07879 [hep-th]].

















\end{thebibliography}
\end{document}